\documentclass[aps,amsmath,amssymb,twocolumn,showpacs,prb]{revtex4}
\usepackage{graphicx,color}
 \graphicspath{{./}{./figures/}}
\usepackage{dcolumn}
\usepackage{bm}

\definecolor{labelkey}{cmyk}{.4,.2,0,0}

\newcommand{\be}{\begin{equation}}
\newcommand{\ee}{\end{equation}}
\newcommand{\bea}{\begin{eqnarray}}
\newcommand{\eea}{\end{eqnarray}}

\newcommand{\nn}{\nonumber }

\newcommand{\fig}[2]{\includegraphics[width=#1]{./figures/#2}}

\begin{document}

\title{More on the Brownian force model: avalanche shapes, tip driven, higher $d$}
\author{Pierre Le Doussal} \affiliation{
Laboratoire de Physique de l'\'Ecole Normale Sup\'erieure, ENS, Universit\'e PSL, CNRS, Sorbonne Universit\'e, Universit\'e de Paris, 75005 Paris, France} 
\date{\today}

\begin{abstract}
The Brownian force model (BFM) is the mean-field model for the avalanches
of an elastic interface slowly driven in a random medium. 
It describes the spatio-temporal statistics of the
velocity field, and, to some extent is analytically tractable.
We extend our previous studies to obtain several 
observables for the BFM with short range elasticity,
related to the local jump sizes $S(x)$ and to 
the avalanche spatial extension in $d=1$, or the avalanche span in $d>1$.
In $d=1$ we consider both driving (i) by an imposed force 
(ii) by an imposed displacement "at the tip" and
obtain in each case the mean spatial shape 
$\langle S(x) \rangle$ at fixed extension, or at fixed seed to edge distance. We find that
near an edge $x_e$, $S(x) \simeq \sigma |x-x_e|^3$ where $\sigma$ has a
universal distribution that we obtain. We also obtain the spatiotemporal shape
near the edge. In $d>1$ we obtain (i) the mean shape $\langle S(x_1,x_\perp) \rangle$
for a fixed span, which exhibits a non-trivial dependence in the transverse distance to the seed $x_\perp$
(ii) the mean shape around a point which has not moved, $\langle S(x) \rangle_{S(0)=0}$, which vanishes at the center 
as $|x|^{b_d}$ with non trivial exponents, $b_1=4$, $b_2=2 \sqrt{2}$ and
$b_3=\frac{1}{2} (\sqrt{17}-1)$. We obtain the probability distributions in any $d$ for 
the maximal radius of an avalanche and the minimal distance of approach to a given point, as well as the probability of not hitting a cone in $d=2$. These results equivalently apply to the continuum limit of some branching diffusions,
as detailed in a companion paper.

\end{abstract}
\maketitle

\begin{widetext}

\medskip

\tableofcontents

\section{Introduction}

\subsection{Definition of the model}

The Brownian force model (BFM) has emerged as the mean-field theory for the space-time statistics of
the instantaneous velocity field in an avalanche, for elastic interfaces driven in a disordered medium
\cite{LeDoussalWiese2011b,LeDoussalWiese2011a,LeDoussalWiese2012a}. It is expected to be an accurate description of this statistics for random media with short range correlations, in the case of interfaces of internal dimension $d$ above or at the upper critical dimension, i.e. $d \geq d_{uc}$.
It is also a starting point to describe models with short range disorder in a dimensional expansions around $d_{uc}$ \cite{LeDoussalWiese2012a,DobrinevskiLeDoussalWiese2014a,ThieryShape,DobrinevskiPhD,ThieryPhD}. Furthermore,
since it is analytically tractable, with non trivial physics, it is also an interesting model of avalanche motion per se, to study in any dimension $d$ \cite{DobrinevskiLeDoussalWiese2011b,ThieryLeDoussalWiese2015,Delorme,GarciaGarcia2017}. It is the natural extension to an elastic manifold of internal dimension $d$, of the celebrated ABBM toy model, which describes the motion of a point driven in a Brownian force landscape, i.e. the $d=0$ limit \cite{AlessandroBeatriceBertottiMontorsi1990,AlessandroBeatriceBertottiMontorsi1990b,
DobrinevskiLeDoussalWiese2011b,Colaiori2008,ZapperiCizeauDurinStanley1998,Records}.

Consider the overdamped 
equation of motion (with friction coefficient $\eta$) for an elastic interface, with internal coordinate $x \in \mathbb{R}^d$,
parameterized by a position field (i.e. height or displacement field) $u(x,t)$ in a quenched random force field $F(u,x)$
\be
\eta \partial_t u(x,t)= \nabla_x^2 u(x,t) + F\!\left(u(x,t),x\right) + m^2(w(x,t)- u(x,t)) \quad , \quad f(x,t)=m^2 w(x,t)
\label{BFMpos1}
\ee
where we focus in this paper on short-ranged elastic forces, modeled by the $d$-dimensional Laplacian $\nabla_x^2$. We have added a confining quadratic potential of curvature $m$, centered at $w(x,t)$, acting as a driving (here and below we use interchangeably $\partial_t u$ or $\dot{u}$ to denote time derivatives). Eq. \eqref{BFMpos1} describes
the "massive" problem, for which the driving force
is $f(x,t)=m^2 w(x,t)$, and below we will also study the related massless problem where one sets $m \to 0$
keeping $f(x,t)$ fixed. Equations of motion such as \eqref{BFMpos1} obey the Middleton property \cite{Middleton1992,middletonMath,RossoPhD}, which ensures that under monotonous forward driving the interface always moves forward, and, e.g. for a uniform driving $w(x,t)=v t$, converges to a unique attractor in configuration space (called the Middleton attractor), independent of the initial condition. 

In the BFM model, the random force field is chosen as a collection of independent one-sided Brownian motions in the $u$ direction
with correlator
\be \label{corrF} 
\overline{F(u,x)F(u',x')}=2 \sigma \delta^d(x-x')\min(u,u')\ .
\ee
This choice makes the BFM more tractable analytically than the case of short-range correlated random force field.

Indeed, taking the temporal derivative $\partial_t$ of Eq.~\eqref{BFMpos1}, and choosing the driving velocity always positive, $\dot w(x,t) \geq 0$, which implies from the Middleton property that $\dot{u}(x,t) \geq 0$ at all $t \geq 0$ if $\dot{u}(x,t=0) \geq 0$, one can show 
\cite{LeDoussalWiese2011a,LeDoussalWiese2012a} 
that \eqref{BFMpos1} implies, for the time evolution of the 
velocity variable $\partial_t u(x,t) \equiv \dot{u}(x,t)$, the following stochastic equation (in the 
Ito sense)\,:
\be
\eta \partial_t \dot{u}(x,t)= \nabla_x^2 \dot{u}(x,t) 
+ \sqrt{2 \sigma \dot{u}(x,t)} \,\xi(x,t) + m^2(\dot{w}(x,t)- \dot{u}(x,t)) \quad , \quad \dot f(x,t)=m^2 \dot w(x,t)
\label{BFMdef1}
\ee
where $\overline{\xi(x,t) \xi(x',t')}=\delta^d(x-x') \delta(t-t')$ is a space time white noise.
The equation \eqref{BFMdef1} can equally well be taken as the definition of the Brownian Force Model (BFM) in space dimension $d$.
The fact that the time evolution of the velocity is Markovian, even in a quenched random medium, is remarkable and results from the properties of the Brownian motion, combined with the Middleton property. The 
details of the correspondence are given in \cite{DobrinevskiLeDoussalWiese2011b,LeDoussalWiese2012a} 
where subtle aspects of the position theory, and its links to the mean-field theory of 
realistic models of interfaces in short-ranged disorder via the Functional Renormalisation Group (FRG),
are  discussed.\\

Here we will study the BFM mainly in dimension $d=1$, with some additional results in dimension
$d=2$ and some for any $d <4$. In fact, even from the solution in $d=1$ one can 
already obtain some (integrated) information for $d>1$. Indeed, the BFM \eqref{BFMdef1} in space dimension $d$ has the property that the velocity field integrated
over $d-d_c$ space dimensions, also satisfies the BFM equation in dimension $d_c$. For instance, 
consider Euclidean $d$-dimensional space, and $\nabla_x^2=\sum_{i=1}^d \partial_{x_i}^2$. 
Then $\dot u(x_1,t) := \int d x_2 \ldots dx_d ~ u(x_1,\ldots x_d,t)$ satisfies exactly the BFM equation \eqref{BFMdef1}
in dimension $d=d_c=1$, in the variable $x=x_1$. Note that there are many possible variants of the BFM (discrete versions, long range elasticity, non-local
in time) but here we study the simplest one \eqref{BFMdef1}.

\subsection{Avalanche observables and driving protocols}

\subsubsection{Finite kick driving, avalanche size distribution, massive and massless units}
\label{subsec:units} 

{\it Kick driving}. Let us define what we want to study. We are interested in the solutions of Eq. \eqref{BFMdef1} 
with an initial condition with zero velocity, $\dot{u}(x,t=0^{-}) =0$, in response to a driving 
function $\dot{f}(x,t)$ which is a kick at time $t=0$, i.e. of the form
\be \label{kick1} 
\dot{f}(x,t) = f(x) ~ \delta(t)  \quad , \quad f(x) = m^2 \delta w(x) \geq 0   
\ee
equivalent to a step $\delta w(x)$ in the driving function $w(x,t)$. This kick, which may be space-dependent, produces an avalanche with $\dot u(x,t \geq 0) \geq 0$,
which starts at time $t=0$ and terminates at a time $t=T$, when all velocities vanish, i.e. 
$\dot u(x,t)=0$ for all $x$. It can be shown that such a duration time $T$ exists for the BFM \eqref{BFMdef1}
with no need to introduce any ad-hoc additional cutoff. If $f(x)=f$ it is called a uniform (or homogeneous) kick. \\

{\it Avalanche size distributions}. For this avalanche one defines the size $S(x)$ of the local jump (local avalanche size) 
and the total size $S$
\bea \label{sizes}
S(x) =\int_{0}^{\infty} dt \, \dot{u}(x,t)  \quad , \quad S = \int d^d x \, S(x)
\eea 
We denote symbolically $P_{\{f(x)\}}[\{S(x)\}] \prod_x dS(x)$ the probability measure for
the field $S(x)$ observed after the kick \eqref{kick1}. There is an {\it exact} formula for that
probability measure for the BFM, Eq. (49) of Ref. \cite{ThieryLeDoussalWiese2015}, 
not reproduced here, as not so useful for our purpose. Let us recall instead the 
probability distribution function (PDF) of the total size $S$
\bea \label{PDFS} 
P_{\{f(x)\}}(S) = \frac{\int d^d x f(x)}{2 \sqrt{\pi} m^2 S_m^{1/2} S^{3/2}} e^{- \frac{(S- \frac{1}{m^2} \int d^d x f(x))^2}{4 S S_m} }
\eea 

{\it Units, massive case}. Here in Eq. \eqref{PDFS} and below, except when explicitly mentionned, 
we use the natural units for the massive problem, i.e. we express 
lengths in units of $x_m=1/m$, total size
$S$ in units of $S_m$, and time in units of $\tau_m$ with
\be
S_m = \frac{\sigma}{ m^4} \ , \qquad  \tau_m = \frac{\eta}{m^2}\ .
\label{units}
\ee
and $w(x) \sim u(x,t) \sim S(x)$ in units $\sigma m^{-\zeta}$ with $\zeta=4-d$,
which is equivalent to setting $m=\sigma=\eta = 1$. We can see from \eqref{PDFS} 
that $S_m$ is the large-size cutoff for avalanches. In the BFM the disorder is
scale invariant, hence the only small scale cutoff is the size of the driving, i.e. from \eqref{PDFS} as $S \to 0$ we see that the smallest
avalanches are of typical size 
$S_{\rm min} \simeq (\int d^d x f(x))^2/\sigma$. In these units Eq. \eqref{PDFS} reads
$P_{\{f(x)\}}(S) = \frac{\int d^d x f(x)}{2 \sqrt{\pi} S^{3/2}} e^{- \frac{(S-  \int d^d x f(x))^2}{4 S} }$.
\\

{\it Units, massless limit}. In the massless limit $m \to 0$ the driving force $f(x,t)$ is kept fixed.
Hence the equations of motion (\ref{BFMpos1}) and (\ref{BFMdef1}) are valid with
\be
\begin{split}
m^2(w(x,t)- u(x,t)) &\to f(x,t) \\
m^2(\dot{w}(x,t)- \dot{u}(x,t)) &\to \dot f(x,t)\ .
\end{split}
\ee
In that limit we see that Eq. \eqref{PDFS} becomes
\be
P_{\{f(x)\}}(S) = \frac{\int d^d x f(x)}{2 \sqrt{\pi} \sigma^{1/2} S^{3/2}} e^{- \frac{(\int d^d x f(x))^2}{4 \sigma S } }
\ee
We can define time in units of $\eta$
and $u \sim w$ in units of $\sigma$, which is equivalent to setting $\eta=\sigma=1$. The results still have an unfixed 
dimension of length. There is no more large scale cutoff size, or cutoff length for the avalanches.
In some cases one can introduce a finite system size $L$ to act as a cutoff and to construct
dimensionless quantities.

\subsubsection{Infinitesimal kick driving, avalanche seed, quasi-static avalanche densities}

The limit of a small kick amplitude $f(x)$ is of special importance to define (i) the
seed of an avalanche (ii) quasi-static avalanches (iii) avalanche densities. 
As can be seen on \eqref{PDFS},
in that limit one has, for any fixed $S>0$ (in the massive case)
\bea \label{expansion1} 
P_{\{f(x)\}}(S)  =  \int d^dx f(x) \rho_x(S) + O(f^2)   \quad , \quad \rho_x(S) =  
\frac{1}{2 \sqrt{\pi} S^{3/2}} e^{- S/4} \quad , \quad \rho_x(S) = \frac{\delta P_{\{f(x)\}}(S) }{\delta f(x)}|_{f=0} 
\eea 
where $\rho_x(S)$ is the so-called local density associated to $S$
(it has unit $1/(w S_m)$, i.e. it is a density per unit length - along $u$).
Note that as $f(x) \to 0$ the weight of the PDF in \eqref{PDFS} 
concentrates on very small
avalanches,
$S \sim [\int d^d x f(x)]^2$, so that \eqref{expansion1} represents the rare 
larger, i.e. $S = O(1)$, avalanches. Densities $\rho_x[{\cal O}]$ for many other 
avalanche observables ${\cal O}$, e.g. ${\cal O} = S(x_0)$, or the avalanche
spatial extension (see below), can be similarly defined  (by the derivative of the probability
as in \eqref{expansion1}). We see from \eqref{expansion1} 
that $\rho_{x_s}(S)$ is the density corresponding to performing a local driving kick
$f(x) = f \delta^d(x-x_s)$. It turns out that
$\rho_x[{\cal O}]$ is also {\it a density associated to avalanches which have their seed at $x$.}
The {\it seed of an avalanche} is the location of the first point which starts moving.
If the driving kick is local, i.e. $f(x) = f \delta^d(x-x_s)$, then the seed is at $x_s$,
and the avalanches spreads from there. Formula \eqref{expansion1} says that
for {\it an extended kick of very small amplitude} $f(x)$, the avalanche will start from
$x=x_s$ with a probability proportional to $f(x_s)$. In particular, consider a uniform
driving $f(x)=f$. In the limit $f \to 0^+$, at fixed interface size $L$, one can reach a "single 
avalanche regime" for avalanches of size $O(1)$
\footnote{Note that in the BFM, even a small uniform kick $f$ always produces 
many very small avalanches of order $O(f^2)$, so the meaning of "single avalanche regime" depends on
the scale, and here we focus on avalanches which are $O(1)$ as $f \to 0^+$.}.
The distribution of location of the seed $x_s$ for this avalanche is uniform in space $x$.
By contrast, for larger kicks $f(x)$, the terms of order $f^2$ and higher in \eqref{expansion1} 
cannot be neglected, and can be interpreted as "several successive avalanches".
One can ask about the correlations of such avalanches. 
As described in details
in \cite{ThieryLeDoussalWiese2015,ThieryDynamicCorrelations}, the avalanches in the BFM are uncorrelated
and form a Levy-type process
(see also \cite{LeDoussalWiese2011b} for a static version, and \cite{ThieryDynamicCorrelations,ThieryStaticCorrelations} for correlations in
short range models). 

Another protocol, a priori different from \eqref{kick1}, is to drive the interface with a fixed uniform driving velocity $v$,
$w(x,t)=v t$, in presence of a mass $m^2>0$. 
Considering the unique (Middleton) steady state, and taking the limit $v \to 0^+$ 
leads to the standard quasi-static avalanches. 
One can show that these have the same statistics (the same densities) as the ones produced by the infinitesimal uniform kick $f(x)=f$ in the limit $f \to 0^+$. \\

None of the above made any reference to the position field $u(x,t)$ of the interface, which could thus
equally well be defined as $u(x,t)-u(x,t=0)=\int_0^t dt' \dot u(x,t')$. There is however an interpretation 
\cite{DobrinevskiLeDoussalWiese2011b,LeDoussalWiese2012a,ThieryLeDoussalWiese2015}
in the position theory, described by Eqs. \eqref{BFMpos1}-\eqref{corrF}, of the avalanches following a kick \eqref{kick1} with the initial condition
$\dot{u}(x,t=0^{-}) =0$. At $t<0$, the interface is driven forward for a sufficient amount of time, with a driving which stops and converges to $w(x,t=0^-)=w_0(x)$, in such a way that at $t=0^-$ the interface is at rest, 
in the (unique) Middleton metastable state 
\footnote{It can be constructed as the rightmost metastable state of the Hamiltonian 
${\cal H}[u,w] = \int d^d x [ \frac{1}{2} (\nabla_x u)^2 + V(u(x),x) + \frac{m^2}{2} (u(x)-w(x))^2]$,
where $F(u,x)=-\partial_u V(u,x)$, 
which leads to the equation of motion \eqref{BFMpos1}, settting $w(x)=w_0(x)$.}
corresponding to $w_0(x)$, denoted $u(x,t=0^-)=u(x;w_0(x))$. 
At $t=0$, the center of the confining potential 
is increased abruptly from $w_0(x)$ to $w(x,t=0^+)= w_0(x)+ \delta w(x)$, with $\delta w(x)=f(x)/m^2$, and remains constant for all $t>0$. As a consequence, the interface moves forward (since $\delta w(x) \geq 0$) up to a new metastable state of the Middleton attractor, specifically $\lim_{t\to +\infty} u(x,t)=u(x;w_0(x)+\delta w(x))$.

Note that the mass $m^2>0$ is important to guarantee a steady state, and well defined statistics for
all avalanches (small and large). On the other hand, one can verify that the
statistics of small avalanches $S \ll S_m$, is well described by the massless limit, and universal. 
This is why we often focus on this limit, where calculations are simpler.

\subsubsection{Avalanche spatial support and extension: recall of previous results} \label{subsubsec:extension}

For the BFM with short range elasticity, as in \eqref{BFMdef1}, it is possible to define the {\it support of the avalanche}
(with no need for any ad-hoc additional cutoff) as the (connected) set of points which have moved during the avalanche
\be
\Omega = \{x \in \mathbb{R}^d \quad s.t. \quad S(x) >0\} 
\ee
In the complementary set, one has $S(x)=0$. It is easy to show that following a 
kick $\dot f(x,t)=f(x) \delta(t)$, all points where $f(x)>0$ will move,
hence $\Omega$ contains the support of $f(x)$.\\

\begin{figure}[h]
\centerline{
   \fig{.6\textwidth}{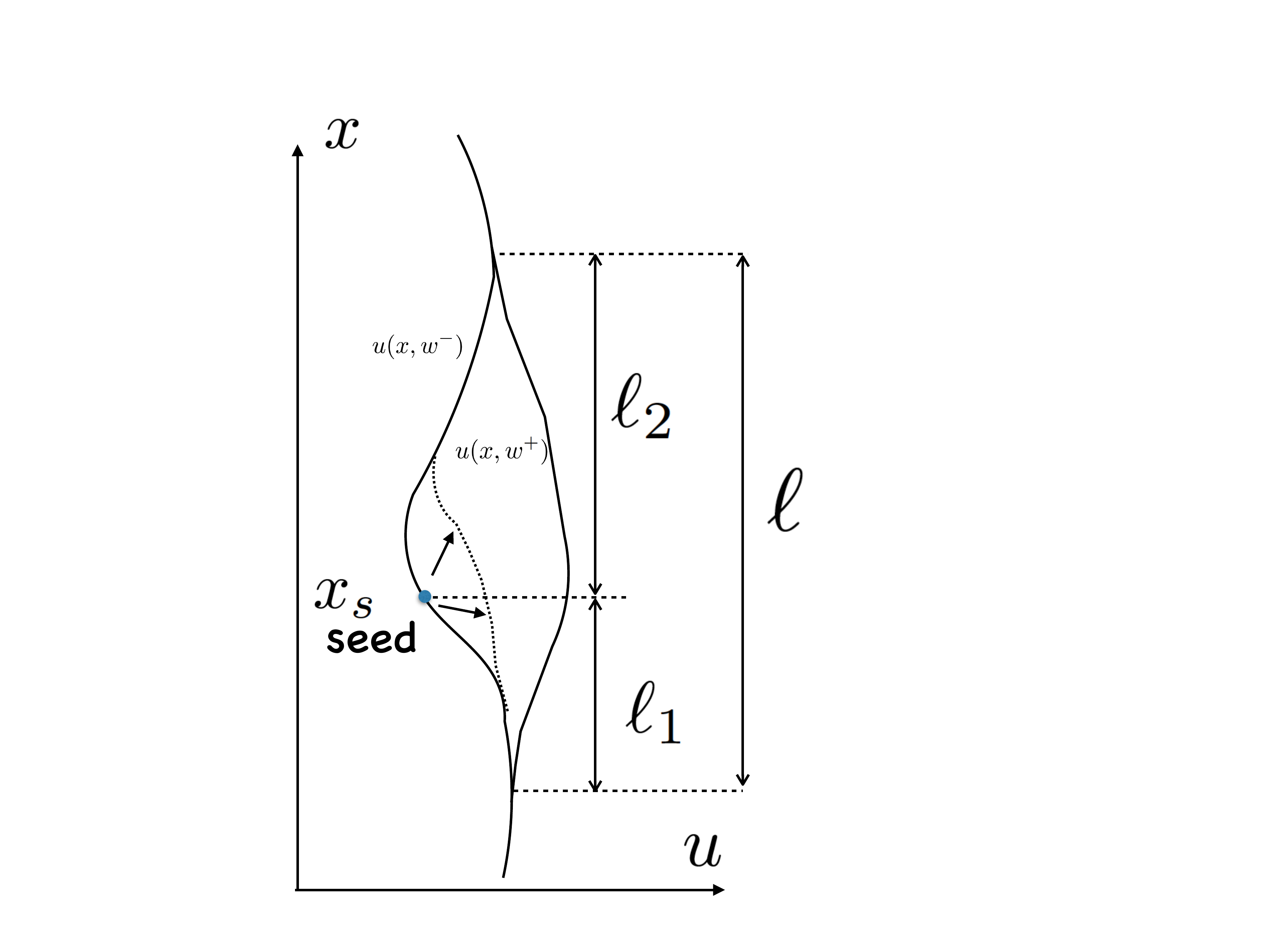} \!\!\!\!\!\!\!\!\!\!\!\!  \!\!\!\!\!\!\!\!\!\!\!\! \!\!\!\!\!\!\!\!\!\!\!\! 
   \!\!\!\!\!\!\!\!\!\!\!\! \!\!\!\!\!\!\!\!\!\!\!\! \fig{.6\textwidth}{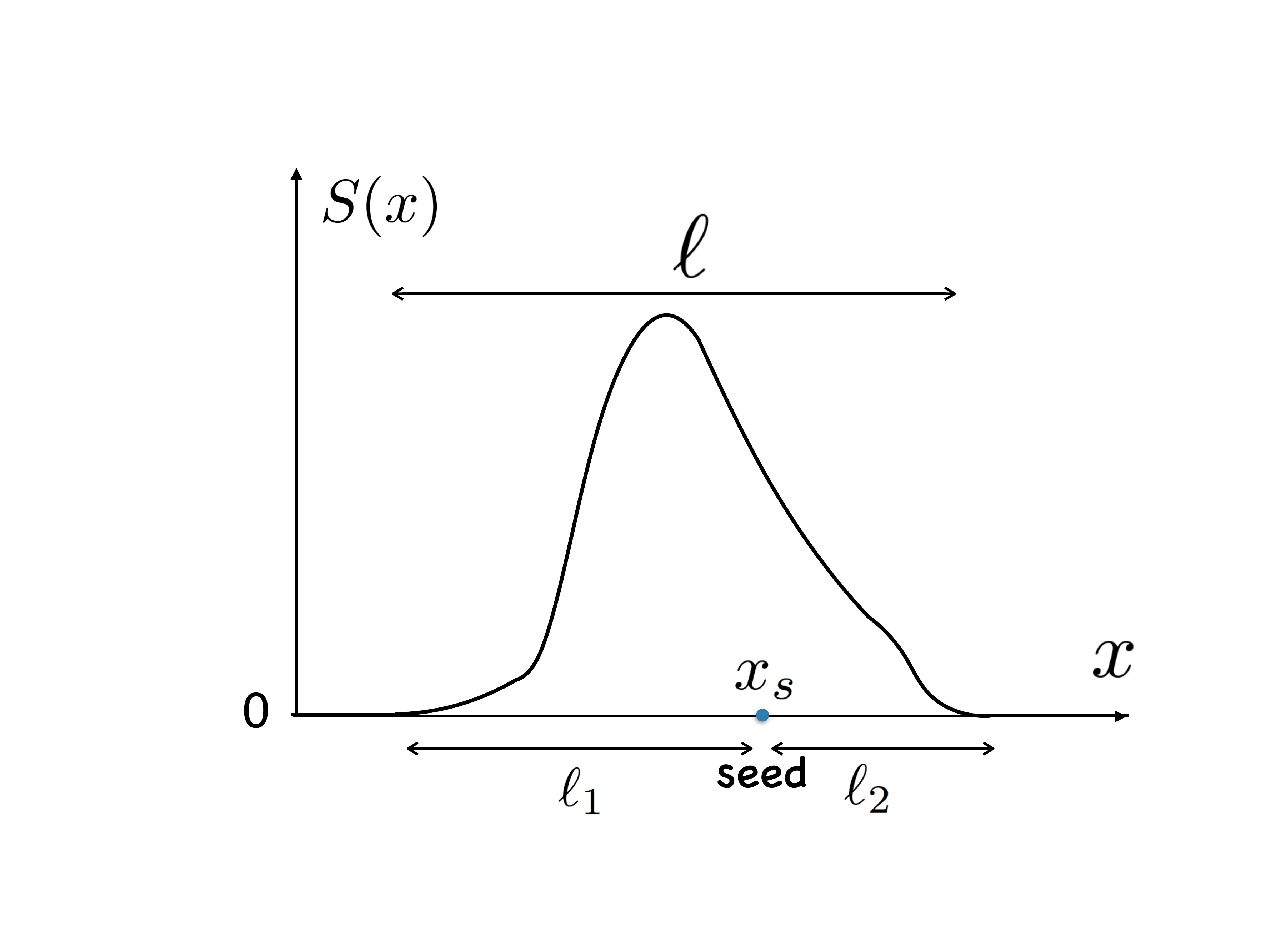}
   }
 \caption{Left: schematic representation of an avalanche for a 1 dimensional interface ($d=1$), between metastable configurations $u(x;w^-)= u(x,t=0)$ and
 $u(x;w^+)=u(x,t=+\infty)$. The avalanche starts at the seed point $x_s$ (the dotted curve represents an
 intermediate time configuration). The set of points which move during the avalanche has a finite extension $\ell$. The distance of the seed from either edges of the avalanche are respectively $\ell_2$ (upper) and $\ell_1$ (lower) with $\ell_2+\ell_1=\ell$. Such an avalanche has aspect ratio $p=\ell_2/(\ell_1+\ell_2)$. Right: resulting total jump size $S(x)= \int_0^{+\infty} dt \dot u(x,t)$ of point $x$ during the avalanche. Beyond the edges it vanishes strictly.}
\label{fig:extensions}
\end{figure}

Consider now a local kick at $x=x_s$, $f(x) = f \delta(x-x_s)$. In $d=1$, for the model
\eqref{BFMdef1}, we have shown \cite{Delorme,ThieryLeDoussalWiese2015} (see also below) that $\Omega$ is an 
interval containing the seed $x_s$, i.e. there are two edges beyond which $S(x)=0$ strictly. 
Hence we can further define the {\it avalanche spatial extension} (also called length) $\ell$ as the length of that interval.
It is also interesting to define the distances $\ell_1,\ell_2$  of each edge to the seed of the avalanche,
see Fig. \ref{fig:extensions}.
That is, for an avalanche with its seed at $x_s$, the lower (respectively upper) extension
$\ell_1$ (respectively $\ell_2$) is defined such that $\Omega=[x_s- \ell_1,x_s+\ell_2]$, and the total extension is 
then $\ell=\ell_1+\ell_2$. Finally one also defines the {\it aspect ratio} 
\bea \label{aspectdef} 
p= \frac{\ell_2}{\ell_1+\ell_2}
\eea
which measures the location of the seed relative to the total length, i.e. $p \in [0,1]$.
The natural units for $\ell$ is $1/m$ in the massive case. In a previous work \cite{Delorme} we have calculated in $d=1$ the joint cumulative distribution function
(CDF) of $\ell_1,\ell_2$ and obtained
\be \label{CFDl1l2}
 {\rm Prob}(\ell_1< l_1, \ell_2< l_2) 
= e^{- 6 f {\cal P}_0(p)/l^2 }  \quad , \quad 
p= \frac{l_2}{l_1+l_2} \quad , \quad  l = l_1+l_2
\ee
where ${\cal P}_0(z)$ is an elliptic Weirstrass function whose definition is as 
follows
\bea \label{defP0}
{\cal P}_0(z) := {\cal P}(z; g_2=0, g_3=g_3^0:= \frac{\Gamma(1/3)^{18}}{(2 \pi)^6} ) 
\eea 
where ${\cal P}(z; g_2, g_3)$ is the standard elliptic Weirstrass function, doubly periodic
in the complex plane, with invariants
$g_2,g_3$, see Appendix \ref{app:W} for definitions. Here the value of $g_3$ is such that
the real period is unity, i.e. $2 \Omega(g_2=0,g_3^0)=1$, see \eqref{halfper}. Its most important properties
for us are, in addition to periodicity ${\cal P}_0(z+1)={\cal P}_0(z)$,
\bea \label{propP0} 
{\cal P}_0(1-z)={\cal P}_0(z)  \quad , \quad {\cal P}_0(z) \simeq_{z \to 0} \frac{1}{z^2} 
\eea 
Taking the limit of small $f$, we obtained from \eqref{CFDl1l2} the avalanche densities for the couple $(p, \ell)$, and, by
integration over the aspect ratio $p$, for the total length $\ell$, as
\be
 \rho(\ell,p)= \frac{8 \pi \sqrt{3} \tilde P(p)}{\ell^3} \quad , \quad \rho(\ell) = \frac{8 \pi \sqrt{3}}{\ell^3}
\ee
where $\tilde P(p)$ is given below, see \eqref{rhoellp}, and is the PDF of the aspect ratio, i.e. it is normalized to unity $\int_0^1 dp \tilde P(p)=1$. Here, for the massless
case, we see that it is independent of the total length (note that in \cite{Delorme} we defined the aspect ratio as $k=(\ell_1-\ell_2)/(2 (\ell_1+\ell_2))$ and the function $R(k)=8 \pi \sqrt{3} \tilde P(p)$). 
The above results were obtained for the massless model (see units above in section \ref{subsec:units}). This corresponds
to the limit $\ell \ll 1$, that is $\ell \ll \ell_m=1/m$, in the massive model. Note that we
have assumed that $\ell \ll L$, where $L$ is the systeme size (or $L \gg \ell_m$ in the massive case),
so that translational invariance in $x$ holds. Finally let us recall that the obtained power law, $\rho(\ell) \sim \ell^{-\kappa}$, with $\kappa=3$ for the extension exponent, is consistent with the general scaling relation 
$\kappa=d+\zeta-1$, with $d+\zeta=4$ for the BFM. \\

The above results for a local kick, $f(x)=f \delta(x-x_s)$, can be extended to any
kick $f(x)$ with bounded support, see section \ref{sec:1boundarymassless}. 
It is important to note that for a {\it uniform kick} $f(x)=f$, and more generally for uniform driving,
one cannot strictly define the avalanche
extension since all points move. However, in the limit of a small $f \to 0^+$, one can access the "single avalanche regime" discussed above, and distinguish
points which move by $O(1)$, and points which move by $O(f^2)$. In that sense one
can still define the extension of a single avalanche. The position
of the seed $x_s$ is then a random variable, equiprobable along the interface
(for $f(x)=f$), and once it is chosen, the above results apply
w.r.t. $x_s$. From the above discussion, this will also hold for quasi-static avalanches
obtained by driving at fixed velocity
in the limit $v\to 0^+$. 

Studying the $d=1$ problem gives some information about the "span" of the
avalanches in the $d$-dimensional BFM. Indeed, in any dimension $d$,
the event "the support is contained in the band $[x_s- \ell_1,x_s+\ell_2] \times \mathbb{R}^{d-1}$"
has the same probability as for $d=1$. Indeed it is easy to see that the statistics of the field $S(x_1)$
defined as $S(x_1):= \int dx_2.. dx_d S(x_1,..x_d)$, 
under the driving $f(x_1)=\int dx_2.. dx_d f(x_1,..x_d)$ in the $d$-dimensional BFM,
is the same as the one of the field $S(x_1)$ under driving $f(x_1)$ in the $d=1$ BFM.
Now clearly if $S(x_1)=0$ is some interval, then all $S(x_1,..x_d)=0$ for all
$x_2,..x_d$ and $x_1$ in that interval. We will return to this below. 

\subsection{Avalanches upon driving by the tip}
\label{sec:intro-tip} 

Above we mentionned the case of a local kick when the driving is applied only
at a point $f(x) = f \delta(x)$. This is driving with an {\it imposed force}. It is also
interesting to study, 
what happens in the case where the displacement of
one point of the interface (or one region) is imposed, i.e. {\it imposed displacement}. This case was
studied numerically in Ref. \cite{Tip}. 
This can be achieved
in the following variant of the equation of motion 
\be
\eta \partial_t u(x,t)= \nabla_x^2 u(x,t) - m^2 u(x,t) + F\left(u(x,t),x\right) + m_0^2 \delta(x) (w(t)- u(x,t))
\label{BFMpos}
\ee
with $x \in \mathbb{R}$, where $F(u,x)$ is again a Brownian force landscape,
leading to
\be
\eta \partial_t \dot{u}(x,t) = \nabla_x^2 \dot{u}(x,t) - m^2 u(x,t)
+ \sqrt{2 \sigma \dot{u}(x,t)} \xi(x,t) + m_0^2 \delta(x) (\dot{w}(t)- \dot{u}(0,t))
\label{BFMdefTip}
\ee
Here we have written the model in $d=1$ and we drive by a point at $x=0$. Then $1/m_0^2$ has
dimension of length. We can either study this massive model (i.e. with $m^2>0$) 
and use the massive units \eqref{units}, or study the massless model setting $m^2=0$. 
In the limit $m_0^2 \to +\infty$ the position of the point $x=0$ is imposed.\\

Finally, one can generalize the model to any dimension $d$, and drive the interface by a subspace of dimension $d_w<d$,
i.e. replace $\delta(x) \to \prod_{i=1}^{d-d_w} \delta(x_i)$ in \eqref{BFMpos}. Again, if we look at a $d$-dimensional
interface, driven by the $d_w=d-1$ dimensional subspace $x_1=0$, then the avalanche sizes
integrated over the coordinates $x_{2},..,x_d$ will be described by the $d=1$ model.

\section{Aim of the present paper, and summary of main results} 

The aim of the present paper is to pursue the investigations started in 
\cite{Delorme,ThieryLeDoussalWiese2015}, and also in \cite{ThieryShape} 
where we calculated the mean shape centered on the seed,
and study many more more observables, not calculated there.
A large part of the results are for $d=1$, but there are also results for $d>1$.
\\

{\it Remark: relations to branching processes and to the super-Brownian motion}.
As detailed in a companion paper \cite{LedouSBM2022}, the BFM is equivalent to
a scaled limit of the branching Brownian motion which is a measure-valued
process called the super-Brownian motion (SBM),
studied in the probability community. There is a one to one correspondence
between the observables in each system, for instance the velocity field of the BFM,
$\dot u(x,t)$, maps onto the density of the SBM measure $\rho(x,t)$ (the limiting density of a large number of 
branching random walks). Although the two fields have developed separately, and the
interest is a bit different in each field, there is some overlap reviewed in \cite{LedouSBM2022}.
The many new results in this paper can thus be easily translated in terms of 
branching diffusions. Some of that translation is discussed in \cite{LedouSBM2022}. 
\\

{\it Note added}. This work was started around 2015 and most results were obtained
around that time. It is independent, and in part complementary, of Ref. \cite{ZhuWiese2017} 
which appeared while this work was almost completed.
That paper also addresses the mean shape at fixed extension, $\langle S(x) \rangle_\ell$, by a slightly different
method, but not the many other observables obtained here. 
\\

We now present our main results. 

\subsection{Density of local jump $S(x)$, versus the distance to the seed at $x=x_s$} 

Let us first describe our results in $d=1$ in the massless case, obtained in section \ref{sec:densitylocal}. For a point $x$ at a distance $|x-x_s|$ from the seed, there is a finite probability $p_f = e^{- 6 f/(x-x_s)^2}$ 
that no motion occured. There is thus a probability $1-e^{- 6 f/(x-x_s)^2} \simeq 6 f/(x-x_s)^2$ (at small $f$), that some motion occured at $x$.
It is possible to calculate the density of the local jump size $S_0 \equiv S(x)$, 
as a function of the distance $|x-x_s|$ to the seed. We obtain the density (and its normalization)
\bea \label{cross1} 
\rho_{x_s}(S_0) = \frac{1}{S_0^{5/3}} \phi(\frac{S_0}{|x-x_s|^3})  \quad , \quad \int_{S_0>0} dS_0 \rho_{x_s}(S_0) = \frac{6}{(x-x_s)^2} 
\eea 
in terms of the scaling function
\bea
\phi(s) := - 12 \times 3^{2/3} s \int_0^{+\infty} dt \, t^2 e^{  - 2 \times 3^{2/3} t s^{1/3}} {\rm Ai}'(t)
\eea 
which also admits the explicit expression \eqref{explicitphi} in terms of exponential-integral functions.
It vanishes linearly as $\phi(s) \simeq \frac{12 \sqrt{3} \times 3^{1/3} \Gamma(2/3)}{\pi} s$ at
small $s$, and goes to a constant $\phi(+\infty) = \frac{1}{3^{2/3} \Gamma(\frac{1}{3})}$
at large $s$. It matches previously known results, i.e. the density at the seed \cite{Delorme}, and the total density \cite{LeDoussalWiese2012a} (with no conditioning to the seed) $\rho(S_0) = \int dx_s \rho_{x_s}(S_0)$
\be
\rho_{x_s}(S_0=S(x_s)) = \frac{1}{3^{2/3} \Gamma(\frac{1}{3}) S_0^{5/3}} \quad , \quad \rho(S_0) =  \frac{2}{{3}^{1/3} \Gamma \left(\frac{2}{3}\right) S_0^{4/3}}
\ee

The result \eqref{cross1} describes the crossover which occurs
for $S_0 \sim |x-x_s|^3$, consistent with the scaling $S_0 \sim \ell^3$ between typical local
jump and typical avalanche extension. As the distance to the seed increases
(equivalently as $S_0$ decreases),
the behavior of the jump density changes from $\rho_{x_s}(S_0) \sim S_0^{-5/3}$ for $S_0 \gg |x-x_s|^3$,
valid within the typical avalanche extension, to 
$\rho_{x_s}(S_0) \sim S_0^{-2/3} |x-x_s|^{-3}$ for $S_0 \ll |x-x_s|^3$
valid outside the typical avalanche extension.\\

We have also calculated the effect of a mass in section \ref{subsec:massivePS0}. In that case, setting $x_s=0$ for notational simplicity, we obtain, in the massive units (we recall that here and above $S_0=S(x)$)
\bea
&& \rho_{x_s=0}(S_0) = \frac{1}{S_0^{5/3}} \frac{4 e^{-|x|}}{(1+e^{-|x|})^2} 
\phi(\frac{S_0}{(\frac{2(1- e^{-|x|})}{1+ e^{-|x|}})^3}, 3^{1/3} S_0^{2/3})
\eea 
where the scaling function $\phi(s,s_0)$ is given in \eqref{phiss0h}. Spatial dependence 
controlled by the mass is now exponential at large distance instead of power law. 

\subsection{Spatial shape: mean shape and fluctuations}

The spatial shape of an avalanche is encoded in the random function $S(x)$ as a function of $x$, and represents 
a fluctuating profile. To define the "mean shape", and study its fluctuations, 
one first needs to center it. There are
several ways to do it, some being easier to handle for calculations. 
One, easy to extract from experimental data, amounts to center it on its
center of mass. An easier one from the point of view of theory is to center it on its seed $x_s$.
In numerics, and sometimes in experiments, it is possible to determine the seed of each avalanche.
In Ref. \cite{ThieryShape} we calculated the mean shape centered on the seed,
and conditioned to its total size $S$ (i.e at fixed fixed total size),
that is $\langle S(x_s + x) \rangle_S$ as a function of $x$, 
in the BFM for any $d$.

We restrict here to $d=1$. Our first set of results concern the mean shape conditioned on the distance, $\ell_2$, from
the seed $x_s$ to one of the edges of the avalanche, see Fig. \ref{fig:extensions}. For a uniform kick we can
then average over the position of the seed and obtain the absolute mean shape centered on one of the edges. Second, we obtain the mean shape of the
avalanche conditioned on its total extension $\ell$. Finally we can also condition on
the aspect ratio of the avalanche, i.e. the parameter $p=\ell_2/(\ell_1+\ell_2)$ defined in \eqref{CFDl1l2}.
We also show that the fluctuations of the shape are universal near the edge and
we calculate the associated probability distribution. Our results are for 
the massless case, unless stated otherwise.

\subsubsection{Mean spatial shape $\langle S(x) \rangle_{\ell_2}$ for fixed edge-to-seed
distance $\ell_2$}
\label{subsec:cumintro} 

The PDF of the distance $\ell_2$ from the seed to one of the edges of the avalanche, see Fig. \ref{fig:extensions}, is given by
\bea
P(\ell_2)  = \frac{12 f}{\ell_2^3} e^{- 6 f/\ell_2^2}  \label{PDFl22} 
\eea 
In practice, it can be read in two ways: either one knows the position of the seed and one
predicts the statistics of the position of one of the two edges, or, for a uniform driving, one measures the position of one edge and one predicts the statistics of the position of the seed (in the single avalanche regime, as discussed above). \\

We can first ask for the mean shape $\langle S(x) \rangle_{\ell_2}$ conditionned on $\ell_2$, i.e. at fixed $\ell_2$. With no loss of generality, let us denote $x_s=0$ the position of the seed, the edge being at $x=\ell_2>0$. The calculation is elementary and amounts to solve a 1D Schrodinger problem in a $1/x^2$ potential. It is performed in Section \ref{subsec:conditioned}. We find that 
if $x$ is between the seed and the edge, the mean shape depends only on the distance to the edge
$\ell_2-x$
\be \label{resms} 
 \langle S(x) \rangle_{\ell_2} 
= \frac{1}{21} (\ell_2-x)^3 
- \frac{1}{28} \frac{(\ell_2-x)^4}{\ell_2}  + \frac{f}{7} \frac{(\ell_2-x)^4}{\ell_2^3}   \quad , \quad x_s=0 < x < \ell_2
\ee
hence at fixed $f$ is vanishes with a cubic law near the edge, $\langle S(x) \rangle_{\ell_2}  \simeq  \frac{1}{21}  (\ell_2-x)^3$,
while in the large $f$ limit the apparent exponent is $4$. If $x$ is on the other side of the seed
the mean shape has a slow, power law decay with exponent $-3$, due to the contribution of large 
avalanches
\be \label{resms2} 
\langle S(x) \rangle_{\ell_2} = 
\frac{1}{21} \frac{(\ell_2)^6}{(\ell_2-x)^3} - \frac{1}{28} \frac{(\ell_2)^7}{(\ell_2-x)^4} 
+ \frac{f}{7} \frac{\ell_2^4}{(\ell_2-x)^3} 
\quad , \quad x < x_s=0 < \ell_2
\ee 
Note that this mean shape is continuous but has a derivative jump $-f$ at the position of the seed $x=x_s$.
For $f \to 0^+$ it remains only a jump in the third derivative, as a signature of the seed.

It is interesting to display the cumulative conditioned version, i.e. the mean shape
conditioned on the event $\ell_2<l_2$, for which it is easier to obtain good statistics in numerics,
and which has a rather compact expression (with $x<l_2$, $l_2>0$)
\bea
\langle S(x) \rangle_{\ell_2<l_2}  = 
\frac{f}{7} \frac{(l_2-x)^4}{l_2^3}  \theta(x> x_s=0) + 
\frac{f}{7} \frac{l_2^4}{(l_2-x)^3}  \theta(x < x_s=0) \label{condmeanint2} 
\eea 
\\
One can also ask about the mean shape, denoted $\bar S(y)$, obtained by centering
the avalanche on the edge (defined as $y=0$), and averaging uniformly over the position of the seed.
This amounts to average $\langle S(x=\ell_2-y) \rangle_{\ell_2}$ over $P(\ell_2)$.
The result is that $\bar S(y) = f^{3/2} g(y/f^{1/2})$ where $g(z)$ is obtained in \eqref{ggz}. It has the universal cubic behavior $g(z) \simeq \frac{z^3}{21}$ for small $z$ and a linear one 
$g(z) \simeq \frac{z}{5}$ for large $z$. This is understood as follows. To 
produce motion at $y$, the avalanche must have an extension larger than $y$.
For small $f$, typical avalanches are small, of extensions $\sim f^{1/2}$, the small scale cutoff.
The cubic behavior is then visible only for distance to
the edge smaller than the extension of these typical avalanches $y \ll f^{1/2}$. 
For larger distances $y \gg f^{1/2}$, only a fraction of order $1/y^2$ of the
avalanche have extension larger than $y$, which upon averaging transforms
the cubic into linear behavior.

Other results are obtained in Section \ref{subsec:JointSl2}, such as the distribution
of the total avalanche size $S$ conditioned on $\ell_2$, which is found to decay as 
$P(S|\ell_2) \sim S^{-5/2}$ for large $S \gg \ell_2^4$. 

\subsubsection{Universal fluctuations of the shape near the edge}
\label{subsec:fluctusumm} 

Since $S(x)$ is a random profile, we can ask how the higher moments of $S(x)$ (here conditioned
to a fixed edge to seed distance $\ell_2$) vanish near the edge as a function of $\ell_2-x$. We have also
obtained these {\it fluctuations} of the shape near the edge in Section \ref{subsec:highermoments}. To leading order in the distance to
the edge, we show that one can write
\bea \label{form0} 
S(x) \simeq (\ell_2-x)^3 \sigma 
\eea 
where $\sigma$ is a random variable of $O(1)$, a form which accounts for the leading asymptotics for all the moments near the edge. We find that the PDF, $p(\sigma)$, of the random variable $\sigma$
is given by its Laplace transform as
\be \label{genfunct0} 
 \int_0^{+\infty}  d\sigma \, p(\sigma) e^{\lambda \sigma}  = \frac{1}{A(\lambda)} - \frac{3  \lambda A'(\lambda)}{A(\lambda)^2} 
\ee
where the function $A(\lambda)$ is solution of the self-consistent equation
\bea \label{cond40} 
A^2 = {\cal P}(1 ; 0, - \frac{\lambda^2}{36} - \frac{2}{3} \lambda A^3) \quad \Leftrightarrow \quad A=A(\lambda) 
\eea 
From this, it is easy to obtain the moments
up to a very high order. The lowest order ones are
\bea \label{sigmom2} 
\langle \sigma \rangle = \frac{1}{21} \quad , \quad 
\langle \sigma^2 \rangle = \frac{1}{273} \quad , \quad
\langle \sigma^2 \rangle^c = \frac{8}{5733} \quad , \quad
\langle \sigma^3 \rangle = \frac{265}{677768} \quad , \quad
\langle \sigma^4 \rangle = \frac{1223}{22874670} 
\eea 
Hence we find that the amplitude of the cubic behavior fluctuates from avalanche
to avalanche, the factor $\frac{1}{21}$ obtained above in \eqref{resms} is only its mean
value. Note for instance the dimensionless ratio for $x \to \ell_2$, i.e. $x$ near the edge 
\bea
\frac{\langle S(x)^2 \rangle}{\langle S(x) \rangle^2} \simeq \frac{\langle \sigma^2 \rangle}{\langle \sigma \rangle^2}= \frac{21}{13} = 1.61538 \label{agaaga} 
\eea 
We have also obtained the exponential tail of $p(\sigma)$ at large $\sigma$ in \eqref{psig}.\\

{\bf Universality}. It turns out that the form \eqref{form0} at the edge of an avalanche,
and the result \eqref{genfunct0}-\eqref{cond40} 
for $p(\sigma)$ are {\it universal} (for the BFM with SR elasticity). For instance, 
in Section \ref{subsec:shapetip} and \ref{subsec:shapetip2} we calculate exactly the
mean shape and "second shape" $\langle S(x)^2 \rangle_{\ell_2}$ for a completely different
type of avalanches (i.e. driven by the tip) and the result near the edge is identical (for all moments,
as we show there). 

The above results, near one edge, are also independent (to leading order in the distance to the edge) of whether one fixes only one edge (as here) or two edges (as below). Indeed, although fixing the edge to seed distance $\ell_2$ is different from fixing the total 
extension $\ell=\ell_1+\ell_2$ of the avalanche, some results do not depend on that, as
we check below, such as the small distance behavior near the edge. Obtaining them here
via the fixed $\ell_2$ calculation, was thus much easier.\\

{\bf Results for the massive case}. Some of the above results have been obtained also in the massive case in 
section \ref{sec:1boundarymass}. The PDF of the seed-to-edge distance
$\ell_2$ is displayed in \eqref{Prhomass}. The mean shape conditioned on
$\ell_2$ is given in \eqref{msm1} and \eqref{msm3}. One sees in \eqref{msm2} that it again vanishes
as $\sim \frac{1}{21} (\ell_2-x)^3$ near the edge, however the 
coefficient of the $O((\ell_2-x)^4)$ term explicitly depend on the
ratio $r=\ell_2/\ell_m=m \ell_2$. We have also calculated the mean size of the avalanche
conditioned on $\ell_2$, 
$\langle S \rangle_{\ell_2}$, in \eqref{Sl2}.

\subsubsection{Mean spatial shape $\langle S(x) \rangle_{\ell}$ for fixed total extension $\ell$}

Until now we have conditioned only to one boundary, i.e. the distance $\ell_2$ between
the seed and the edge. One can also condition to the total extension $\ell=\ell_1+\ell_2$,
see Fig. \ref{fig:extensions}. The calculation
is more involved and is performed in several steps in Sections 
\ref{subsec:aspect}, \ref{subsec:jointsizeext} 
and \ref{subsec:shapext}. It is related to the solution of the so-called Lam\'e equation. The final result,
obtained and discussed in Section \ref{subsec:shapext} reads
\bea \label{scal1}
&& \langle S(x)  \rangle_{\ell} = \ell^3 {\sf s}(z) \quad , \quad z = \frac{x+\ell_1}{\ell} 
\eea
where $z$ is the distance to the lower edge normalized to the total extension (so that $z=0,1$ at the two edges). The scaling function has the explicit expression, from \eqref{mean4}, 
\bea \label{sexpl}
&& {\sf s}(z) = \frac{1}{4 \pi \sqrt{3}}  ( 2 g(z) + (1-2 z) g'(z) + z (z-1) g''(z) ) 
\eea
in terms of the function $g(z)$ 
\bea \label{defg2} 
&& g(z)=g(1-z) = \frac{1}{6 g_3^0} 
( 2 {\cal P}_0(z)^2 -\frac{4 \pi  {\cal P}_0(z) }{\sqrt{3}}+ {\cal P}'_0(z)
 (\zeta \left(z;0,g^0_3\right)-\frac{2 \pi  z}{\sqrt{3}}) )
\eea 
where ${\cal P}_0(z)$ and $g_3^0$ were defined in \eqref{defP0} and $\zeta \left(z;g_2,g_3\right)$
is the Weirstrass Zeta function, a primitive of ${\cal P}_0(z)$. The mean shape function
is symmetric around $z=1/2$ and has the following expansion near the edge (at small $z$)
\bea \label{mse} 
&& {\sf s}(z) = \frac{z^3}{21}-\frac{z^4}{28}-\frac{3  \sqrt{3} z^7 \Gamma
   \left(\frac{1}{3}\right)^{18}}{98560 \pi ^7}+\frac{3 \sqrt{3} z^8 \Gamma
   \left(\frac{1}{3}\right)^{18}}{112640 \pi ^7}+O\left(z^9\right) 
\eea
and near its maximum at $z=\frac{1}{2}$
\bea
&&  {\sf s}(z) =  0.00182091 - 0.0314909 (z - \frac{1}{2})^2 + .. 
\eea 
Integrating the mean shape we obtain the mean total avalanche size for a fixed
extension $\ell$ 
\bea
\langle S \rangle_{\ell} =  \ell^4 \int_0^1 dz {\sf s}(z) 
= 0.00073657566 \, \ell^4
\eea 
which we also verify through a distinct, more direct
and simpler
calculation in Section \ref{subsec:jointsizeext}.

\subsubsection{Mean spatial shape $\langle S(x) \rangle_{\ell,p}$ for fixed total extension $\ell$ {\it and}
aspect ratio $p$}

We now consider the mean shape conditionned to the couple $(\ell_1,\ell_2)$ 
equivalently to the couple $(\ell, p)$ where $p$ is the aspect ratio defined in \eqref{aspectdef}.
The calculation is performed in Section \ref{subsec:aspect}.

First let us recall \cite{Delorme} that the avalanche density $\rho(\ell,p)$ is given by (see also Section \ref{subsec:PDF12}, e.g. Eq. \eqref{end}, using ${\cal P}_0''(p)=6 {\cal P}_0(p)^2$). 
\bea \label{rhoellp} 
\rho(\ell,p) = \frac{\tilde R(p)}{\ell^3} \quad , \quad 
\tilde R(p) = 8 \pi \sqrt{3} \tilde P(p) = - 36 \left((p-1) p {\cal P}_0(p)^2 +(p- \frac{1}{2}) {\cal P}_0'(p)
+ {\cal P}_0(p) \right)
\eea 
and $\tilde P(p) \simeq_{p \ll 1} \frac{3 \sqrt{3}  \Gamma \left(\frac{1}{3}\right)^{18}}{896 \pi ^7} p^3$. Hence the aspect ratio $p$ and the total length $\ell$ are {\it independent random variables}
and $\tilde P(p)=\tilde P(1-p)$ is the PDF of the aspect ratio in the limit of small applied force.
Its lowest moments are $\langle p \rangle=\frac{1}{2}$ and $\langle p^2 \rangle=0.275664$.  \\

Let us first display the result for the cumulative conditioning, which are
simpler. We find
\bea \label{meansh0} 
&& \langle S(x)  \rangle_{\ell_1<l_1, \ell_2 < l_2} =
\frac{f l}{3 g_3^0} ( \psi_1(z)  \psi_1(p) \theta(z<1-p) + \psi_1(1-z)  \psi_1(1-p) \theta(z>1-p) ) \\
&& z = \frac{x+l_1}{l}  \quad , \quad l=l_1+l_2 \quad , \quad p=\frac{l_2}{l_1+l_2} \label{meansh00} 
\eea
where $z=p$ is the position of the seed ($x=0$ is the seed), and $\psi_1(z)$ is a solution to the 
Lam\'e equation, see \eqref{Lame}, given by
\bea \label{psi1first} 
&& \psi_1(z)=2 {\cal P}_0(z) + z {\cal P}'_0(z) 
\eea
Note that \eqref{meansh0} is symmetric in the variables $z,p$. Note that the notation $p$ is a slight abuse
of notation here in \eqref{meansh00}, since the aspect ratio is really $\frac{\ell_2}{\ell_1+\ell_2}$.  
One can compare with the behavior near a single edge obtained above in Section \ref{subsec:cumintro}, 
and one finds 
$\langle S(x) \rangle_{\ell_1<l_1, \ell_2<l_2}  = \frac{f}{7} \frac{(l_2-x)^4}{r_2^3} 
\phi(\frac{l_2}{l_1+l_2})$ where $\phi(z)$ is a smoothly decaying function from $\phi(0)=1$ 
to $\phi(1)=0$,
given in \eqref{phiphi}. Finally, the formula for the (cumulative) conditional mean total size, 
$\langle S \rangle_{\ell_1<l_1, \ell_2<l_2}$ is given in
\eqref{cumcondtot}.\\

The mean shape for fixed extension $\ell$ and aspect ration $p$ is obtained in \eqref{shapeboth}-\eqref{shapeboth2} and reads
\bea \label{shapeboth00} 
&& \langle S(x) \rangle_{\ell,p}  = \ell^3 \, {\sf s}_p(z) \\
&& {\sf s}_p(z) =  \frac{1}{3 g_3^0 \tilde R(p)} 
( G(p,z) \theta(z<1-p) + G(1-p,1-z) \theta(z>1-p) ) \quad , \quad
z = \frac{x+\ell_1}{\ell}  \\
&& G(p,z) = 
 (p-1) p \psi_1\left(z\right) \psi_1''(p)+\left(z-1\right) z \psi_1(p) \psi_1''\left(z\right)+\left(p \left(2
   z-1\right)-z+1\right) \psi_1'(p) \psi_1'\left(z\right)
\eea
where $\psi_1(z)$ is given in \eqref{psi1first} and $\tilde R(p)$ in \eqref{rhoellp}.
Since $\tilde R(p)=\tilde R(1-p)$ one has ${\sf s}_p(z)={\sf s}_{1-p}(1-z)$ as expected
from parity invariance. Note also that $G(p,z)$ is a symmetric function of its arguments
which gives an expected symmetry $\tilde R(p) {\sf s}_p(z)=\tilde R(z) {\sf s}_z(p)$. 
Expanding \eqref{shapeboth00} at small $z$ (for fixed $p$) we find that the
behavior of ${\sf s}_p(z)$ near the edge $z=0$ ($x=-\ell_1$) is
\be
{\sf s}_p(z) = \frac{z^3}{21} + c_4(p) z^4 + \dots \quad , \quad c_4(p)=- \frac{1}{12} + \frac{p}{48} + O(p^2)
\quad , \quad c_4(p)=\frac{1}{28 (1-p)} - \frac{5}{48} - \frac{1-p}{64} + O((1-p)^2) 
\ee 
with the same behavior near the other edge $z=1$ ($x=\ell_2$) with $p \to 1-p$. We have indicated the
behavior of the coefficient of the subleading term $c_4(p)$ for $p$ near $p=0$ (i.e. $\ell_2 \ll \ell_1$) and near $p=1$ (i.e. $\ell_2 \gg \ell_1$). One can check that the expectation of $c_4(p)$ over $p$ is $\int_0^1 \tilde P(p) c_4(p)=- \frac{1}{28}$ consistent with the subleading term in \eqref{mse}.
Remarkably, the leading term is again cubic with the coefficient $\frac{1}{21}$ independently of the value of the fixed aspect ratio, the same coefficient as in \eqref{mse}. This confirms, once again, the universality of the leading behavior near the edge, 
as discussed above in Section \ref{subsec:fluctusumm}.

%

Finally by integration over $x$ we obtain the result \eqref{mts} for the mean total size at fixed extension and aspect ratio, which reads
\bea
&& \langle S  \rangle_{\ell,p} = \frac{{\sf s}(p)}{\tilde P(p)} \ell^4 
\eea

\subsection{Spatiotemporal shape near the edge of the avalanche}
\label{subsec:spatiotemporal}

\begin{figure}[h]
\centerline{
   \fig{.5\textwidth}{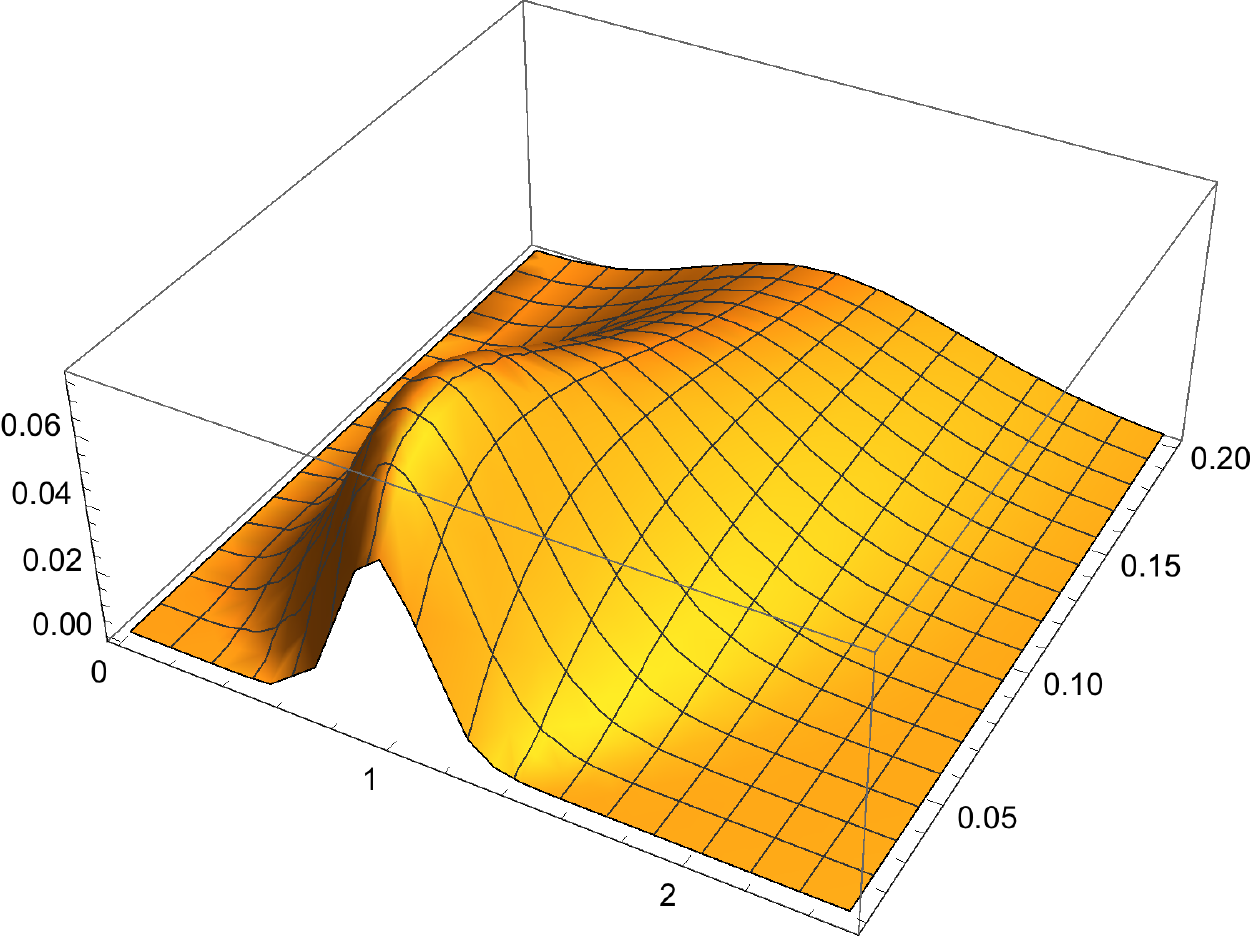} }
   \caption{
  Plot of the spatio-temporal shape: mean local instantaneous 
  velocity $v(z,t)=t^{-1} F(z,1/\sqrt{t})$, i.e. Eq. \eqref{spatiotemp} 
  for $z=\frac{\ell_2-x_0}{\ell_2} \in [0,2.5]$, and $t \in [0,0.2]$.
  We use units such that the seed-to-edge distance is $\ell_2=1$.
  The mean velocity is larger near the seed at $x_0=x_s=0$, i.e. $z=1$.
  It vanishes exactly for all times on the edge of the avalanche, 
  i.e. the line $z=0$, as given in \eqref{edgedyn}. }
\label{fig:velocity}
\end{figure}

The spatiotemporal shape is defined by the instantaneous velocity, $\dot u(x,t)$,
for an avalanche started at time $t=0$ with its seed at $x=0$. We study
the mean spatiotemporal shape near an edge of the avalanche in Section 
\ref{sec:spatiotemp}. To calculate the mean we will condition on a single edge, since it an easier calculation. We first compute the (cumulative) mean 
$\langle \dot u(x,t) \rangle_{\ell_2<l_2}$ conditioned to the upper extension $\ell_2$
being smaller that $l_2$. It amounts to compute the euclidean Green function of a quantum particle
in a $1/x^2$ potential, a well known textbook problem. We find
\bea \label{mvel1} 
\langle \dot u(x,t) \rangle_{\ell_2<l_2} =  
\frac{\sqrt{l_2(l_2-x)}}{2 t} I_{\frac{7}{2}}(\frac{l_2(l_2-x)}{2 t}) e^{- \frac{l_2^2+(l_2-x)^2}{4 t}} 
\eea 
in terms of a modified Bessel function.
Similarly we find that the mean spatiotemporal shape at fixed avalanche upper extension $\ell_2$ is
given by the scaling form 
\bea \label{spatiotemp} 
&& \langle \dot u(x,t)\rangle_{\ell_2} 
= \frac{\ell_2^3}{t} F(\frac{\ell_2-x}{\ell_2},\frac{\ell_2}{\sqrt{t}}) 
\eea
where the spatiotemporal shape scaling function $F(z,r)$ is given in terms of Bessel functions in \eqref{spatiotempscal}, 
and is plotted in Fig. \ref{fig:velocity}. We find that the mean instantaneous local velocity vanishes with the distance to the edge of the avalanche, with the same exponent $3$, as does the mean static shape. 
The associated amplitude ${\cal A}_{t}$ now depends on time, and we obtain its exact time dependence
\bea
\langle \dot u(x,t) \rangle_{\ell_2} \simeq {\cal A}_{t} \left(l_2-x\right){}^3 \quad , \quad 
 {\cal A}_{t} = \frac{1}{t} C(\frac{t}{\ell_2^2}) \quad , \quad C(\tau)= \frac{e^{-\frac{1}{4 \tau}}}{5040 \sqrt{\pi } \tau^{7/2}} \label{edgedyn} 
\eea 
As time increases, the amplitude ${\cal A}_{t}$ reaches a maximum and then decays.
One can check that by integration over time 
$\int_0^{+\infty} dt {\cal A}_{t} =  \int_0^{+\infty} \frac{d\tau}{\tau} C(\tau) = \frac{1}{21}$
one recovers the universal static amplitude.

\subsection{Higher dimension} 

The BFM in higher spatial dimension $d>1$ is more complicated to analyze since the 
instanton equation admits fewer analytical solutions. However one can still obtain 
results about some properties of the support of the avalanche. We denote this support $\Omega(x_s)$, 
where $x_s$ is the position of the seed, and it consists in the set of points $x$ which have moved during
the avalanche, i.e. with $S(x)>0$. For the SR BFM that we are studying we can surmise that the
avalanche is connected, but for $d>1$ it can contain "holes". Our study is limited to the BFM in dimension $d<4$. The observables that we study are illustrated in the Figure \ref{fig:2dspan}.

\begin{figure}[h]
\centerline{
   \fig{.6\textwidth}{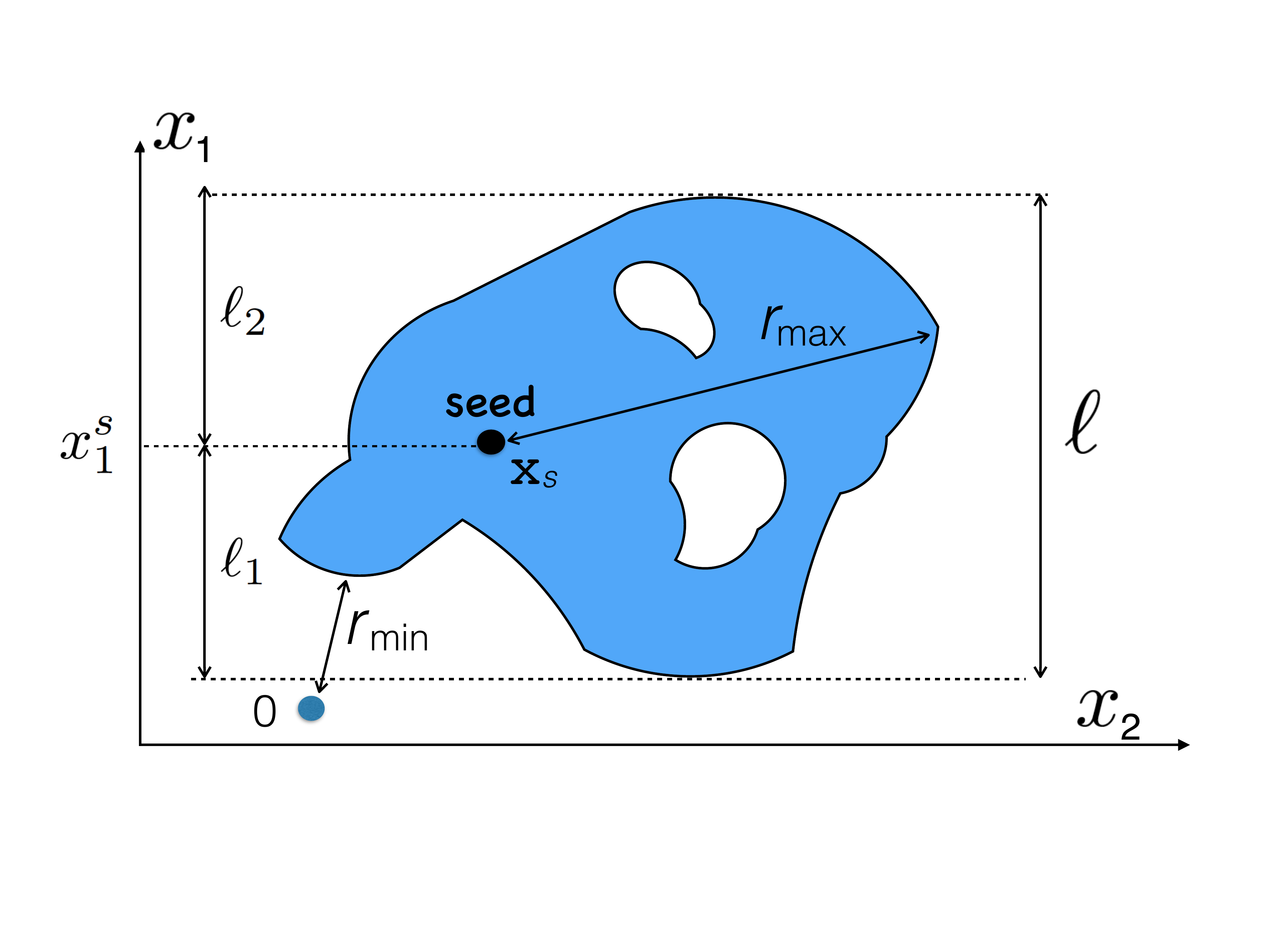} \!\!\!\!\!\!\!\!\!\!\!\!  \!\!\!\!\!\!\!\!\!\!\!\! \!\!\!\!\!\!\!\!\!\!\!\! 
   \!\!\!\!\!\!\!\!\!\!\!\! \!\!\!\!\!\!\!\!\!\!\!\! 
   }
 \caption{Schematic representation of the support of an avalanche for a 2-dimensional interface
 in the internal plane $x=(x_1,x_2)$ (the direction of motion - not shown - being transverse to the 
 plane of the figure).
 The region in blue is the support $\Omega$, i.e. it contains all points that have moved ($S(x)>0$). The region in white is the complementary, and contains all points that have not moved ($S(x)=0$). 
 The picture is a cartoon, as one rather expects a fractal structure. Here $x_s$ is the position of the seed
 (the first point which moves).
 One studies here the following observables shown in the Figure:  (i) the {\it maximal distance reached by the avalanche} 
 $r_{\rm max}$ (ii) {\it the minimal distance of approach} of a given point $0$, $r_{\rm \min}$ (iii) 
 {\it the total span} $\ell$ associated to a direction $x_1$ as the
 $\ell= \max_{x \in \Omega} (x_1) - \min_{x \in \Omega}(x_1)$, as well as the
 {\it upper span} $\ell_2= \max_{x \in \Omega} (x_1) - x_s$ and the {\it lower span}  
 $\ell_1= x_s - \min_{x \in \Omega} (x_1)$. }
\label{fig:2dspan}
\end{figure}

\subsubsection{Probability that a given point has not moved and mean shape around it}

In Section \ref{sub:point} we obtain the probability that a given point $x_0$ does not belong to
the support of the avalanche. It reads, for any $d<4$
\bea \label{probanot} 
{\rm Prob}(S(x_0) =0) = {\rm Prob}(x_0 \notin \Omega(x_s)) = e^{- f \frac{2 (4-d)}{(x_0-x_s)^2} }
\eea 
Note that it means that either the avalanche maximal radius (studied below) is smaller than $|x_s-x_0|$, or
(for $d>1$) it is larger, and the point $x_0$ belongs to a "hole" inside the avalanche.

Now we can compute some observables conditioned to the event that $S(x_0)=0$. 
The first one is the PDF of the variable $\Sigma= \int d^dx \frac{S(x)}{(x-x_0)^4}$. 
It turns out that this variable is finite and its PDF is obtained in \eqref{PSigma}.
It exhibits a $\Sigma^{-3/2}$ power law regime, reminiscent of the one for the total size $S$,
with some cutoffs (see discussion there).\\

The second observable is the mean shape $\langle S(x) \rangle_{S(x_0)=0}$ conditioned to $S(x_0)=0$. 
It amounts to solve for the Green function of a Schrodinger problem in a $1/r^2$ potential in dimension $d$. The complete result in any $d$ is given in \eqref{avmean5}, and in $d=2$ (which is simpler)
it is given in \eqref{avmean3}. Note that in $d=1$ this observable corresponds to the
cumulative conditional shape with a single boundary, see \eqref{avmean2}. The most interesting property 
is that the shape vanishes near the point $x_0$ as a power law of the distance with a quite non-trivial exponent $b_d^+$ (we have chosen $x_0=0$ and denote the radial coordinates around that point 
as $r=|x|$ and $r_s=|x_s|$)
\bea \label{avmean66}
\langle S(x) \rangle_{S(0)=0} \simeq_{r \to 0} f C_d  r_s^{b_d^-} \, r^{b_d^+} \quad , \quad b_d^\pm =  \frac{1}{2} \left(2-d \pm \sqrt{d^2-20 d+68}\right)
\label{exponent2} 
\eea
where $C_d$ is a dimension dependent constant and with $b_2^+=2 \sqrt{2}$, $b_3^+=\frac{1}{2} (\sqrt{17}-1)$ and $b_d^+ \to 0$ as $d \to 4^-$ (and
one recovers $b_1^+=4$ as found in $d=1$).

\subsubsection{Maximal distance reached by an avalanche and minimal distance of its support to a point}

Let us define the {\it maximal radius} of an avalanche as $r_{\max}=\max_{x \in \Omega(x_s)} |x-x_s|$
the maximal distance that it reaches counted from the seed, see Figure \ref{fig:2dspan}.
It is a random variable and we obtain in Section 
\ref{subsec:reach} its CDF and its density, which read
\be
{\rm Prob}(r_{\max} < R) = e^{- c_d f/R^2} \quad , \quad 
\rho(r_{\max}) = \frac{2 c_d}{r_{\max}^3}
\ee
with $c_{d=2}  \approx 12.5634$, $c_{d=3}  \approx 15.7179$. In $d=1$
this observable identifies with ${\rm Prob}(\ell_1<R,\ell_2<R)$ and one finds
$c_1= \frac{3}{2} {\cal P}_0(\frac{1}{2},\Gamma(1/3)^{18}/(64 \pi^6))=8.8475159..$.\\

Another interesting observable is the {\it minimal distance of approach}, i.e. 
$r_{\min}=\min_{x \in \Omega(x_s)} |x|$, $0 < r_{\min}<r_s=|x_s|$, the minimal distance 
between the origin $x=0$ and any point of the
avalanche (started at $x_s$), see  Figure \ref{fig:2dspan}.
We denote its PDF as $P_{r_s}(r_{\min})$.
Of course, we know from \eqref{probanot} that there is a non zero probability,
equal to $p_0=1-e^{-  \frac{2 (4-d) f}{r_s^2}}$, that $r_{\min}=0$, i.e. that the point $x=0$ belongs to
the avalanche. So $P_{r_s}(r_{\min})$ has a delta function piece $p_0 \delta(r_s)$ at zero. 
We have calculated $P_{r_s}(r_{\min})$ in \eqref{pdfrmin}, \eqref{pdfrmin2}. Let us 
indicate here only the behavior of the associated density $\rho_{r_s}(r_{\min}=R)= \partial_f P_{r_s}(r_{\min}=R)|_{f=0^+}$ when $r_{\min} \ll r_s$
\be \label{pdfrmin25} 
 \rho_{r_s}(r_{\min}=R) 
 \simeq  \frac{B_d}{R^{3+b_d^-} r_s^{-b_d^-}}  \quad , \quad 0 < r_{\min} \ll r_s 
\ee 
where $B_d$ is a dimension dependent constant, and the exponent $b_d^-$ is
defined in \eqref{avmean66}. Note that $3 + b_d^- \geq 0$ and
increases from $0$ to $1$ as $d$ increases from $d=1$ to $d=4$.

\subsubsection{Probability of not hitting a cone in two dimension} 

\begin{figure}[h]
\centerline{
   \fig{.5\textwidth}{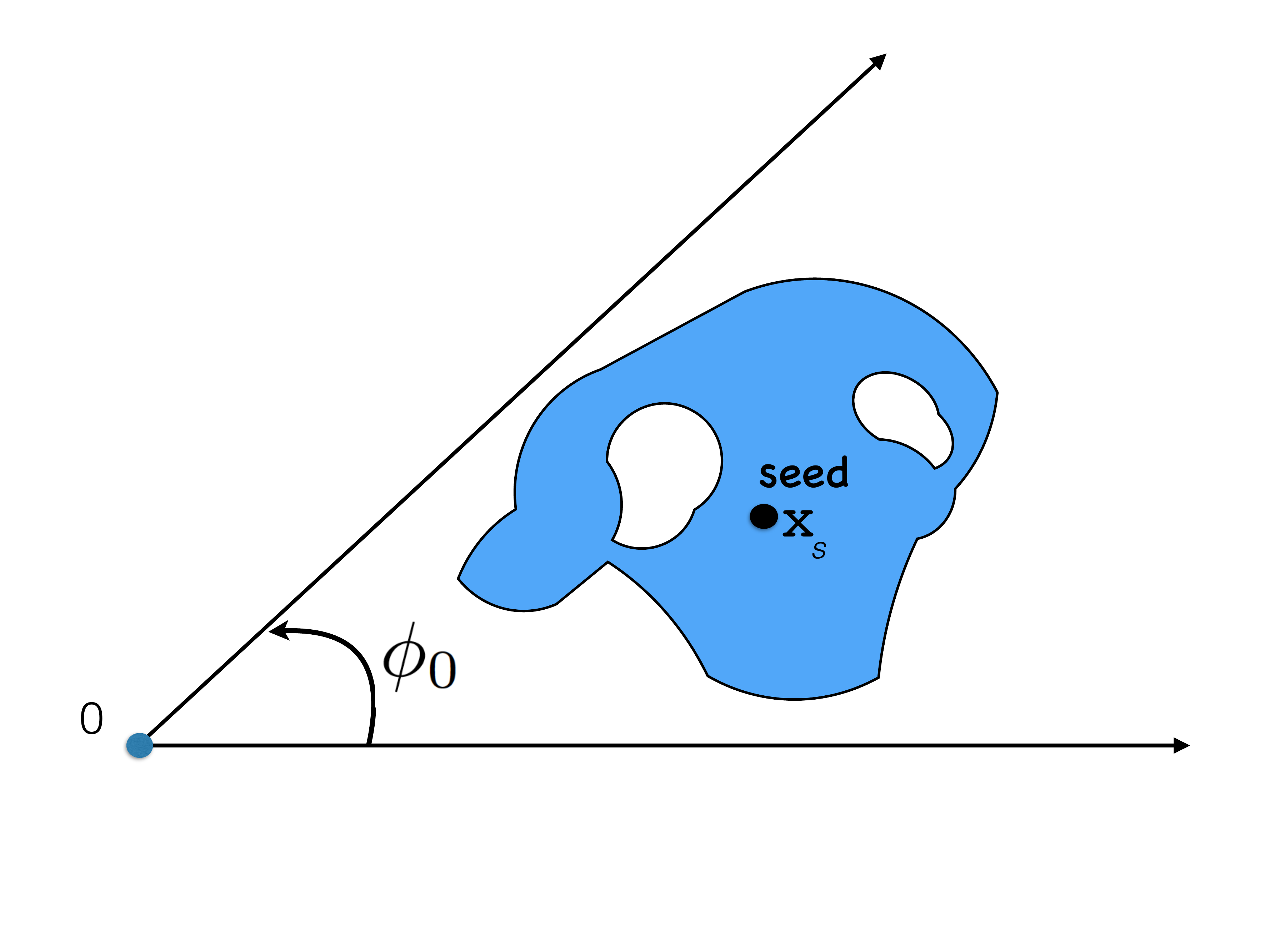} \!\!\!\!\!\!\!\!\!\!\!\!  \!\!\!\!\!\!\!\!\!\!\!\! \!\!\!\!\!\!\!\!\!\!\!\! 
   \!\!\!\!\!\!\!\!\!\!\!\! \!\!\!\!\!\!\!\!\!\!\!\! 
   }
 \caption{Avalanche of seed $x_s$ and support $\Omega$ (in blue) in two dimensions. 
 We obtain the probability that it remains confined within the conical domain of apex in $0$ 
 and opening angle $\phi_0$, as represented schematically. For $\phi_0=\pi$ it is related to
 the distribution of $\ell_1$ defined in Fig. \ref{fig:2dspan}. For $\phi_0 \to 2 \pi$ it becomes the
 probability that the avalanche does not intersect the positive half-axis. 
 }
\label{fig:2dcone}
\end{figure}

Consider the conical domain in $d=2$, in polar coordinates  
${\cal D} = (r,\phi) \in ]0,+\infty[ \times ]0,\phi_0[$
with an opening angle $0< \phi_0 \leq 2 \pi$. We ask
what is the probability that an avalanche $\Omega(x_s)$, with a seed inside the wedge
at $x=(r_s,\phi_s)$, 
remains strictly confined inside the wedge (i.e. is not allowed to touch the
two radial lines at $\phi=0,\phi_0$). We find in Section \ref{subsec:cone} that it equals
\be
 {\rm Prob}(S(r,0)=S(r,\phi_0)=0, \forall r \in [0,+\infty[) 
= \exp \left(- \frac{6 f}{r_s^2} (\frac{1}{3} + h_{\phi_0}(\phi_s)) \right)
\ee
where $h_{\phi_0}(\phi)$ is a Weierstrass function with parameter $g_3=g_3(\phi_0)$ determined by the condition
that its period, denoted $2 \Omega(\frac{4}{3},g_3)$, equals $\phi_0$, namely
\be
h_{\phi_0}(\phi) = {\cal P}(\phi; g_2=\frac{4}{3} , g_3(\phi_0))  \quad , \quad g_3=g_3(\phi_0) \quad \Leftrightarrow \quad 
2 \Omega(\frac{4}{3}, g_3) = 2 \int_{e_1}^{+\infty} \frac{dt}{\sqrt{4 t^3 - \frac{4}{3} t-g_3}} = \phi_0
\ee
where $t=e_1$ is the largest positive root of $4 t^3 - \frac{4}{3} t-g_3=0$, a condition
equivalent to asking that ${\cal P}'(\phi_0/2; g_2=\frac{4}{3} , g_3(\phi_0)) =0$.
One finds that $g_3$ decreases as $\phi_0$ increases, with $g_3 \in [g_3^c,+\infty[$, where
$g_3 \simeq g_3^0/\phi_0^6$ for a small cone opening angle $\phi_0 \ll 1$ and
$g_3^c \approx - 0.295402$ corresponds to $\phi_0=2 \pi$, i.e. the probability
that the avalanche does not intersect the positive half axis.

From these results, we also obtain in Eq. \eqref{jcdf} the joint CDF of the "angular extensions"
$\varphi_2= \max_{x=(r,\phi) \in \Omega} \phi$ and
$\varphi_1= - \min_{x=(r,\phi) \in \Omega} \phi$ (for $\phi_s=0$),
i.e. the angular analogs of $\ell_1,\ell_2$, as well as the PDF of the total
angular extension $\varphi=\varphi_1+\varphi_2$ by differentiation.

\subsubsection{Avalanche span and shape in $d>1$} 

In dimension $d$, as we discussed above, one can define the span $\ell$ of the
avalanche along one given axis, say $x_1$, 
and the distances $\ell_1$ and $\ell_2$ from the seed to the lines (in general
hyperplanes) as defined in the caption of Fig. \ref{fig:2dspan}.

With this definition we know that all joint statistics of $\ell_1,\ell_2$ and 
$S_{tot}(x_1) = \int d^{d-1} x_\perp S(x_1,x_\perp)$ is {\it identical}
to the one for the $d=1$ BFM that we have already calculated.
For instance, the mean shape integrated over the transverse space $x_\perp$,
$\langle S_{tot}(x_1) \rangle_{\ell}$, is given by the same formula.

One can also ask about the local mean shape at a point, $\langle S(x_1,x_\perp) \rangle_{\ell}$,
i.e. not integrated. It is a related but different question which contains non trivial information
also in the transverse direction. This is studied in Section \eqref{subsec:spanhigherd}, where we obtain the following result
\bea \label{resSy1} 
&& \langle S(x_1,x_\perp) \rangle_{\ell_2<l_2} 
= f \int \frac{d^{d-1}q}{(2 \pi)^{d-1}} e^{i q (x_\perp-x^s_\perp )}  F_{|q|}(l_2+ x_1^s -x_1,l_2) \\
&& F_q(a,b) = \sqrt{a b} \big( I_{7/2}(q a) K_{7/2}(q b) \theta(0<a<b)
+ K_{7/2}(q a) I_{7/2}(q b) \theta(0<b<a) \big) \nn
\eea 
in terms of modified Bessel functions.
If one integrates this equation over $x_\perp$, it selects $q=0$, in which
limit one has $F_{q \to 0}(a,b) = \frac{a^4}{7 b^3} \theta(0<a<b) + \frac{b^4}{7 a^3} \theta(0<b<a)$,
recovering \eqref{resms} and \eqref{resms2} as expected (see discussion above).\\

One obtains similarly the mean shape at fixed upper span $\ell_2$. We display
its expression in $d=2$, choosing the origin $x_s=(x_1^s,x_\perp^s)=0$ as the position of the seed
\bea \label{shape2d} 
&& \langle S(y_1,y_\perp) \rangle_{\ell_2} 
=  \frac{\ell_2 (2 \ell_2 -y_1)}{24 (2 \pi) (\ell_2-y_1)^2} \, (y_1^2  + y_\perp^2)  \,
\Sigma\left( \frac{y_1^2  + y_\perp^2}{2 \ell_2 (\ell_2-y_1)} \right) \\
&& \Sigma(z) =  
\frac{(z+1) (15 z (z+2)+2)}{2 z (z+2)}
-\frac{3}{4} (5 z (z+2)+4) \log (1+ \frac{2}{z}) 
\eea 
From the asymptotics of the scaling function $\Sigma(z)$, one finds again that the mean shape vanishes near the upper edge with a cubic
power, but with an amplitude which decreases with $|x_\perp|$, the transverse
distance to the seed, on a scale $\ell_2$, and with a high power, as $1/|x_\perp|^8$, more
precisely
\be
\langle S(x_1,x_\perp) \rangle_{\ell_2} =
\frac{16  \ell_2^7}{105 \pi  \left(\ell_2^2+ x_\perp^2 \right)^4} \,  (\ell_2-x_1)^3
+ O(  (\ell_2-x_1)^4 ) 
\ee
Integration of the amplitude of the cubic power over $x_\perp$ recovers the universal coefficient
$\frac{1}{21}$. So, for a given position of the seed, there is a non-trivial shape structure in the transverse direction near the edge of the avalanche. 

\subsection{Tip driving PDF of local and total sizes, extension distribution and spatial shape} 

We now consider driving by the tip, as described in Section \ref{sec:intro-tip}, and focus on
$d=1$. The equation of motion is \eqref{BFMdefTip} contains two mass scales, a local one, $m_0^2$, equal to 
the curvature of the local well which drives $u(x=0,t)$ and a uniform mass $m^2$ which, when present,
acts as a cutoff for large avalanches. We study the avalanche produced by
the driving kick at $x=0$, $f(x,t)=\delta(x) m_0^2 w \delta(t)$ in \eqref{BFMdefTip}.

We first present the results for $m=0$. Since $\hat L_{m_0}=m_0^{-2}$ has the dimension of a lenght scale (in the internal $x$ direction) a crossover occurs as $m_0$ is varied. In the limit $m_0 \to 0$ one recovers
the results for "imposed local force" driving, while for $m_0 \to + \infty$ we have "imposed local displacement". The crossover between the two was not studied in \cite{Delorme} and
is obtained here exactly. We obtain the joint PDF, $P_w(S,S_0)$, of the total size $S$, and of the local jump $S_0=S(x=0)$ at the tip $x=0$, and its marginal distributions, as a function of $m_0$.
There is a crossover scale at $S = \hat S_{m_0}$ for avalanche sizes, such that below
and above this scale the avalanche look different 
\bea
&& S \ll \hat S_{m_0} \quad \Leftrightarrow \quad \text{imposed local force} \\
&& S \gg \hat S_{m_0} \quad \Leftrightarrow \quad \text{imposed local displacement} 
\eea 
with, and similarly for the crossover scale in the local size at $S_0 = \hat S^0_{m_0}$ (in units $\sigma=\eta=1$)
\be
\hat S_{m_0} = m_0^{-8} \quad , \quad \hat S^0_{m_0} = m_0^{-6} 
\ee
(noted with a hat to distinguish it from $S_m$ the 
upper cutoff scale of avalanches in presence of a uniform mass). 
So large avalanches always feel the "imposed displacement". 
This scaling is expected, more generally one expects $\hat S_{m_0} \sim \hat L_{m_0}^{d+\zeta}$ 
and $S_0 \sim \hat L_{m_0}^{\zeta}$ with here $\zeta=\zeta_{\rm BFM}=3$ and more generally $d+\zeta_{\rm BFM}=4$. Note that in the limit $m_0 \to +\infty$ the local jump is exactly constrained,

\subsubsection{Distribution of local jumps at the tip}

Let us start with $P_w(S_0)$, we find in Section \ref{subsec:localtip} 
\bea \label{resPS0m0} 
P_w(S_0) &=& \frac{1}{3^{1/3} S_0^{5/3}} w m_0^2 \, e^{\frac{2}{3} y^3} 
(- y {\rm Ai}(y^2) - {\rm Ai}'(y^2) ) \quad , \quad y = \frac{m_0^2 (S_0-w)}{2 \times 3^{1/3} S_0^{2/3}} 
\eea 
One can check that the first moment exists with $\langle S_0 \rangle=w$, and that in the limit $m_0 \to +\infty$, $P_w(S_0)$ converges to $\delta(S_0-w)$
as expected, since then the displacement is fully imposed. Note that at 
fixed $w$, there is a cutoff for small avalanches $S_0^{\min} \sim m_0^3 w^{3/2}$. 
From \eqref{resPS0m0} one obtains the density $\rho(S_0)=\partial_{w} P_w(S_0)|_{w=0^+}$
in \eqref{denstip0},
and one finds that exhibits the single crossover scale $S_0 \sim \hat S_{m_0} = m_0^{-6}$. 
Using Airy function asymptotics (see \eqref{Airyas2}) one finds that the density crosses 
over between the two power laws characteristic respectively of the force imposed and
the displacement imposed regimes
\bea
&& \rho(S_0) \simeq \frac{m_0^2}{3^{2/3} \Gamma(\frac{1}{3}) S_0^{5/3}}  
\quad , \quad S_0 \ll \hat S_{m_0} = m_0^{-6} \label{fi} \\
&& \rho(S_0)  \simeq \sqrt{\frac{3}{2 \pi }} \frac{1}{m_0^3 S_0^{5/2}}
\quad , \quad S_0 \gg \hat S_{m_0} = m_0^{-6} \label{dens02} 
\eea 
and our result \eqref{denstip0} describes the full crossover.  
The first limiting case (force imposed) \eqref{fi} was already obtained in \cite{Delorme}
but the behavior \eqref{dens02} is new. Note however that it holds only
in some regime at finite $m_0$, since for $m_0=+\infty$ the displacement at
the tip is fixed by the driving and equal to $S_0=w$. This crossover is
discussed in Section \ref{subsec:localtip}.


\subsubsection{Distribution of total sizes}

The distribution of the total size $S$ of the avalanche, $P_{w}(S)$, is obtained 
in Section \eqref{sub:totalcross} in the Laplace transform form \eqref{PwSLapl}.
On this form one shows that it exhibits a crossover
between two limits. For $S \ll \hat S_{m_0}$, equivalently for $m_0 \to 0$ at fixed $S,w$
\bea
P_{w}(S) = \frac{m_0^2 w e^{-\frac{m_0^4 w^2}{4 S}}}{2 \sqrt{\pi } S^{3/2}}  
\eea 
which is normalized to unity. It is exactly the massless limit 
$P_f(S)=\frac{f e^{-\frac{f^2}{4 S}}}{2 \sqrt{\pi } S^{3/2}}$
of the standard size distribution with driving at imposed force
$f= \int dx dt m_0^2 w \delta(t) \delta(x)$. In the opposite limit 
$S \gg \hat S_{m_0}$, equivalently $m_0^2 \to +\infty$ at fixed $S,w$, we obtain
\bea
P_{w}(S) = LT^{-1}_{p \to S} e^{- w \sqrt{\frac{2}{3}} (4 p)^{3/4}}  = \frac{3^{2/3} }{(4w)^{4/3}} 
{\cal L}_{3/4}(3^{2/3} S/ (4w)^{4/3})
\eea 
with ${\cal L}_{3/4}$ being the stable Levy law of index $3/4$, see Appendix \ref{app:Levy}
where its explicit expression is given in terms of hypergeometric functions. The associated
size density $\rho(S)=\partial_{w}|_{w=0^+} P_w(S)$ has the explicit expression 
\bea
\rho(S) = \frac{\sqrt{3}}{\Gamma(1/4) S^{7/4}} 
\eea 
which was also found in \cite{Delorme}. 

\subsubsection{Joint distribution of total size and of local jump at the tip}

Although it is not easy to obtain directly an explicit expression for the
full crossover form $P_w(S)$, we have able to obtain the explicit expression of 
the JPDF $P_w(S,S_0)$ for both the massive and massless cases in Section
\ref{subsec:crossPP}. The result is bulky and displayed in \eqref{jointPw}. Here we
show only its associated density, $\rho(S,S_0) = \partial_w|_{w=0} P_w(S,S_0)$, 
and for the massless case $m^2=0$. It takes the scaling form 
\be
 \rho(S,S_0) =  m_0^{20} \hat \rho(m_0^8 S, m_0^6 S_0)
\ee
where the scaling function reads
\bea
&& \hat \rho(S,S_0) = 
\frac{2\ 3^{2/3} S_0^{1/3}  \exp \left(-\frac{3 S_0^3}{S^2}
+\frac{S_0}{36}-\frac{S_0^2}{2 S}
-\frac{6 S_0^4}{S^3} 
\right)}{\sqrt{\pi } S^{5/2}} \left(y \text{Ai}\left(y^2\right) - \text{Ai}'\left(y^2\right) \right)
\quad , \quad  y = S_0^{1/3}  \frac{S +6 S_0}{2 \sqrt[3]{3} S}
\eea

Finally, note that we have also given the lowest moments of $S,S_0$ and joint moments 
in the massive case in Section \ref{sec:moments}.

\subsubsection{Distributions of extensions for tip-driven avalanches}
 
In tip-driven avalanches the seed is always at $x_s=x=0$. One is thus
interested in the distribution of $\ell_2$ the distance of the seed to the upper edge at
$x=\ell_2$. In Section \ref{subsec:tipext}
we obtain its CDF as
\bea
{\rm Prob}(\ell_2 < r) = e^{ - 6 m_0^6 w \, \phi(r m_0^2) }
\eea 
where the function $\phi(r)$ is defined by inversion
\bea \label{invert1} 
\phi = \phi(r) \quad \Leftrightarrow \quad r=\int_{\phi}^{+\infty} \frac{dy}{\sqrt{4 y^3 + \phi^2 + 4 \phi^{5/2}  }} = 
\, _2F_1\left(\frac{1}{6},\frac{1}{2};\frac{7}{6};-\frac{\phi^2 + 4 \phi^{5/2}}{4 \phi^3}\right) \frac{1}{\sqrt{\phi}}
\eea 
which can also be written as $\phi = {\cal P}(r,0,g_3=-\phi^2 - 4 \phi^{5/2})$. 
The detailed asymptotics of $\phi(r)$ and of the CDF are given in \eqref{detailedphi}, \eqref{detailedphi2} and \eqref{detailed1}. 
From the CDF one obtains 
the PDF $P(\ell_2)$ and the corresponding density, which reads
\bea
\rho(\ell_2) =  \partial_{w}|_{w=0^+} P(\ell_2)  = - 6 m_0^8 \phi'(m_0^2 \ell_2) 
\eea 
Its asymptotics are obtained in Section \ref{subsec:tipext} as
\bea
&& \rho(\ell_2) \simeq \frac{12 m_0^2}{\ell_2^3}   \quad , \quad   \ell_2 \ll m_0^{-2} \\
&&  \rho(\ell_2) \simeq \frac{36 \Gamma \left(\frac{1}{3}\right)^3 \Gamma \left(\frac{7}{6}\right)^3}{\pi ^{3/2}
   \ell_2^4}  = \frac{99.2466..}{ \ell_2^4}   \quad , \quad   \ell_2 \gg m_0^{-2} 
\eea 
The first regime of small $\ell_2$ also corresponds to the limit of small $m_0$, i.e. $m_0 \to 0$, 
and one recover $\rho_f(\ell_2) = \partial_f|_{f=0^+} P(\ell_2) \simeq \frac{12}{\ell_2^3}$, i.e.
\eqref{PDFl22}, the result (for a single boundary) in the
regime of "imposed local force" $f=m_0^2 w$. The second
regime, of large extension $\ell_2$, also corresponds to the limit of large $m_0$, i.e. 
$m_0 \to +\infty$: it is the regime of "imposed local position" and leads to
a new result, with a different decay exponent, $\rho(\ell_2) \sim 1/\ell_2^4$. 

Now, for finite $m_0$ the upper extension $\ell_2$ and the lower one $\ell_1$
will be correlated non-trivially. In the limit $m_0 \to +\infty$ of strict imposed displacement,
however, they become two independent random variables. We now study the avalanche shape in this limit. \\

\subsubsection{Mean shape of a tip-driven avalanche and its fluctuations near the edge}

%

We have obtained in Section \ref{subsec:shapetip} the mean shape of an avalanche
of a given extension $\ell_2$, in the limit $m_0 \to +\infty$ of strict imposed displacement,
for $0 \leq x \leq r$
\bea
&& \langle S(x) \rangle_{\ell_2<r} = w \, {\sf s}_{>}(\frac{x}{r})  \quad , \quad {\sf s}_{>}(0) =1  \quad , \quad {\sf s}_{>}(1) =0 \\
&& {\sf s}_{>}(z) = 
\frac{1}{{\cal P}'(1,0,g_3^*)}
  (2 {\cal P}(1-z,0,g_3^*)
+ (1-z) {\cal P}'(1-z,0,g_3^*) )
\eea
where ${\cal P}'(1,0,g_3^*)=-(-g_3^*)^{1/2}$ from the properties of the Weierstrass function,
and $g_3^* = -\frac{4 \Gamma \left(\frac{1}{3}\right)^6 \Gamma \left(\frac{7}{6}\right)^6}{\pi ^3}=
- 30.400905$ is the negative root of the equation ${\cal P}(1,0, g_3) = 0$. The condition
${\sf s}_{>}(0) =1$ is correctly recovered, as it should since for
strictly imposed displacement (limit $m_0 \to +\infty$), at the tip $S(x=0)=w$ with no fluctuations. 
Near the edge of the avalanche
\bea
{\sf s}_{>}(z) \simeq \frac{3 \Gamma \left(\frac{1}{3}\right)^3 \Gamma \left(\frac{7}{6}\right)^3}{7 \pi ^{3/2}}
(1-z)^4 = 1.18150748.. \times (1-z)^4
\eea 

The mean shape conditioned to the extension $\ell_2$ is given by the exact formula (also compatible with
$S(x=0)=w$)
\bea
\langle S(x) \rangle_{\ell_2}  = w {\sf s}_{>}(z) - \frac{w}{\ell_2 P(\ell_2)} z \, {\sf s}_{>}'(z) \quad , \quad z=\frac{x}{\ell_2}
\eea 
which, in the limit of infinitesimal driving has the finite limit, for $0 \leq x \leq \ell_2$
\bea
&& \langle S(x) \rangle_{\ell_2}  =  - \frac{1}{\ell_2 \rho(\ell_2)} z {\sf s}'_{>}(z) = \ell_2^3 \, \tilde {\sf s}(z) \quad , \quad z=\frac{x}{\ell_2} \\
&& \tilde {\sf s}(z) = \frac{1}{6 g_3^*}
z ( 2 (1-z) {\cal P}(1-z,0,g_3^*)^2 + {\cal P}'(1-z,0,g_3^*) ) 
\eea 
It is plotted in Fig. \ref{fig:shapes} (note that there ${\sf s}(y_0)=\tilde {\sf s}(z=1-y_0)$ here).
It vanishes near the tip (since $w \to 0$) as
\bea
 \tilde {\sf s}(z)  = \frac{18 \pi ^{3/2} z}{\Gamma \left(\frac{1}{6}\right)^3 \Gamma
   \left(\frac{1}{3}\right)^3}+O\left(z^2\right) = 0.0302277 z + O(z^2) 
\eea 
and near the edge $z \to 1^-$ as
\bea
 \tilde {\sf s}(z)  =
\frac{1}{21} (1-z)^3-\frac{1}{21} (z-1)^4+O\left((z-1)^9\right)
\eea
i.e. with the same cubic power and the same prefactor as for the force driven avalanches obtained in
\eqref{mse}, a manifestation of the aforementioned universality of the edge behavior.
More detailed series expansions are given in \eqref{se1} and \eqref{st1}.

Integrating $\int_0^{\ell} dx_0$ one obtains the mean total size at fixed extension
\bea
&& \langle S \rangle_{\ell_2}  
= \ell_2^4 \frac{- \zeta(1;0,g_3^*)}{18 g_3^*} = 0.0022097287584 \, \ell_2^4
\eea 
which is also obtained by an equivalent calculation in Section \ref{subsec:conddirect},
together with the higher moments $\langle S^p \rangle_{\ell_2}$ in \eqref{cc}
and their generating function. \\

\begin{figure}[h]
\centerline{
   \fig{.45\textwidth}{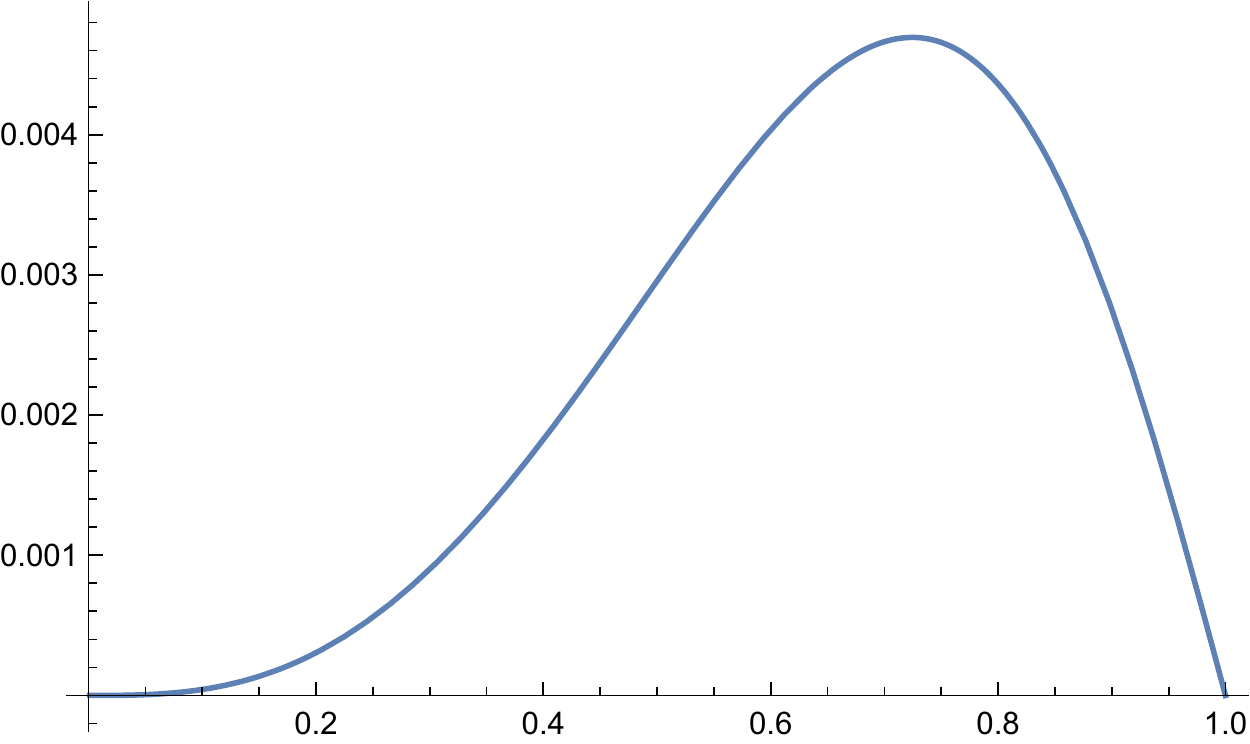} ~~~
  \fig{.45\textwidth}{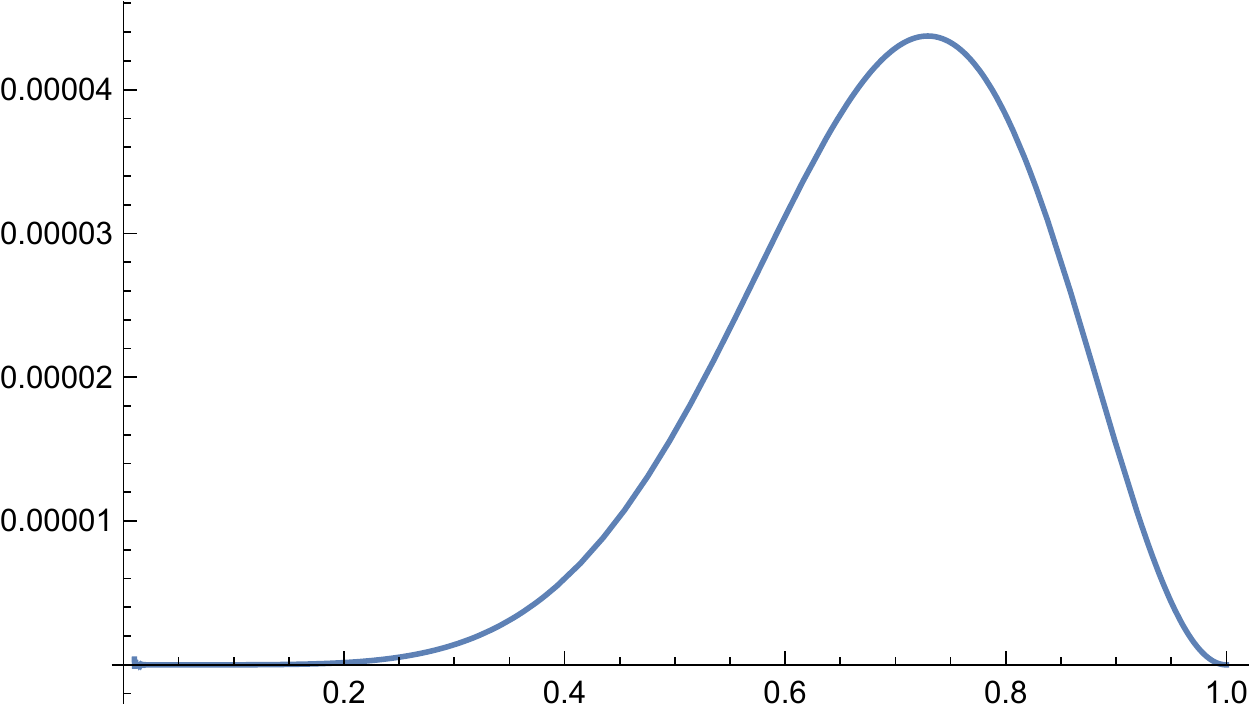}
   }
 \caption{Left: Plot of the scaling function describing the mean shape of an avalanche with imposed
 displacement at the tip ($m_0=+\infty$). What is plotted is 
 ${\sf s}(y_0)=\tilde s(z=1-y_0)$  as a function of $y_0$. 
 The mean shape vanishes with a cubic power at the edge of
 the avalanche $y_0=0$ ($z=1$) and linearly at the driving tip $y_0=1$ 
 ($z=0$). Right: idem for the second shape (second moment of the shape),
 here is a plot of ${\sf s}_2(y_0)$ defined in the text. It vanishes with the sixth power at the edge,
 and we show that the $n$-th moment vanishes with power $3 n$, with a prefactor
 that can be extracted from our general formula.}
\label{fig:shapes}
\end{figure}

\begin{figure}[h]
\centerline{
   \fig{.4\textwidth}{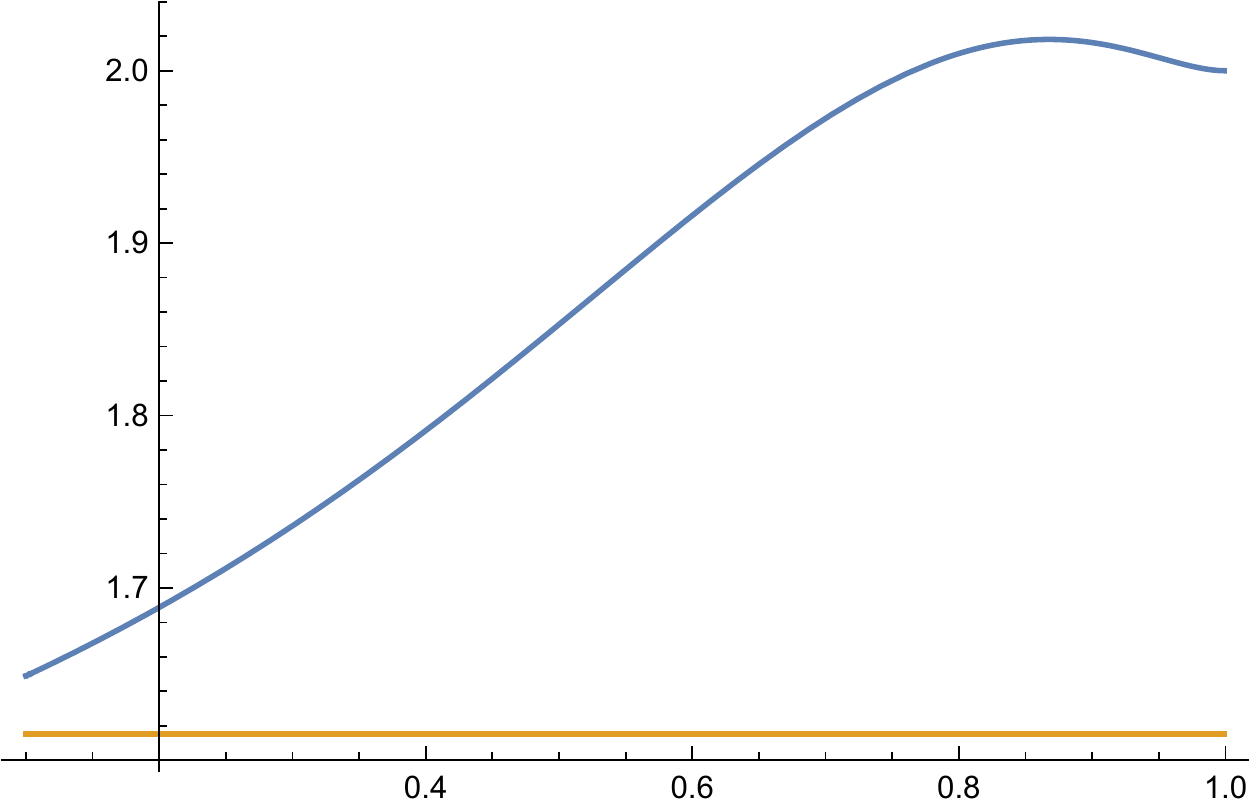}}
   \caption{Plot of the ratio 
   ${\sf s}_2(y_0)/{\sf s}(y_0)^2$ for $y_0 \in [0,1]$, with the same notations as in Fig. \ref{fig:shapes}. }
\label{fig:ratio}
\end{figure}

{\bf Second shape and full mean distribution near the edge}. In Section \ref{subsec:shapetip2} 
we calculate the second shape, that is $\langle S(x)^2 \rangle_{\ell_2}$. It is given in the form
$\langle S(x)^2 \rangle_{\ell_2} = \ell_2^6 \, {\sf s}_2(1- \frac{x}{\ell_2})$ where the
scaling function ${\sf s}_2(y)$, $0 \leq y \leq 1$ is given in \eqref{s2s2} and plotted in Fig. \ref{fig:shapes}.
The ratio $\frac{\langle S(x)^2 \rangle_{\ell_2} }{\langle S(x)\rangle_{\ell_2}^2}$ is plotted 
in Fig. \ref{fig:ratio},
varies between $\frac{21}{13}=1.61538$ for $x$ near $\ell_2$, as in \eqref{agaaga},
and $2$ for $x$ near the tip $x=0$.
One finds that near the edge of the avalanche 
\be
\langle S(x)^2 \rangle_{\ell_2} \simeq \frac{1}{273} (\ell_2-x)^6
\ee 
exactly the same leading order result, up to the amplitude, as for the force imposed
avalanches. We show in Section \ref{subsec:shapetip2} (see Remark about universality at the end)
that for the tip driven avalanches, one can also write $S(x) \simeq \sigma \times (\ell_2-x)^3$ near the edge,
with the same probability distribution $p(\sigma)$ for the random variable $\sigma$
as given in \eqref{genfunct0}, \eqref{cond40} (with lowest order moments given in 
\eqref{sigmom2}), which is thus
universal (and reproduces all the moments of the shape near the edge).


\section{Method: generating function and instanton equation}
\label{sec:method} 

In the BFM model \eqref{BFMdef1} it is possible to express averages of exponentials of linear
functions of the velocity field in terms of solutions of a non-linear partial differential
equation, see Refs. \cite{DobrinevskiLeDoussalWiese2011b,LeDoussalWiese2012a} for details.
Since here we mostly study observables based on avalanche sizes,
we mostly need the "static" version of this property. It states 
that for an initial condition $\dot u(x,t=0^-)=0$ and any forward
driving $\dot f(x,t) \geq 0$, as well as any "source" function $\lambda(x)$ so
that the l.h.s exists, one has 
\be
G[\lambda(x)] : = \overline{ \exp \left(\int d^d x \lambda(x) S(x) \right) } = 
e^{ \int d^d x f(x) \tilde{u}^\lambda(x) }\ .
\label{generating}
\ee
where here $f(x) = \int_0^{+\infty} dt \dot f(x,t)$. In this paper $\overline{\cdots}$ denotes the average over 
the disorder, i.e. the Brownian random force field, equivalently over $\xi(x,t)$ in \eqref{BFMdef1}.
The function $\tilde u^\lambda(x)$ is the solution of the (time independent) "instanton equation"
\be
\nabla_x^2 \tilde{u}(x)- A \tilde{u}(x) + \tilde{u}(x)^2 =- \lambda(x) \ .
\label{instanton}
\ee
where $A=1$ for the massive case and $A=0$ for the massless case,
in the units defined in Section \ref{subsec:units} in each case. 

To calculate the total avalanche size one needs a uniform source in the instanton equation: $\lambda(x) = \lambda$, and for the local size we need a local source, i.e. in $d=1$, $\lambda(x) =  \lambda \delta(x)$. To obtain information about the extension of the avalanches, we consider a source localized at two points in space, in $d=1$, $\lambda(x) = \lambda_1 \delta(x-r_1) + \lambda_2 \delta(x-r_2)$.\\

The dynamical version of \eqref{generating} and \eqref{instanton} is
recalled when we need it in Section \ref{sec:spatiotemp} below. In Sections
\ref{sec:densitylocal} to 
\ref{sec:tip} we specialize to 
$d=1$, and consider $d>1$ in Section \ref{sec:higherd}.

\section{Bulk driving: density of local size versus the distance to the seed} 
\label{sec:densitylocal}

We study here various properties of the local size of an avalanche in $d=1$. We use the dimensionless units. The Laplace transform (LT) of the PDF, $P_{\{f(x)\}}(S_0)$, of the local avalanche size $S_0=S(x=0)$ following a kick \eqref{kick1} of amplitude $f(x)$ 
 is given from \eqref{generating} as
\bea \label{lt1} 
\overline{e^{\lambda S_0}} = \int dS_0 e^{\lambda S_0} P_{\{f(x)\}}(S_0) 
= e^{\int dx f(x) u^\lambda(x) }
\eea 
where $u^\lambda(x)$ is the solution of the 1D equation \eqref{instanton}
\bea \label{instanton-tip} 
&& \tilde u''(x) - A \tilde u(x) + \tilde u(x)^2 =  - \lambda \delta(x) 
\eea 
which vanishes at $\pm \infty$, with $A=0$ in the massless case and $A=1$ in the massive case.
The explicit solution reads (see e.g. \cite{Delorme} and references therein) 
\bea  \label{solu10} 
&& \tilde u^{\lambda}(x)  =   \frac{6 (1-z_\lambda^2) e^{- |x|} }{(1 + z_\lambda + (1-z_\lambda) e^{- |x|})^2}  \quad , \quad A=1 \\
&& \tilde u^{\lambda}(x)  = - \frac{6}{(|x|+x_\lambda)^2} \quad , \quad 
\quad , \quad A=0
\eea 
where, ($A=0$) from the discontinuity of the derivative at $x=0$, $x_\lambda^3 = - \frac{24}{\lambda}$, and 
($A=1$), $z=z_\lambda$ is the solution of
\bea \label{eqz}
&& \lambda = 3 z (1-z^2) 
\eea 
whose branch is continuously connected to $z=1$ for $\lambda=0$. 
Note that $\tilde u(x=0)=\frac{3}{2}  (1- z^2)$ ($A=1$) and 
$\tilde u(x=0)=-6/x_\lambda^2=\frac{3^{1/3}}{2} (-\lambda)^{2/3}$. \\

\subsection{Massless case}

In the massless case the general result reads, from \eqref{lt1} and \eqref{solu10}
\bea
P_{\{f(x)\}}(S_0) = LT^{-1}_{- \lambda \to S_0} e^{ - 6  \int dx f(x) \frac{\lambda^{2/3}}{(|x| (-\lambda)^{1/3} + (24)^{1/3} )^2} } 
\eea 
where $LT^{-1}$ denotes the Laplace inversion, which can be obtained explicitly in a few cases. The first two are known 
\cite{LeDoussalWiese2008c,LeDoussalWiese2012a,Delorme},
and we recall them for completeness, the third is new.

\begin{itemize} 

\item {\bf Uniform kick} $f(x)=f$. 
Using $\int dx \tilde u^{\lambda}(x)=  - 12/x_\lambda$ we obtain
\bea
P_{\{f\}}(S_0) := P_f(S_0)  = LT^{-1}_{- \lambda \to S_0}  e^{ - 2 \times 3^{2/3} (- \lambda)^{1/3}  f } 
\eea 
Denoting $\alpha=2 \times 3^{2/3} f$ and introducing $z:=(-\lambda)^{1/3}$, we can write
\bea
S_0 P_f(S_0) &=& \int_C \frac{d \lambda}{2 i \pi} e^{- \lambda S_0} \partial_\lambda e^{ - \alpha (- \lambda)^{1/3}  } =  \int_C \frac{d z}{2 i \pi} e^{ S_0 z^3} \partial_z e^{ - \alpha z  } \\
&=&  - \alpha \Phi(a,b,c)|_{a=3 S_0, b=0, c=-\alpha} 
\eea
Following Appendix A of \cite{Delorme} we have defined $\Phi(a,b,c)$, for $a \neq 0$, as
\bea \label{Phi}
 \Phi(a,b,c) := \int_C \frac{dz}{2 i \pi} e^{a \frac{z^3}{3} + b z^2 + c z} = |a|^{-1/3} e^{\frac{2 b^3}{3 a^2} - \frac{b c}{a}} 
{\rm Ai} \!\left(\frac{b^2}{|a|^{4/3} } - \frac{c ~ {\rm sgn}(a)}{|a|^{1/3}} \right)\ ,
\eea 
in terms of the Airy function. Here and below $C$ generically denotes appropriate contours in the complex plane. As in \cite{Delorme}, our use of inverse LT formula and change in variables therein
are heuristic here and below, an approach validated by careful checks of the final formulae. This gives
\bea
&& P_f(S_0) = \frac{2 \times 3^{1/3}  f}{S_0^{4/3}} {\rm Ai}(\frac{2 \times 3^{1/3} f}{S_0^{1/3}}) 
\eea  
correctly normalized to unity, using $\int_0^{+\infty} dy {\rm Ai}(y)=1/3$. One has the two limiting
behaviors
\bea
P_f(S_0) &\simeq& \frac{\sqrt[4]{\frac{3}{2}} f^{3/4} e^{-\frac{4 \sqrt{\frac{2}{3}} f^{3/2}}{\sqrt{S_0}}}}{\sqrt{\pi }
   S_0^{5/4}} \quad , \quad S_0 \ll f^3 \\
   &\simeq& f \rho(S_0) + O(S_0^{-5/3}) \quad , \quad S_0 \gg f^3  \quad , \quad \rho(S_0)=
    \frac{2}{\sqrt[3]{3} \Gamma \left(\frac{2}{3}\right) S_0^{4/3}}   \label{dens0} 
\eea 
where $\rho(S_0)$ is the total density of local size, which exhibits the exponent $4/3$. 

\item {\bf Local kick and local size at the point of the kick}  $f(x)=f \delta(x)$, and we recall, $S_0=S(x=0)$. 
Using that $\tilde u(x=0)=-6/x_\lambda^2=- \frac{3^{1/3}}{2} (-\lambda)^{2/3}$
\bea
P_{\{f \delta(x)\}}(S_0)  = LT^{-1}_{- \lambda \to S_0}  e^{ - \frac{3^{1/3}}{2} (-\lambda)^{2/3} f } 
\eea 
Denoting $\alpha=\frac{3^{1/3}}{2} f$ and introducing $z:=(-\lambda)^{1/3}$ we can write
\bea
 S_0 P_{\{f \delta(x)\}}(S_0) &=& \int_C \frac{d \lambda}{2 i \pi} e^{- \lambda S_0} \partial_\lambda e^{ - \alpha (- \lambda)^{2/3}  } =  \int_C \frac{d z}{2 i \pi} e^{ S_0 z^3} \partial_z e^{ - \alpha z^2  } \\
& =& - 2 \alpha \partial_c \Phi(a,b,c)|_{a=3 S_0, b=-\alpha,c=0} 
\eea
leading to
\bea \label{localforcelocal1}
&& P_{\{f \delta(x)\}}(S_0) =  \frac{f e^{-\frac{f^3}{36 S_0^2}}}{\sqrt[3]{3} S_0^{5/3}}  ( y  {\rm Ai}(y^2) - {\rm Ai}'(y^2) ) \quad , \quad y=\frac{f}{2 \times 3^{1/3} S_0^{2/3}} 
\eea  
The associated density exhibits the exponent $5/3$ and reads
\bea \label{dens1} 
\rho_0(S_0) = \frac{1}{3^{2/3} \Gamma(\frac{1}{3}) S_0^{5/3}}
\eea 

\item {\bf Local kick, and local size at the point at a distance $|x_s|$ from the kick}

The formula for the PDF of the local jump at $x=0$, $S_0=S(x=0)$, for a kick at $x=x_s$, can be written as above as a contour integral by introducing $z:=(-\lambda)^{1/3}$ 
\bea \label{Pcontour1} 
S_0 P_{\{f \delta(x-x_s)\}}(S_0) =  \int_C \frac{d z}{2 i \pi} e^{ S_0 z^3} \partial_z 
e^{- 6 f \frac{z^2}{(|x_s| z + (24)^{1/3} )^2} } = 3 S_0 \int_C \frac{d z}{2 i \pi} e^{ S_0 z^3} z^2 
e^{- 6 f \frac{z^2}{(|x_s| z + (24)^{1/3} )^2} } 
\eea 
A simpler expression can be obtained for the density. Note that there is
now a non-zero probability that the avalanche started at $x_s$ has not reached $x=0$.
The probability that $x=0$ has moved is 
\be
p_f = 1- e^{f \tilde u^{\lambda=-\infty}(x_s)} = - f \tilde u^{\lambda=-\infty}(x_s) + O(f^2)  \quad , \quad 
\tilde u^{\lambda=-\infty}(x_s) = -6/x_s^2
\ee
hence one can write
\bea
&& P_{f \delta(x-x_s)}(S_0) = (1-p_f)  \delta(S_0) + p_f \tilde P_{f \delta(x-x_s)}(S_0) 
\eea 
where the regular part, $\tilde P$, is also normalized to unity,
$\int_{S_0>0} \tilde P_{f \delta(x-x_s)}(S_0)=1$. 
We obtain the density by expanding in $f$ for fixed $S_0>0$ as in \eqref{expansion1} (which gets rid of
a $\delta(S_0)$ piece which does not enter in the density). Taking a derivative w.r.t. $f$ 
of \eqref{Pcontour1} we have
\bea
 \rho_{x_s}(S_0) &=& - 18 \int_C \frac{d z}{2 i \pi} e^{ S_0 z^3} 
\frac{z^4}{(|{x_s}| z + (24)^{1/3} )^2}   = - 18 \int_0^{+\infty} t dt \int_C \frac{d z}{2 i \pi} z^4 e^{ S_0 z^3 - t (|{x_s}| z + (24)^{1/3} )} \\
& =& - 18 \int_0^{+\infty} t dt  e^{  - t  (24)^{1/3} } 
 \partial_b^2 \Phi(a,b,c)|_{a=3 S_0, b=0,c=- t |{x_s}|} \\
&=&
 \frac{4 \times 3^{1/3}}{S_0^{5/3}} 
  \int_0^{+\infty} t dt  e^{  - 2 \times 3^{1/3} t}  ( \frac{1}{2} (\frac{t {|x_s|}}{3^{1/3} S_0^{1/3}})^2 {\rm Ai}(\frac{t {|x_s|}}{3^{1/3} S_0^{1/3}}) 
 + {\rm Ai}'(\frac{t {|x_s|}}{3^{1/3} S_0^{1/3}}) )
\eea
It can be put in a scaling form
\bea
 \rho_{x_s}(S_0) &=& \frac{1}{S_0^{5/3}} \phi(\frac{S_0}{|{x_s}|^3}) \\
 \phi(s) 
&=& - 12 s^{2/3} \int_0^{+\infty} dt e^{  - 2 \times 3^{2/3} t s^{1/3}} \frac{d}{dt}  (\frac{t^2}{2} {\rm Ai}'(t) ) \\
&=&  - 12 \times 3^{2/3} s \int_0^{+\infty} dt t^2 e^{  - 2 \times 3^{2/3} t s^{1/3}} {\rm Ai}'(t)
\eea 
This integral can be carried out exactly leading to
\bea \label{explicitphi} 
 && \phi(s) = \frac{3^{1/6} 6 }{\pi}  s e^{-24 s} \bigg(4 (18 s-1) \big(2 s^{1/3} \Gamma
   (\frac{1}{3}) E_{\frac{4}{3}}(-24 s)-3^{2/3} \Gamma(\frac{5}{3}) E_{\frac{5}{3}}(-24 s) \big) \\
   && ~~~~~~~~~~~~~~~~~~~~~~~~~~~~ +3 e^{24 s} \big(2 s^{1/3}
   \Gamma(\frac{1}{3})-3^{2/3} \Gamma(\frac{5}{3})\big)\bigg)
\eea 
where $E_n(z)=\int_1^{+\infty} e^{-z t} dt/t^n$ is the exponential integral.
Using that ${\rm Ai}'(0)=- 1/(3^{1/3} \Gamma(1/3))$
and $\int_0^{+\infty} dt t^2 {\rm Ai}'(t) = 3^{2/3}/\Gamma(-2/3)$ one can write respectively the large
and small $s$ asymptotics 
\bea
&& \phi(s) = \frac{1}{3^{2/3} \Gamma(\frac{1}{3})}  -\frac{\left(\frac{1}{s}\right)^{2/3} \Gamma \left(\frac{1}{3}\right)}{4
   \left(3^{5/6} \pi \right)}+\frac{5 \Gamma \left(\frac{2}{3}\right)}{36 \sqrt[6]{3} \pi
    s}+O\left(\left(\frac{1}{s}\right)^{4/3}\right) \\
&& \phi(s) = \frac{12 \sqrt{3} \times 3^{1/3} \Gamma(2/3)}{\pi} s + O(s^{4/3}) 
 \eea
The first is consistent with the formula \eqref{dens1} for $x_s=0$, 
and the second gives the leading behavior for fixed $S_0$ and the distance to the seed 
$|x_s| \to +\infty$
\bea
\rho_{x_s}(S_0) \simeq_{|x_s| \to +\infty} \frac{12 \sqrt{3} \times 3^{1/3} \Gamma(2/3)}{\pi}  \frac{1}{S_0^{2/3} |x_s|^3} 
\eea 
which is cutoff at $S_0 \sim |x_s|^3$. We also check that 
\bea
\int_{S_0>0} dS_0 \rho_{x_s}(S_0) = \int_0^{+\infty}  \frac{dS_0}{S_0^{5/3}} \phi(\frac{S_0}{|x_s|^3}) 
= \frac{1}{x_s^2} \int_0^{+\infty}  \frac{ds}{s^{5/3}} \phi(s) = \frac{6}{x_s^2} = - \tilde u^{\lambda=-\infty}(x) 
\eea 
i.e. $\partial_f p_f|_{f=0}$, the probability density per unit force that motion has occured at $x=0$, 
in agreement with the discussion above. 
We also check that it agrees with the result \eqref{dens1} for the uniform kick, indeed
\bea
\int dx_f \rho_{x_f}(S_0) = \int dx_f \frac{1}{S_0^{5/3}} \phi(\frac{S_0}{|x_f|^3})
= \frac{2}{3} \frac{1}{S_0^{4/3}} \int_0^{+\infty} \frac{ds}{s^{4/3}} \phi(s)
= \frac{2}{\sqrt[3]{3} \Gamma \left(\frac{2}{3}\right)} \frac{1}{S_0^{4/3}} 
\eea 

\end{itemize}

\subsection{Massive case} \label{subsec:massivePS0} 

We now consider now the massive case. We only give the results in the massive dimensionless
units and present the details of the derivation in Appendix \ref{app:massivePS0}.

We first recall the known results \cite{LeDoussalWiese2008c,LeDoussalWiese2012a,Delorme} 
(whose derivation is also given in the Appendix \ref{app:massivePS0}).
For a uniform kick $\dot w(x,t)=w \delta(t)$ one has
\bea \label{mass1}
P_{w}(S_0) = \frac{2 \times 3^{1/3} w e^{6 w}}{S_0^{4/3}} {\rm Ai}( \frac{3^{1/3} (S_0  + 2 w)}{S_0^{1/3}} ) 
\eea 
Note that the joint PDF $P_w(S,S_0)$ for a uniform kick was calculated in \cite{Delorme}.
For a local kick at $x=0$, $w(x,t)=w \delta(x) \delta(t)$,
the PDF of $S_0=S(x=0)$ at the position of the kick reads
\bea \label{mass2}
P_{\{w \delta(x)\}}(S_0) &=& \frac{w}{3^{1/3} S_0^{5/3}}  e^{w - \frac{w^3}{36 S_0^2}  } 
(y {\rm Ai}(y^2+ 3^{1/3} S_0^{2/3}) - {\rm Ai}'(y^2+ 3^{1/3} S_0^{2/3}) ) \quad , \quad 
y = \frac{w}{2 \times 3^{1/3} S_0^{2/3}} \nn \\
&&
\eea 
and the associated density reads
\bea \label{rhoS0mass} 
\rho_0(S_0)= - \frac{1}{3^{1/3} S_0^{5/3}}  {\rm Ai}'(3^{1/3} S_0^{2/3}) 
\eea 

For a local kick at $x_s$, we obtain the following scaling form for the
density of $S_0=S(x=0)$
\bea \label{scalingS0}
&& \rho_{x_s}(S_0) = \frac{1}{S_0^{5/3}} \frac{4 e^{-|x_s|}}{(1+e^{-|x_s|})^2} 
\phi(\frac{S_0}{(\frac{2(1- e^{-|x_s|})}{1+ e^{-|x_s|}})^3}, 3^{1/3} S_0^{2/3}) 
\eea
with the scaling function defined by the following integral
\bea
&& \phi(s,s_0) 
= - 12 s^{2/3}  \times  \int_0^{+\infty} dt e^{  - 2 \times 3^{2/3} t s^{1/3}} [ 3^{2/3} s^{1/3} t^2 {\rm Ai}'(t+s_0)
- \frac{3}{2} s_0 t^2 {\rm Ai}(t+s_0) ] \label{phiss0h} 
\eea 
One sees that the result \eqref{rhoS0mass} for $x_s=0$ is correctly recovered using that $\phi(+\infty,s_0) = \frac{1}{3^{1/3}} {\rm Ai}'(s_0)$ as easily shown by rescaling $t$ in \eqref{phiss0h}. The check
that 
\bea \label{sumrule1} 
\int dx_s \rho_{x_s}(S_0) = \frac{2 \times 3^{1/3}}{S_0^{4/3}} {\rm Ai}(3^{1/3} S_0^{2/3}) 
\eea 
is performed in the Appendix \ref{app:massivePS0}. Finally one can check
that the total weight
\bea
\int dS_0 \rho_{x_s}(S_0) = - \tilde u^{\lambda=-\infty}(x)  =  \frac{6 e^{- |x_s|} }{(1- e^{- |x_s|})^2}  
\eea
equals the probability density (per unit $w$) that $x=0$ belongs to the avalanche. 
Another limit is $|x_s| \to +\infty$ then
\bea
&& \rho_{x_s}(S_0) = \frac{4}{S_0^{5/3}} e^{-|x_s|}
\phi(\frac{S_0}{8}, 3^{1/3} S_0^{2/3}) 
\eea
We can check that it is consistent with the moments (by a numerical integration over $t$)
\be
\langle S_0 \rangle_{\rho_x} = \frac{1}{2} e^{-x} \, \text{ ,} \,\langle S_0^2 \rangle_{\rho_x} = \frac{1}{3} e^{-x} \, \text{ , } \, \langle S_0^3 \rangle_{\rho_x} = \frac{23}{48} e^{-x} \, \text{ and } \, \langle S_0^4 \rangle_{\rho_x} = \frac{13}{12} e^{-x} 
\ee
which are predicted by the small $\lambda$ expansion of $\tilde{u}^{\lambda}(x)$ at large $x$\,:
\be
\tilde{u}^{\lambda}(x) \underset{x \rightarrow \infty , \lambda \rightarrow 0}{\sim} \frac{1}{2} \lambda e^{-x} + \frac{1}{6} \lambda^2 e^{-x} + \frac{23}{288} \lambda^3 e^{-x} + \frac{13}{288} \lambda^4 e^{-x} + O(\lambda^5)
\ee

\section{Bulk driving: spatial shape conditioned to one boundary}
\label{sec:1boundary}

In this Section we study the spatial shape conditioned to the position of
one edge of the avalanche, more precisely on $\ell_2$ the seed-to-edge distance.
The calculations are much simpler than for the two boundary problem (conditioning 
on the total extension $\ell$) but still very instructive. 

\subsection{Massless case}
\label{sec:1boundarymassless}

\subsubsection{PDF of seed-to-edge distance $\ell_2$} 

Consider the massless case in $d=1$, and ask whether an avalanche under the applied
force kick \eqref{kick1} of amplitude $f(x)$ reaches the point of the interface at position $r$. The probability that this point has not moved at all is
\bea
{\rm Prob}( S(r) =0) = \lim_{\lambda_1 \to -\infty} \overline{e^{\lambda_1 S(r)}} = e^{ \int dx f(x) \tilde u_r(x)} 
\eea 
where $\tilde u_r(x)$ is the solution of
\bea
\tilde u''(x) + \tilde u(x)^2 = 0  \quad , \quad \tilde u(r)=-\infty \label{eq1n} 
\eea 
and vanishing at $x=\pm \infty$. Indeed the corresponding 
source term $- \lambda_1 \delta(x-r)$, when inserted in the r.h.s. of \eqref{eq1n} and
taking $\lambda_1 \to -\infty$ simply imposes that $\tilde u(r)=-\infty$. The solution of \eqref{eq1n} with such boundary conditions is simply 
\be
\tilde u_r(x) = - \frac{6}{(r-x)^2}
\ee
hence we have
\bea
{\rm Prob}( S(r) =0) =  e^{ - \int dx \frac{6}{(x-r)^2} f(x) } = 
e^{ - \int_{x<r} dx \frac{6}{(x-r)^2} f(x) } \times e^{ - \int_{x>r} dx \frac{6}{(x-r)^2} f(x) } 
\label{ps0} 
\eea
We see on this formula that (i) all points inside the support of the force will move (since for $r$ in the support, the integral diverges) (ii) if the applied kick has components on both sides of $x=r$
the probability of $S(r)=0$ splits in two independent probabilities: the events of the (multiply seeded) 
avalanche not reaching $r$ on each side are independent. In the following we will thus 
restrict to driving included in the half line $x<r$. In that case, if a second condition $S(r')=0$ is added,
with $r'>r$, one finds that the solution of \eqref{eq1n} with a second condition $\tilde u(r')=-\infty$
is $\tilde u_{r,r'}(x)=u_r(x)$ for $x<r$ hence ${\rm Prob}( S(r) =0, S(r')=0)={\rm Prob}( S(r) =0)$.
As pointed out in \cite{Delorme} this implies that the support of the avalanche is an interval.

Let us now consider a local kick at $x=x_s$, i.e. $f_x=f \delta(x-x_s)$, and recall
that the point $x_s$ is also called the seed, since the avalanche will start there. 
For notational simplicity we take the seed as the origin, i.e. we choose $x_s=0$. 
From \eqref{ps0} we obtain the CDF of the edge-to-seed distance, $\ell_2$, defined in Section 
\ref{subsubsec:extension}, see also
Fig. \ref{fig:extensions} (the other segment $\ell_1$ has the same distribution by symmetry,
however $\ell_1$ and $\ell_2$ are correlated, see next Section)
\bea
{\rm Prob}( S(r) =0) = {\rm Prob}(\ell_2 < r) = e^{ - 6 f/r^2  } \label{CDFl2} 
\eea 
for any $r>0$, which by differenciation gives the PDF of the one-sided extension $\ell_2$
\bea
P(\ell_2)  = \frac{12 f}{\ell_2^3}
e^{- 6 f/\ell_2^2}  \label{PDFl2} 
\eea
It is interesting to see that this is obviously distinct from the distribution $P(\ell)$ 
for the total extension $\ell=\ell_1+\ell_2$ (we use the same letter but the argument
indicates that it is a different distribution) but is has the same exponent 
$P(\ell_2)  \sim \ell_2^{-\kappa}$ with $\kappa=3$ for the BFM, the mean-field value.
Using that for the BFM $\zeta=4-d$, this value is in agreement with the general expected exponent relation is $\kappa=1 + (\tau-1) (\zeta+d) = d+\zeta-1$ where the first
equality comes from the scaling $S \sim \ell^{d+\zeta}$ and the second from the Narayan-Fisher
conjecture \cite{NarayanDSFisher1993a,ZapperiCizeauDurinStanley1998,DobrinevskiLeDoussalWiese2014a}
$\tau=2 - \frac{2}{d+\zeta}$. \\

\subsubsection{Mean shape conditioned on the seed-to-edge distance $\ell_2$} 
\label{subsec:conditioned} 

Let us now ask about the shape of the avalanche near its upper edge. 
More precisely we consider for $0,x_0<r$ the Laplace transform of the
joint PDF $P( S(r) =0 , S(x_0))=P( \ell_2<r , S(x_0))$ of the event that the local size at $x_0$ is $S(x_0)$ and that $\ell_2 < r$. It is
given by 
\bea \label{start1} 
\int dS(x_0) e^{\lambda S(x_0)} P( \ell_2<r, S(x_0)) = e^{f \tilde u_r^\lambda(x=0)} 
\eea 
where $\tilde u_r^\lambda(x)$ is the solution of the equation
\bea
\tilde u''(x) + \tilde u(x)^2 = - \lambda \delta(x-x_0)  \quad , \quad \tilde u(r)=-\infty  \label{eqlambda1} 
\eea 
for $x \in ]-\infty,r]$, 
i.e. with $\tilde u^\lambda_r(r)=-\infty$ and vanishing at $x=- \infty$. \\

We give the solution for arbitrary $\lambda$ below, which determines arbitrary moments of $S(x_0)$. 
As a starter, we first calculate the first moment of $S(x_0)$, the mean shape, which is elementary. To this aim,
we solve \eqref{eqlambda1} to first order in $\lambda$, i.e. we write its solution as
$\tilde u_r(x) = \frac{-6}{(r-x)^2} + \lambda  \tilde u_r^{(1)}(x) + O(\lambda^2)$, 
where $\tilde u_r^{(1)}(x)$ is the solution of the linear equation
\bea
\tilde u^{(1) \prime \prime}(x)  - \frac{12}{(r-x)^2} \tilde u^{(1)}(x)  = -  \delta(x-x_0)  \label{eqlambda2} 
\eea 
It can be seen as the zero-energy Green function of a Schrodinger problem in a repulsive inverse square potential $V(x)$
\bea
u_r^{(1)}(x) = G(x,x_0) = \langle x | \frac{1}{H} | x_0 \rangle \quad , \quad H= -\partial_x^2 + V(x) 
\quad , \quad V(x) = -2 \tilde u^{\lambda=0}(x) = \frac{12}{(r-x)^2}
\eea 
It is easy to see that the general solution of the homogeneous equation is
a sum of two powers $(r-x)^a$ with $a=-3,4$. Imposing continuity at $x_0$ 
and the proper discontinuity from \eqref{eqlambda2}, i.e. 
$[u^{(1)}(x)]_{x_0^-}^{x_0^+} = -1$ we find that the 
proper solution
is 
\bea \label{proper} 
&& u_r^{(1)}(x) = \frac{1}{7} (r-x_0) (\frac{r-x_0}{r-x})^{3}  \theta(x < x_0) + 
\frac{1}{7} (r-x_0) (\frac{r-x}{r-x_0})^{4}  \theta(x_0 < x < r)
\eea 
which satisfies $u_r^{(1)}(r)=0$ and $u_r^{(1)}(-\infty)=0$. Note that
$\int dx u_r^{(1)}(x)  = \frac{1}{10} (r-x_0)^2$. \\

Expanding \eqref{start1} to first order in $\lambda$, we obtain
\bea
\int dS(x_0) S(x_0) P( \ell_2 <r , S(x_0)) = f u_r^{(1)}(x) e^{ - f \frac{6}{r^2}  }   \label{exp1} 
\eea 
Dividing by the CDF \eqref{CDFl2}, we obtain the 
mean avalanche local size at $x_0$, $\langle S(x_0) \rangle_{\ell_2<r}$, 
with its seed at $x_s=0$ and
conditioned to the avalanche not
reaching $r$ (i.e. to $\ell_2<r$) as
\bea
\langle S(x_0) \rangle_{\ell_2<r} = f \tilde u_r^{(1)}(x=0)
= 
\frac{f}{7} \frac{(r-x_0)^4}{r^3}  \theta(0 < x_0 < r) + 
\frac{f}{7} \frac{r^4}{(r-x_0)^3}  \theta(x_0 < 0 < r) \label{condmeanint} 
\eea 
Note that as a function of $x_0$ it is continuous at the position of the seed
but has a jump derivative: one has $\frac{d}{dx_0} \langle S(x_0) \rangle_{\ell_2<r}|_{x_0=x_s^+=0^+}=- 4 f/7$ while $\frac{d}{dx_0} \langle S(x_0) \rangle_{\ell_2<r}|_{x_0=x_s^-=0^-}=3 f/7$. The total
jump is thus $-f$.\\

Taking instead first a derivative of \eqref{exp1} w.r.t. $r$, and then dividing by $P(\ell_2)$ 
given in \eqref{PDFl2}, we obtain the 
mean avalanche local size at $x_0$, $\langle S(x_0) \rangle_{\ell_2}$, i.e.
the mean spatial shape, conditionned to the seed-to-edge distance being $\ell_2>0$
\bea \label{res1b} 
\langle S(x_0) \rangle_{\ell_2} & = & 
 \frac{\ell_2^3}{12}  \partial_r|_{r=\ell_2}  u_{r}^{(1)}(x=0) + f \tilde u_{r=\ell_2}^{(1)}(x=0) \\
&=&   \frac{1}{21} (\ell_2-x_0)^3 
- \frac{1}{28} \frac{(\ell_2-x_0)^4}{\ell_2} + \frac{f}{7} \frac{(\ell_2-x_0)^4}{\ell_2^3}  
\quad , \quad 0 < x_0 < \ell_2 \label{res1b2}  \\
&=&
  \frac{1}{21} \frac{(\ell_2)^6}{(\ell_2-x_0)^3} - \frac{1}{28} \frac{(\ell_2)^7}{(\ell_2-x_0)^4} 
+  \frac{f}{7} \frac{\ell_2^4}{(\ell_2-x_0)^3}
\quad , \quad x_0 < 0 \label{res1b3} 
\eea 
where here the seed position has been chosen as $x_s=0$. The general case is obtained
by a change of coordinate $\ell_2 \to \ell_2- x_s$, $x_0 \to x_0 - x_s$. Again it is continuous
at the seed with $\langle S(x_0=x_s=0) \rangle_{\ell_2} = \frac{1}{84} \ell (\ell^2 + 12 f)$, and has exactly
the same left/right derivatives and jump in derivative as the cumulative mean above, i.e.
$\frac{d}{dx_0} \langle S(x_0) \rangle_{\ell_2}|_{x_0=x_s^+=0^+}=- 4 f/7$ while $\frac{d}{dx_0} \langle S(x_0) \rangle_{\ell_2}|_{x_0=x_s^-=0^-}=3 f/7$, i.e. a total 
jump $-f$. The total derivative jump at the seed being $-f$ is always the case, since
it simply comes from the $- \lambda \delta(x-x_0)$ in the instanton equation. 
Note that in the limit $f \to 0$ the first derivative jump vanishes and one finds that
the second derivative is continuous at the seed, and the right and left third derivatives are respectively
$4/7$ and $-10/7$ and represent the leading non-analyticity. These singularities are
smoothed out when averaging over the seed position (see below). \\

Hence we find that the mean shape of the avalanche vanishes near its edge as 
\bea \label{leadingD} 
\langle S(x_0) \rangle_{\ell_2} \simeq \frac{1}{21} D^3 \quad , \quad D=\ell_2-x_0
\eea 
Hence, expressed as a function of the distance $D$ to the edge, the leading behavior
\eqref{leadingD} is
fully independent on $\ell_2$ and of the position of the seed at $x=x_s$.
It turns out (see next section) that \eqref{leadingD} is also independent of whether the shape is
conditioned to a single edge, or to two edges. 
In all cases only the next order term $O(D^4)$ depends on $\ell_2$. 
More generally we see that in the limit $f \to 0^+$ the mean
shape is only a scaling function of the ratio of distances
\bea
\langle S(x_0) \rangle_{\ell_2} = D^3 F(D/\ell_2) \quad , \quad F(0)=\frac{1}{21}
\eea 
If this remains a general property beyond the BFM, then 
from a simple dimensional argument (since $S(x_0) \sim \ell_2^\zeta$) 
one then expects $\langle S(x_0) \rangle_{\ell_2} \simeq c_d D^\zeta$ where
$\zeta_{BFM,d=3}=4-d=3$.  \\

It is also interesting to center the avalanche on the edge, and average over the position of the seed.
This average can be calculated from \eqref{res1b2}, \eqref{res1b3} and one obtains
\be
 \bar S(y) = \int_0^{+\infty} \langle S(\ell_2-y) \rangle_{\ell_2} P(\ell_2) d\ell_2  =  f^{3/2} g(y/f^{1/2}) 
 \ee
with 
\bea \label{ggz} 
 g(z) =
\frac{432 \sqrt{6 \pi } \text{erfc}\left(\frac{\sqrt{6}}{z}\right)+5
   z^7-e^{-\frac{6}{z^2}} \left(5 z^6+9 z^4-36 z^2+432\right) z}{105 z^4}
&=& \frac{z^3}{21} - e^{-6/z^2} ( \frac{3}{35} z + O(z^2)) ~ , ~ z \ll 1 \nonumber
 \\
   &= & \frac{z}{5}-\frac{6}{z^3}
   +O(z^{-4}) 
     \quad  , \quad z \gg 1      
\eea
\\

It is also interesting to calculate the mean of the total avalanche size (i.e. the total mean area swept by the avalanche) conditioned to $\ell_2$, $\langle S \rangle_{\ell_2}$,
as well as the mean of the total size below, $S_{<}$, and above, $S_{>}$, the seed. They read
\bea
&& \langle S_{<} \rangle_{\ell_2} = \int_{-\infty}^0 dx_0 \langle S(x_0) \rangle_{\ell_2}  
= \frac{1}{84} \ell_2^4 + \frac{f}{14} \ell_2^2  
\quad , \quad   \langle S_{>} \rangle_{\ell_2} =  \int_{0}^{\ell_2} dx_0 \langle S(x_0)\rangle_{\ell_2}  = \frac{1}{210} \ell_2^4 + \frac{f}{35} \ell_2^2 \\
&& \langle S \rangle_{\ell_2} = \int_{-\infty}^{\ell_2} dx_0 \langle S(x_0) \rangle_{\ell_2}  =  \langle S_{<} \rangle_{\ell_2}  + \langle S_{>} \rangle_{\ell_2} =
\frac{1}{60} \ell_2^4 + \frac{f}{10} \ell_2^2 \label{sizem} 
\eea 
Note that the scaling $S \sim \ell_2^{d+\zeta} = \ell_2^4$ is expected for the BFM. 

\subsubsection{joint PDF of total size $S$ and $\ell_2$} 
\label{subsec:JointSl2} 

One can also calculate directly the mean total size conditionned to $\ell_2$ to recover the
above results. The same calculation gives the joint PDF of the total size $S$ and $\ell_2$.

To study the distribution of the total size $S$ we need to add a space-uniform source and solve the
instanton equation
\bea \label{eqmu} 
\tilde u''(x) + \tilde u(x)^2 = - \mu \quad , \quad \tilde u(r) =-\infty
\eea 
on $x \in ]-\infty,r]$ which has the limit $- \sqrt{-\mu}$ at $x=- \infty$. 
Let us denote $\tilde u_r^\mu(x)$ its solution, which
reads
\bea
&& \tilde u^\mu_r(x) =- c  - \frac{3 c }{ \sinh^2(\sqrt{c} (r-x)/\sqrt{2})} \quad , \quad c=\sqrt{-\mu} 
\eea 
We consider a kick at $x=x_s=0$ (position of the seed) and obtain
\bea \label{idmu} 
\int_0^{+\infty} dS e^{\mu S} P(\ell_2<r ,S) = e^{ f \tilde u^\mu_r(x=0)}
\eea 
with $c=\sqrt{-\mu}$, from which we can write the joint PDF $P(\ell_2<r ,S)$ as
\bea \label{ILT1} 
P(\ell_2<r ,S) = LT^{-1}_{p \to S} ~ \exp( - f (\sqrt{p}+ \frac{3 \sqrt{p} }{ \sinh^2(p^{1/4} r/\sqrt{2})}) )
\eea 
as well as $P(\ell_2,S)$ by derivation w.r.t. $r$. Expanding to first order in $p$
we obtain the conditional means
\bea \label{momS1} 
&& \langle S \rangle_{\ell_2<r} = \frac{1}{10} f r^2 \\
&& \langle S \rangle_{\ell_2} = \frac{1}{10} f \ell_2^2  + \frac{1}{60} \ell_2^4
\eea 
which correctly recovers the result (\ref{sizem}) of the previous section (obtained by integrating
the spatial shape). 

The above formula contain additional information about the fluctuations of $S$ and $\ell_2$.
Note that the LT in \eqref{ILT1} has a non analyticity $\sim p^{3/2}$ at small $p$, which
implies that the second moment for $S$ diverges (which is possible since it is the massless case) and that the following conditional probabilities decay
at large $S$ as
\bea
&& P(S|\ell_2<r) \simeq_{S \gg \ell_2^4} \frac{A}{S^{5/2}} \quad , \quad     A=\frac{f \ell_2^4}{168 \sqrt{\pi }} \\
&& P(S|\ell_2) \simeq_{S \gg \ell_2^4} \frac{A}{S^{5/2}} \quad , \quad     A=  \frac{f \ell_2^4}{168 \sqrt{\pi }}+\frac{\ell_2^6}{504 \sqrt{\pi }}
\eea 
It means that the constraint to fix one seed-to-edge distance, $\ell_2$, does not provide an operational cutoff on the other
seed-to-edge distance, $\ell_1$, which can still be very large. From the limit $f \to 0^+$ we also 
obtain the joint density as
\bea
\rho(S,\ell_2) = \partial_{f}|_{f=0} P(\ell_2,S) = \frac{1}{S^{7/4}} R(S/\ell_2^4) \quad , \quad 
R(s) = LT^{-1}_{p \to s} ~ 3 \sqrt{2} p^{3/4} \frac{\cosh(\frac{p^{1/4}}{\sqrt{2}})}{\sinh^3(\frac{p^{1/4}}{\sqrt{2}})}
\eea 
The inverse LT can be performed and expressed as a series (we will not give its expression here). 
Its finite moments w.r.t. $S$ are 
\bea \label{52} 
&& \int dS \rho(S,\ell_2) = \rho(\ell_2) = \frac{12}{\ell_2^3} \quad , \quad \int dS S \rho(S,\ell_2) = \frac{\ell_2}{5} 
\eea 
consistent with $\langle S \rangle_{\ell_2} \simeq_{f \to 0^+} = \frac{1}{60} \ell_2^4$ obtained as
the ratio of the second to the first result in \eqref{52}.
Finally, the joint density also decays with the exponent $5/2$ at large $S$, i.e. one has $\rho(S,\ell_2) \simeq \ell_2^3/(42 \sqrt{\pi} S^{5/2})$ for $S \gg \ell_2^4$.

%
%
%
%

\subsubsection{Higher moments of the shape} 

\label{subsec:highermoments}

We now give the solution of Eq. \eqref{eqlambda1} for arbitrary $\lambda$ which
allows to obtain the higher moments of the shape. \\

The general solution of \eqref{eqlambda1} can be written as
\bea \label{solu1side} 
\tilde u^\lambda_r(x) = - \frac{6}{(\rho - x)^2} \theta(x_0-x) - 6 {\cal P}(r-x;0,g_3) \theta(x-x_0)
\eea 
where $\mathcal{P}(z;g_2,g_3)$ is the Weierstrass elliptic function, whose properties
are recalled in the Appendix \ref{app:W} (here the parameter $g_2=0$). 
We will denote $\mathcal{P}'(z;g_2,g_3) = \partial_z \mathcal{P}(z;g_2,g_3)$.
Sometimes we use the shorthand $\mathcal{P}(z)=\mathcal{P}(z;g_2,g_3)$ (with here
$g_2=0$ implicit, unless said otherwise). 
The unknown parameters $\rho$ and $g_3$ (which depend on $x_0$ and $\lambda$) are determined by the following conditions of continuity and proper derivative jump at $x=x_0$
\bea
&& \frac{1}{(\rho - x_0)^2} =  {\cal P}(r-x_0;0,g_3) \label{cond1} \\
&& -  \frac{2}{(\rho - x_0)^3} - {\cal P}'(r-x_0;0,g_3) = \frac{\lambda}{6} \label{cond2}
\eea 
where we recall that ${\cal P}(z)$ is an even function. 
We can use that, see \eqref{eqdiff1}, that 
${\cal P}'(z)^2 = 4 {\cal P}(z)^3 - g_3$ (for $g_2=0$) to simplify one of the
relations as
\bea
g_3= - \frac{2}{3 (\rho-x_0)^3} \lambda - \frac{\lambda^2}{36}  \label{res1g3} 
\eea 
For $\lambda=0$ one has $g_3=0$ and one uses that ${\cal P}(z,{g_2=0,g_3=0})= \frac{1}{z^2}$ 
to obtain that $\rho=r$, recovering as required $\tilde u(x) = - \frac{6}{(r - x)^2}$. 

We can now expand the system \eqref{cond1}, \eqref{cond2} in powers of $\lambda$, for $\rho$ near $r$ and
$g_3$ near $0$. An efficient way to do it is as follows. One can use the expansion of the Weierstrass function in powers of $g_3$, whose coefficients obey a simple recursion relation
\be
{\cal P}(z,{g_2=0,g_3})= \sum_{p=0}^{\infty} c_p g_3^p z^{6 p - 2} \quad , \quad c_p = \frac{1}{(p-1)(6 p +1)} \sum_{k=1}^{p-1} c_k c_{p-k}  \label{Pexp3t} 
\ee
with initial conditions $c_0=1$, $c_1=1/28$ (some $c_p$ are displayed in \eqref{Pexp1}).
We rewrite \eqref{cond1},  \eqref{cond2}, denoting $z=r-x_0$ as
\bea \label{great} 
\lambda = - 12 {\cal P}(z,g_3)^{3/2} - 6 {\cal P}'(z,g_3)
\eea 
and substitute \eqref{Pexp3t} to obtain $\lambda$ as a series in $g_3$ at fixed $z$.
Then we invert the series and we find $g_3$ as a series in $\lambda$ at fixed $z$, e.g.
\be \label{g3ser} 
g_3=-\frac{2 \lambda }{3 z^3}-\frac{\lambda ^2}{252} + O(\lambda^3) 
\ee
higher orders being given in the Appendix
\ref{app:lambda}. We now calculate $\rho$, using \eqref{cond1}, and again \eqref{Pexp3t} and
\eqref{g3ser}, leading to
\bea
&& \rho =r - z+ 1/\sqrt{{\cal P}(z,g_3)}= r +\frac{\lambda  z^4}{84}+\frac{\lambda ^2 z^7}{3822} + 
+ O(\lambda^3)
\eea 
which leads, using \eqref{solu1side}, to the expansion in power of $\lambda$ of the desired solution of the instanton equation at $x=x_s=0$ in power of $\lambda$ as (we recall that $z=r-x_0$) 
\bea \label{bothbranches} 
&& \tilde u^\lambda_r(0) = - \frac{6}{\rho^2} = -\frac{6}{r^2}+\frac{\lambda  z^4}{7 r^3}+\frac{\lambda ^2 \left(16 r z^7-13
   z^8\right)}{5096 r^4} + O(\lambda^3) \quad , \quad 0 < x_0 < r \\
&&  \tilde u^\lambda_r(0)  = - 6 {\cal P}(r,0,g_3)=-\frac{6}{r^2}+\frac{\lambda  r^4}{7 z^3}+\lambda^2
 \left(\frac{r^4}{1176}-\frac{r^{10}}{3822 z^6}\right) \quad , \quad x_0 < 0 
\eea
which is continuous at $x_0=0$, i.e. $z=r$. It is given up to $O(\lambda^4)$ in Appendix \ref{app:lambda}.

We now obtain the conditional moments. Proceeding as in Section \ref{subsec:conditioned} above, or see also Appendix \ref{sec:rel}, we can write
\bea
e^{f (\tilde u_r^\lambda(0)- \tilde u_r^0(0))}  = \langle e^{\lambda S(x_0)} \rangle_{\ell_2<r} =  
1 + \sum_{p \geq 1} \frac{\lambda^p}{p!} \langle S(x_0)^p \rangle_{\ell_2<r}
\eea 
and expand both sides in powers of $\lambda$. 
For $p=1$ we then recover \eqref{condmeanint}, obtained there via a much simpler calculation.
For $p=2$ we obtain
\bea
&& <S(x_0)^2>_{\ell_2<r} = \frac{4
   f \left(r-x_0\right){}^7}{637 r^3} -\frac{f \left(r-x_0\right){}^8}{196 r^4}+  \frac{f^2 \left(r-x_0\right){}^8}{49 r^6} \quad , \quad 0<x_0<r \\
&& <S(x_0)^2>_{\ell_2<r} = \frac{f r^4}{588}  -\frac{f r^{10}}{1911
   \left(r-x_0\right){}^6}+ \frac{f^2 r^8}{49 \left(r-x_0\right){}^6}
\quad , \quad x_0<0
\eea 
As shown in Appendix \ref{sec:rel}
\bea
\langle e^{\lambda S(x_0)} \rangle_{\ell_2=r} 
= \frac{\partial_r u_r^\lambda(0) }{\partial_r u_r^0(0)} e^{f (u_r^\lambda(0)- u_r^0(0))}
\simeq_{f \to 0^+} \frac{\partial_r u_r^\lambda(0) }{\partial_r u_r^0(0)} 
\eea 
Expanding in $\lambda$ on both sides gives the conditional moments, e.g. we give here the
second moment in the limit $f \to 0^+$
\bea
&& \langle S(x_0)^2 \rangle_{\ell_2} \simeq \frac{\left(28 \ell_2^2-38 \ell_2 \left(\ell_2-x_0\right)+13 \left(\ell_2-x_0\right){}^2\right)
   \left(\ell_2-x_0\right){}^6}{7644 \ell_2^2} + O(f) \quad , \quad 0<x_0<\ell_2 \\
&& \langle S(x_0)^2 \rangle_{\ell_2} = \frac{\ell_2^6 \left(6 \ell_2^7-10 \ell_2^6 \left(\ell_2-x_0\right)+13 \left(\ell_2-x_0\right){}^7\right)}{22932
   \left(\ell_2-x_0\right){}^7} + O(f) \quad , \quad x_0<0
\eea 
which is continuous at the seed, at $x_0=x_s=0$, 
where one finds $<S(x_0=x_s=0)^2>_{\ell_2} = \frac{\ell_2^2 \left(156 f^2+35 f \ell_2^2+3 \ell_2^4\right)}{7644}$.\\

{\bf Scaling of higher moments near the edge}. From the above results we see that near the edge, i.e.
for $\ell_2 - x_0 \to 0$, the first
and second moment scale as
\bea
&& \langle S(x_0) \rangle_{\ell_2} \simeq \frac{1}{21} (\ell_2-x_0)^3 \quad , \quad  \langle S(x_0)^2 \rangle_{\ell_2} \simeq \frac{1}{273} (\ell_2-x_0)^6
\eea 
i.e. they scale in a way consistent with the following ansatz: in any given realization of the disorder
conditioned to a given value of $\ell_2$
\bea \label{sigmadef} 
S(x_0) \simeq (\ell_2-x_0)^3 \sigma 
\eea 
where $\sigma$ is a positive random variable of $O(1)$, with $ \langle \sigma \rangle= \frac{1}{21}$ and 
$ \langle \sigma^2 \rangle= \frac{1}{273}$. One can ask if there is an efficient method to obtain the higher 
moments of $\sigma$.

It is indeed the case. One can surmise (and check a posteriori) that near the edge $\rho-x_0$ and 
$r-x_0$ vanish similarly, i.e. guess that the following scaling form holds
\be \label{ansatz1} 
\frac{1}{\rho-x_0} \simeq \frac{A(\tilde \lambda)}{r-x_0} \quad , \quad  \tilde \lambda=(r-x_0)^3 \lambda
\ee
Then using \eqref{cond1}, \eqref{res1g3} and the homogeneity property
of the Weirstrass function $a^2 \mathcal{P}(a y; a^{-4} g_2, a^{-6} g_3) = \mathcal{P}(y; g_2, g_3)$
(see \eqref{homo}) with $y=1$, $a=r-x_0$, we obtain the condition
\bea \label{cond4} 
A(\tilde \lambda)^2 = {\cal P}(1 ; 0, - \frac{\tilde \lambda^2}{36} - \frac{2}{3} \tilde \lambda A(\tilde \lambda)^3)
\eea 
which allows to determine $A(\tilde \lambda)$ in an expansion in $\lambda$ using the expansion
\eqref{Pexp3t}. It gives for lowest order terms
\bea \label{At} 
A(\tilde \lambda)=1-\frac{\lambda }{84}-\frac{11 \lambda ^2}{91728}-\frac{289 \lambda
   ^3}{146397888}+O\left(\lambda ^4\right)
\eea

Then we obtain
\bea
&& \langle e^{\lambda S(x_0)} \rangle_{\ell_2=r}|_{f=0^+}= \frac{r^3}{12} \partial_r u_r^\lambda(0)  = f(\lambda (r-x_0)^3) 
\eea
where the function $f(\tilde \lambda)$ is given by
\bea \label{deff} 
f(\tilde \lambda) = \frac{r^3}{12} \partial_r \tilde u^\lambda_r(0) = - \frac{r^3}{2} \partial_r \frac{1}{\rho^2} 
= - \frac{r^3}{2} \partial_r \frac{1}{(x_0 + \frac{(r-x_0)}{A((r-x_0)^3 \lambda)})^2}
= \frac{1}{A(\tilde \lambda)} - \frac{3 \tilde \lambda A'(\tilde \lambda)}{A(\tilde \lambda)^2}
\eea 
discarding subdominant terms of order $O(r-x_0)$. From \eqref{sigmadef}, this gives the generating function of the moments of $\sigma$, i.e.
\be \label{genfunct} 
\langle e^{\tilde \lambda \sigma} \rangle = f(\tilde \lambda)
\ee
Explicit calculation gives
\bea
f(x) = \frac{x}{21}+\frac{x^2}{546}+\frac{265 x^3}{4066608}+\frac{1223
   x^4}{548992080}+\frac{3223267 x^5}{43363786415040}+\frac{407163391
   x^6}{166863850125073920}+O\left(x^7\right)
\eea 
which gives the six lowest moments as
\bea \label{momom} 
\left\{\frac{1}{21},\frac{1}{273},\frac{265}{677768},\frac{1223}{22874670},\frac{3223267}
   {361364886792},\frac{407163391}{231755347395936}\right\}
\eea 
while the cumulant are obtained from $\ln f(\tilde \lambda)$ as
\bea
\left\{\frac{1}{21},\frac{8}{5733},\frac{1531}{18299736},\frac{2623}{346932495},\frac{989
   243}{1084094660376},\frac{7165892365}{52014590781175386}\right\}
\eea
Examination of $f(x)$ shows that the coefficients of $\lambda^p$ behaves as $ \approx 3.4 (0.031)^p$
which suggests that $f(\tilde \lambda)$ blows up at a finite $\lambda^*$. It is easy to see from \eqref{deff} that such a $\lambda^*$ corresponds to $A(\tilde \lambda^*)=0$, which is also in agreement
with examination of the series \eqref{At}. Comparing with \eqref{cond4} we see that indeed
such a root exists since we know that the
Weirstrass function has a zero 
\bea
{\cal P}(1 ; 0, g_3^*) = 0
\eea 
for the negative value (see also discussion below in Section \ref{subsec:cdf}) 
\bea
g_3^* = - \frac{256 \pi ^3 \Gamma \left(\frac{1}{3}\right)^6}{\Gamma \left(-\frac{1}{6}\right)^6}
=  -\frac{4 \Gamma \left(\frac{1}{3}\right)^6 \Gamma \left(\frac{7}{6}\right)^6}{\pi ^3} =  - 30.40090507702060
\eea
Hence we see that 
\bea
A(\tilde \lambda^*)=0 \quad , \quad \lambda^* = 6 \sqrt{-g_3^*} = 
\frac{12 \Gamma \left(\frac{1}{3}\right)^3 \Gamma \left(\frac{7}{6}\right)^3}{\pi ^{3/2}} = 33.08220946
\eea 
We can further expand the equation \eqref{cond4} near $\lambda^*$. Using the following properties of the Weierstrass function
\bea
&& \partial_{g_3} {\cal P}(z ; 0, g_3) = \frac{1}{6 g_3} (2 {\cal P}(z ; 0, g_3) + {\cal P}'(z ; 0, g_3) ) \\
&& {\cal P}'^2=4 {\cal P}^3 - g_2 {\cal P} - g_3 
\eea
we see that at $g_3=g_3^*$ one has ${\cal P}'(1,0,- |g_3^*|)  = - (- g_3^*)^{1/2}$, hence
\bea
\partial_{g_3} {\cal P}(z ; 0, g_3)|_{g_3=g_3^*} = \frac{1}{6 (- g_3^*)^{1/2} } = \frac{1}{\tilde \lambda^*}
\eea
Hence \eqref{cond4} implies the leading behavior near the singularity
\bea
A(\tilde \lambda) \simeq \frac{1}{3 \sqrt{2}} (\tilde \lambda^*- \tilde \lambda)^{1/2}
\eea 
which in turn leads to
\bea \label{fls} 
f(\tilde \lambda) \simeq \frac{9 \tilde \lambda^*}{\sqrt{2} (\tilde \lambda^*- \tilde \lambda)^{3/2}}
\eea 
We have obtained in Appendix \ref{app:lambdastar} three more orders in this expansion
around $\lambda^*$. The leading order in \eqref{fls} 
implies the following large $\sigma$ behavior of the PDF of $\sigma$
\bea \label{psig} 
p(\sigma) \simeq_{\sigma \to +\infty} \frac{9 \sqrt{2}}{\sqrt{\pi}} \tilde \lambda^* \sigma^{1/2} e^{- \tilde \lambda^* \sigma} 
\eea
Thus, although we have not obtained an explicit form for the exact PDF of $\sigma$
we have been able to extract a lot of information from the implicit form (moments, tail behavior).  


\subsection{spatiotemporal shape near a boundary}
\label{sec:spatiotemp} 

Here we will study the dynamics near the edge of the avalanche.
We consider an avalanche (started at time zero) which has its upper edge at $r$. Then we
ask about the PDF of the instantaneous local velocity $\dot u(x_0,t_0)$ at some space time point 
$(x_0,t_0)$, for such an avalanche (i.e. conditioned to have its edge at $r$). To address this question we use the source
\bea
&& \lambda(x,t)=\lambda_r \delta(x-r) +  \lambda \delta(x-x_0) \delta(t-t_0) 
\eea
in the limit $\lambda_r \to -\infty$. 
Note that the source
\bea
&& \lambda(x,t)=\lambda_r \delta(x-r) \theta(t_0-t) +  \lambda \delta(x-x_0) \delta(t-t_0) 
\eea
in the limit $\lambda_r \to -\infty$ would address a different question, by only imposing 
that the avalanche has not reached point $r$ up to time $t_0$ (it could reach it afterwards).
Here we impose that the avalanche will never reach point $r$, i.e. for all times, which turns out to
be a simpler question. 

The general method of \cite{LeDoussalWiese2011a,LeDoussalWiese2012a,DobrinevskiLeDoussalWiese2011b}
implies that, for any (monotonous forward) driving $\dot f(x,t) \geq 0$, the LT of the
following joint probability $P(S(r)=0, \dot u(x_0,t_0))=P(\ell_2<r,\dot u(x_0,t_0))$ involving the instantaneous local velocity 
\bea
\int_0^{+\infty} d\dot u(x_0,t_0)P(S(r)=0, \dot u(x_0,t_0)) e^{\lambda \dot u(x_0,t_0)}
= e^{ \int_0^{+\infty} dt \int dx \dot f(x,t) \tilde u_r^\lambda(x,t) } 
\eea 
where $\tilde u_r^\lambda(x,t)$ is the solution of
\bea
\partial_t \tilde u(x,t) + \partial_x^2 \tilde u(x,t) + \tilde u(x,t)^2 = - \lambda \delta(x-x_0) \delta(t-t_0) \label{eq1n}
\eea 
on $x \in ]-\infty,r]$, with $\tilde u(r,t)= - \infty$, vanishing at $x=- \infty$. We will only determine the mean of $\dot u(x_0,t_0)$, hence we can expand in $\lambda$ and look for a solution
of the form
\bea
\tilde u_r^\lambda(x,t) = - \frac{6}{(r-x)^2} + \lambda \tilde u^{(1)}_r(x,t) + O(\lambda^2)
\eea 
where $\tilde u_1(x,t) =0$ for $t>t_0^+$ (a consequence of causality, see \cite{LeDoussalWiese2011a,LeDoussalWiese2012a}). Inserting in \eqref{eq1n} we see that
$\tilde u^{(1)}_r(x,t)$ is solution of the linear
equation
\bea
\partial_t \tilde u^{(1)}(x,t) + \partial_x^2 \tilde u^{(1)}(x,t) - \frac{12}{(r-x)^2} \tilde u^{(1)}(x,t) = -  \delta(x-x_0) \delta(t-t_0) \label{eq1n}
\eea 
Here one recognizes the equation for the propagator $G=G(x,y;\tau)=\langle x |e^{- H \tau}|y\rangle$ of the 1D Schrodinger equation in imaginary time with
inverse square potential, solution of 
\bea
\partial_\tau G = - H G = \partial_x^2 G - \frac{\gamma^2-1/4}{x^2} G \quad , \quad G(x,y,\tau=0)=\delta(x-y)
\eea 
which reads
\bea
G(x,y,\tau) = \frac{\sqrt{x y}}{2 \tau} I_\gamma(\frac{x y}{2 \tau}) e^{- \frac{x^2+y^2}{4 \tau}} 
\eea 
The correspondence takes the form
\bea
\tilde u^{(1)}_r(x,t)  = G(r-x,r-x_0,t_0-t) \quad , \quad \gamma= \frac{7}{2} 
\eea 

Let us now consider a kick $\dot f(x,t)=f \delta(x) \delta(t)$, i.e. for which the seed of the
avalanche occurs at $x_s=0,t_s=0$. Using the same method as in previous Sections (see 
Appendix \ref{sec:rel}) we obtain the conditional mean velocity
\bea \label{mvel1} 
\langle \dot u(x_0,t_0) \rangle_{\ell_2<r} = f \tilde u_r^{(1)}(0,0) = f 
\frac{\sqrt{r(r-x_0)}}{2 t_0} I_{\frac{7}{2}}(\frac{r(r-x_0)}{2 t_0}) e^{- \frac{r^2+(r-x_0)^2}{4 t_0}} 
\eea 
There should be a sum rule that integrating over time one recovers the mean spatial shape, 
since $\int_0^{+\infty} dt_0 \dot u(x_0,t_0)= S(x_0)$. Indeed, integrating \eqref{mvel1} over time, one can
check that one recovers \eqref{condmeanint} using that
\bea
\int_0^{+\infty} d\tau G(x,y,\tau) =\frac{1}{2 \gamma} x^{\frac{1}{2}-\gamma} 
y^{\frac{1}{2}+\gamma} \theta(x>y) + \frac{1}{2 \gamma} x^{\frac{1}{2}+\gamma} 
y^{\frac{1}{2}-\gamma} \theta(x<y) 
\eea

We can now calculate the mean spatio-temporal profile at fixed seed-to-egde distance $\ell_2$ as
(see Appendix \ref{sec:rel})
\bea
\langle \dot u(x_0,t_0) \rangle_{\ell_2} = \frac{1}{\rho(\ell_2)} \partial_r u_r^{(1)}(0,0)|_{r=\ell_2} + \langle \dot u_{x_0t_0} \rangle_{\ell_2<r}
\simeq_{f \to 0^+} \frac{\ell_2^3}{12} \partial_r u_r^{(1)}(0,0)|_{r=\ell_2} 
\eea 
Performing the derivatives the final result takes the following scaling form
\bea \label{spatiotemp} 
&& \langle \dot u(x_0,t_0)\rangle_{\ell_2} 
= \frac{\ell_2^3}{t_0} F(\frac{\ell_2-x_0}{\ell_2},\frac{\ell_2}{\sqrt{t_0}}) 
\eea
where the spatiotemporal shape scaling function takes the two equivalent forms 
\bea \label{spatiotempscal} 
&& F(z,r)=
\frac{z+1}{48 \sqrt{z}} e^{-\frac{1}{4} r^2 \left(z^2+1\right)} \left(r^2 z
   I_{\frac{5}{2}}\left(\frac{r^2 z}{2}\right)-\left(6+r^2 z\right)
   I_{\frac{7}{2}}\left(\frac{r^2 z}{2}\right)\right) \\
&& = \frac{(z+1) e^{-\frac{1}{4} r^2 (z+1)^2} }{24 \sqrt{\pi } r^7 z^4}
\left(12 e^{r^2 z} \left(r^2 z \left(r^2
   z-10\right)+30\right)
- r^2 z \left(r^2 z \left(r^2 z \left(r^2
   z+12\right)+72\right)+240\right)-360\right) 
\eea
using explicit forms of half-integer Bessel functions, see Appendix \ref{sec:Bessel}.
The spatiotemporal velocity profile is plotted in Fig. \ref{fig:velocity}. \\

Near the avalanche edge the spatiotemporal shape scaling function exhibits the following
behavior
\be
F(z,r)= \frac{e^{-\frac{r^2}{4}} r^7 z^3}{5040 \sqrt{\pi }}-\frac{\left(e^{-\frac{r^2}{4}} r^7
   \left(r^2-8\right)\right) z^4}{40320 \sqrt{\pi }}+\frac{e^{-\frac{r^2}{4}} r^9
   \left(r^2-18\right) z^5}{241920 \sqrt{\pi }}+O\left(z^6\right) 
\ee
which leads to 
\bea
&& \langle \dot u(x_0,t_0) \rangle_{\ell_2} =
\frac{l_2^{7} e^{-\frac{l_2^2}{4 t_0}} \left(l_2-x_0\right){}^3}{5040
   \sqrt{\pi } t_0^{9/2}}  -\frac{l_2^{6} e^{-\frac{l_2^2}{4 t_0}} \left(l_2^2-8
   t_0\right) \left(l_2-x_0\right){}^4}{40320 \sqrt{\pi }
   t_0^{11/2}} + O\left((l_2-x_0){}^5\right)
\eea 
Thus the mean instantaneous local velocity vanishes near the edge of the avalanche 
with the same exponent $3$, as the mean shape. One can define the time 
dependent amplitude ${\cal A}_{t_0}$ as
\bea
\langle \dot u(x_0,t_0) \rangle_{\ell_2} \simeq {\cal A}_{t_0} \left(l_2-x_0\right){}^3 \quad , \quad 
 {\cal A}_{t_0} = \frac{1}{t_0} C(\frac{t_0}{\ell_2^2}) \quad , \quad C(\tau)= \frac{e^{-\frac{1}{4 \tau}}}{5040 \sqrt{\pi } \tau^{7/2}} \label{edgedyn} 
\eea 
One finds that as time increases, the amplitude ${\cal A}_{t_0}$ reaches a maximum and then decays, 
the maximum being reached 
at $t_0=t_0^*=\ell_2^2/18$ with value $\frac{2187 \sqrt{\frac{2}{\pi }}}{35 e^{9/2} \ell_2^2} = 0.553854/\ell_2^2$.\\

One can now check that 
integrating over $t_0$ one recovers, order by order in $z= \frac{\ell_2-x_0}{\ell_2}$, the mean spatial shape
obtained above in \eqref{res1b} 
\bea
\int_0^{+\infty} dt_0 \langle \dot u(x_0,t_0) \rangle_{\ell_2} = 
\langle S(x_0) \rangle_{\ell_2} \simeq \frac{1}{21} (\ell_2-x_0)^3 
- \frac{1}{28 \ell_2} (\ell_2-x_0)^4
\eea 
It contains only two terms, i.e. one checks that all coefficients in the expansion, beyond $z^4$ integrate to zero. The sum rule 
\bea
\int_0^{+\infty} dt_0 {\cal A}_{t_0} =  \int_0^{+\infty} \frac{d\tau}{\tau} C(\tau) = \frac{1}{21}
\eea 
leads the static leading amplitude obtained in \eqref{leadingD}.


\subsection{Massive case}
\label{sec:1boundarymass}

In this section we study the effect of a finite mass, and generalize some of
the previous results to the massive case. We use the massive units.
We only sketch the derivations, which are very similar to the ones of the
previous sections.\\

\subsubsection{PDF of seed-to-edge distance $\ell_2$}

In the massive case, to condition to one boundary, i.e. to calculate $P(S(r)=0)$, one needs to solve the instanton equation 
\bea \label{instmass} 
\tilde u(x)'' - \tilde u(x) + \tilde u(x)^2 = 0 \quad , \quad \tilde u(r)=-\infty
\eea 
with the condition that $\tilde u(x)$ vanishes at $x=- \infty$. The solution is
\bea
\tilde u_r(x) =  - \frac{3}{2 \sinh(  \frac{1}{2} (r-x))^2}
\eea 
Considering a kick of amplitude $f$ (we recall $f=m^2 w$ in massive case dimensionful units)
applied at $x_s=0$, it
immediately leads to the CDF and PDF of the seed-to-edge distance $\ell_2$ as
for 
\bea \label{Prhomass} 
&& {\rm Prob}(\ell_2<r ) =  e^{ - f \frac{3}{2 \sinh(  \frac{1}{2} r)^2} } \\
&& P(\ell_2) = \frac{3 f}{4} \frac{\sinh \ell_2}{\sinh^4(\ell_2/2)} e^{ - f \frac{3}{2 \sinh(\ell_2/2)^2} } \quad , \quad
\rho(\ell_2) = \frac{3}{4} \frac{\sinh \ell_2}{\sinh^4(\ell_2/2)} 
\eea 
The behavior for small $\ell_2 \ll 1$ (that is $\ell_2 \ll 1/m$) is the same as the one of the massless case, Eq.
\eqref{PDFl22},
and more generally all small distance behavior is described by the massless case. The
mass only provides a cutoff for the large avalanches. \\

\subsubsection{Mean shape conditionned to the seed-to-edge distance $\ell_2$}

To calculate the mean spatial shape at the point $x_0$, let us add the source term $- \lambda \delta(x-x_0)$ on the l.h.s of \eqref{instmass}, and look for the solution in a small $\lambda$ expansion as $\tilde u_r^\lambda(x) = \tilde u_r(x) + \lambda 
\tilde u^{(1)}_r(x) + O(\lambda^2)$. Then the first order term,
$\tilde u^{(1)}_r(x)$, satisfies the linear equation
\bea
\tilde u^{(1) \prime \prime}(x) - (1 + \frac{3}{\sinh^2((r-x)/2)} ) \tilde u^{(1)}(x) 
=  -  \delta(x-x_0) \label{greenmass1} 
\eea 
i.e. it is again the zero energy Green function $\tilde u^{(1)}_r(x)=\langle x| \frac{1}{H} | x_0 \rangle$
of the linear "Schrodinger" equation, with $H=-\partial_x^2 + V(x)$ and potential $V(x)=1 + \frac{3}{\sinh^2((r-x)/2)}$.

It is convenient to introduce the variable $t=\tanh((r-x)/2)$, and write $\tilde u^{(1)}(x)=g(t)$. 
In these coordinates the homogenous part of the differential equation \eqref{greenmass1} 
becomes 
\bea
\left(t^2-1\right)^2 g''(t)+2 t \left(t^2-1\right) g'(t)+g(t)
   \left(-\frac{12}{t^2}+8\right) = 0
\eea 
and we find two solutions of this equation. The first one blows up at $x=r$ and
decays at large negative $x$
\bea
u_1(x) = g_1(t) = \frac{1-t^2}{t^3} = \frac{4 e^{r-x} \left(e^{r-x}+1\right)}{\left(e^{r-x}-1\right)^3} \simeq \frac{8}{(r-x)^3} 
\eea 
where the last equation is the small $r-x$ behavior.
The second solution vanishes at $x=r$ and blows up at large negative $x$
\bea
u_2(x)=g_2(t)= \frac{15 \left(1-t^2\right) (r-x)}{8 t^3}  - \frac{8 t^4-25 t^2+15}{4 t^2(1-t^2)}
\simeq \frac{(r-x)^4}{56} 
\eea 
where the last equation is the small $r-x$ behavior.
Each of these solutions match the two power laws that we found in the massless case
in \eqref{proper}.\\

The solution of \eqref{greenmass1} is thus
\bea
\tilde u^{(1)}_r(x) = u_1(x) u_2(x_0) \theta(x<x_0) + u_2(x) u_1(x_0) \theta(x_0<x<r) 
\eea 
which is continuous, and with the required derivative discontinuity $u_1(x) u_2'(x)- u'_1(x) u_2(x)=-1$, at $x=x_0$. 

We give now the explicit expressions for the mean shape conditioned to the 
seed-to-edge distance $\ell_2$ in the massive case. One finds, for a seed at $x_s=0$
\bea
&& \langle S(x_0) \rangle_{\ell_2<r} = f \tilde u^{(1)}_r(0) \\
&& = \frac{f}{32} 
\frac{\sinh (r-x_0)}{\sinh^2\left(\frac{r}{2}\right)
\sinh^4\left(\frac{r-x_0}{2}\right) }
   \left(-14 \cosh (r)+\cosh (2 r)+30 r \coth
   \left(\frac{r}{2}\right)-47\right) \quad , \quad x_0<0 \\
   &&
  = \frac{f}{16} \frac{\coth \left(\frac{r}{2}\right)}{ \sinh^2\left(\frac{r}{2}\right)
   \sinh^2\left(\frac{r-x_0}{2}\right) } \left(-14 \cosh (r-x_0)+\cosh (2 (r-x_0))+30
   (r-x_0) \coth \left(\frac{r-x_0}{2}\right)-47\right) \quad , \quad x_0>0
 \nn
\eea 
One also can write, in the limit $f \to 0^+$
\bea
\langle S(x_0) \rangle_{\ell_2=r} = \frac{1}{\rho(\ell_2=r)} \partial_r \tilde u^{(1)}_r(0) 
=  \frac{4}{3} \frac{\sinh^4(r/2)}{\sinh r} \partial_r \tilde u^{(1)}_r(0) 
\eea 
using \eqref{Prhomass}. This leads to, the mean shape in the massive case, for $x_s=0<x_0<\ell_2$
\bea \label{msm1} 
&& \langle S(x_0) \rangle_{\ell_2}= -\frac{1}{96 \sinh\left(\ell_2 \right) 
   \sinh^4\left(\frac{\ell_2-x_0}{2}\right)} \bigg(\cosh (2 \ell_2-3 x_0)-10 \cosh (\ell_2-2 x_0)-6 \cosh (3
   \ell_2-2 x_0) \nn \\
   && -130 \cosh (\ell_2-x_0)-32 \cosh (2 (\ell_2-x_0))+2 \cosh (3 (\ell_2-x_0)) +80 \cosh (\ell_2)  \\
   && +5 (4 (3 (\ell_2-x_0) (2 (\sinh
   (\ell_2-x_0)+\sinh (\ell_2))+\sinh (2 \ell_2-x_0))+3 \cosh (x_0)+8)-25 \cosh (2 \ell_2-x_0)) \bigg) \nn \\
&& = \frac{1}{21} (\ell_2-x_0)^3-\frac{\left(4 e^{\ell_2}+e^{2 \ell_2}+1\right) (\ell_2-x_0)^4}{84 \left(e^{2
   \ell_2}-1\right)} +\frac{(\ell_2-x_0)^7}{9240}+O\left((\ell_2-x_0)^8\right) \label{msm2} 
\eea 
One can check that for small $r,x_0 \ll 1$ (in the massive units)
one recovers exactly \eqref{res1b2}. Comparing with \eqref{res1b2}
one also sees that the leading behavior of the mean shape near the edge for $\ell_2-x_0 \to 0$ is the same as in the massless 
case, however the coefficient of the 
next order term now depends on $\ell_2$ in a different way: it crosses over from
$- \frac{1}{28 \ell_2}$ for small $\ell_2$ to $- \frac{1}{84}$ for $\ell_2 \to +\infty$. 
Note also the term $O((\ell_2-x_0)^7)$ which, although its amplitude is $\ell_2$ independent,
does not exist in the massless case. In fact all terms of order higher than $4$ disappear
in this limit. One finds, for the mean shape for $x_0<0$ 
\bea 
&&\!\!\!  \langle S(x_0) \rangle_{\ell_2}= 
   \frac{1}{96 \sinh (\ell_2) \sinh ^4\left(\frac{\ell_2-x_0}{2}\right)}
   (-20 (3 \ell_2 (2 (\sinh (\ell_2-x_0)+\sinh (\ell_2))+\sinh (2
   \ell_2-x_0))+8) \nn \\
   && -80 \cosh (\ell_2-x_0)+125 \cosh (2 \ell_2-x_0)+6 \cosh (3 \ell_2-x_0)+10 \cosh (\ell_2+x_0)-\cosh (2
   \ell_2+x_0) \nn \\
   && +130 \cosh (\ell_2)+32 \cosh (2 \ell_2)-2 \cosh (3 \ell_2)-60 \cosh (x_0))  = A(\ell_2) e^{x_0} + O(e^{2 x_0}) \label{msm3}  \\
   &&  A(\ell_2) = \frac{40 (3 \ell_2+2)+e^{-3 \ell_2}-10 e^{-2 \ell_2}+60 e^{-\ell_2}-6 e^{2 \ell_2}+5 e^{\ell_2 }(12 \ell_2-25)}{6-6 e^{2 \ell_2}}
   = 1 + (\frac{125}{6} - 10 \ell_2) e^{-\ell_2} + O(\ell_2 e^{-2 \ell_2}) \nn \\
   && ~~~~~~
   = \frac{\ell_2^6}{42}-\frac{5 \ell_2^7}{168}+\frac{\ell_2^8}{56}+O\left(\ell_2^9\right) \nn
      \eea 
Hence the mean shape $\langle S(x_0) \rangle_{\ell_2}$ now decays exponentially at large $x_0 \to -\infty$, instead as a power law for the massless case. The prefactor of this exponential decay, $A(\ell_2)$, depends on $\ell_2$: it grows as $\ell_2^6$ at small $\ell_2$ and converges to unity at large $\ell_2$.
 
\subsubsection{Joint PDF of total size $S$ and $\ell_2$}

Finally we can ask about the the joint PDF of the total size $S$ and $\ell_2$
in the massive case. We need to solve the massive version of \eqref{eqmu}
\bea \label{eqmu2} 
\tilde u''(x) - \tilde u(x) + \tilde u(x)^2 = - \mu \quad , \quad \tilde u(r) =-\infty
\eea 
on $x \in ]-\infty,r]$. The solution $\tilde u(x)=\tilde u^\mu_r(x)$ reads
\bea
\tilde u^\mu_r(x) = \frac{1}{2} (1- \sqrt{1-4 \mu}) - \frac{3 \sqrt{1-4 \mu} }{2 \sinh(  \frac{1}{2} (1-4 \mu)^{1/4} (r-x))^2}
\eea 
Comparing with \eqref{idmu}-\eqref{ILT1} we see that
\bea
P_{\rm massive}(\ell_2<r ,S) = e^{f/2} P_{\rm massless}(\ell_2<r ,S) e^{-S/4}
\eea 
where $P_{\rm massless}(\ell_2<r ,S)$ is given in \eqref{ILT1}. A similar
relation holds for the densities, taking the first order in small $f$. 

Hence, all moments of $S$ now exist. Let us give here the lowest conditional moments. 
We use the relation (see Appendix \ref{sec:rel})
\bea
\langle e^{\mu S} \rangle_{\ell_2<r} = e^{f(\tilde u^\mu_r(0)- \tilde u^0_r(0))}
\eea
Expanding in powers of $\mu$ one finds the mean conditional size, together with its small $\ell_2$ and
large $\ell_2$ behavior
\bea
 \langle S \rangle_{\ell_2<r} = f \,
\frac{\cosh (r)-3 r \coth \left(\frac{r}{2}\right)+5}{\cosh (r)-1} &=& \frac{f r^2}{10}-\frac{f r^4}{168}+O\left(r^5\right) \\
&= & f ( 1 - 6 (r-2) e^{-r} + O(e^{-2 r})) 
\eea 
Comparing with \eqref{momS1} we see that the leading small $r$ term is the same 
as the massless case,
but the next order is different. Similarly for the conditional second cumulant we find
\bea
 && \langle S^2 \rangle^c_{\ell_2<r} =
\frac{f}{4 \sinh^4\left(\frac{r}{2}\right)} \left(-6 r^2+\left(8-3 r^2\right) \cosh (r)+3 r
   \sinh (r)+\cosh (2 r)-9\right) = \frac{f r^4}{84}-\frac{f r^6}{900}+O\left(r^8\right) \\
 && ~~~~~~~~~~~~~~~~~~~~~~ ~~~~~~~~~~~~~~~~~~~~~~ ~~~~~~~~~~~~~~~~~~~~~~~~~~~~~~ ~~~~~~~~~~~~~~ = f \left( 2 - 6  e^{-r} (r^2-r-4) + O(e^{-2 r}) \right) \nn
\eea
Here we cannot compare with the massless case, since the second moment does not exist in that case.

Next we can use the relation (see Appendix \ref{sec:rel})
\bea
\langle e^{\mu S} \rangle_{\ell_2=r} = \frac{\partial_r \tilde u^\mu_r(0)}{\partial_r \tilde u^0_r(0)} 
e^{f(\tilde u^\mu_r(0)- \tilde u^0_r(0))}
\eea
Expanding in powers of $\mu$ one finds the mean conditional size, together with its small $\ell_2$ and
large $\ell_2$ behavior
\bea \label{Sl2} 
 \langle S \rangle_{\ell_2} &=& \ell_2 ( 2 \coth
   \left(\frac{\ell_2}{2}\right)-\coth (\ell_2)) -3 + f \,\frac{ \left(\cosh (\ell_2)-3 \ell_2 \coth \left(\frac{\ell_2}{2}\right)+5\right)}{\cosh (\ell_2)-1}
   = \frac{f \ell_2^2}{10}+\left(\frac{1}{60}-\frac{f}{168}\right) \ell_2^4+O\left(\ell_2^5\right) \nn \\
   & = & \ell_2 -3 +  4 e^{-\ell_2} \ell_2  + f \left(1-6 e^{-\ell_2} (\ell_2-2)\right)+ O(e^{-2 \ell_2}) 
\eea
Comparing with \eqref{momS1} we see that the $\ell_2^4$ and $f \ell_2^2$ terms are the same
as the massless case, but the next orders are different. 
We give the second conditional moment in the limit $f \to 0$ and its asymptotics
\bea
 \langle S^2 \rangle_{\ell_2, f \to 0^+} &=& 
\ell_2^2+\frac{3}{2} \ell_2 \left(\tanh \left(\frac{\ell_2}{2}\right)-3 \coth \left(\frac{\ell_2}{2}\right)+ \frac{2 \ell_2}{
   \sinh^2\left(\frac{\ell_2}{2}\right)} \right)-3 = \frac{\ell_2^6}{252}-\frac{\ell_2^8}{1800}+O\left(\ell_2^{10}\right) \\
   &=&  \ell_2 (\ell_2-3)-3 + 12 \ell_2 (\ell_2-1) e^{-\ell_2} + O(e^{-2 \ell_2}) 
\eea
Note that, restoring dimensions, one has $ \langle S^2 \rangle_{\ell_2} \simeq \frac{\sigma^2}{252 m^2} \ell_2^6$ at small $\ell_2 \ll 1/m$, with a prefactor which indeed diverges as $m \to 0$ (since the
second moment does not exist in the massless case).
%

\section{Bulk driving: spatial shape conditioned to two boundaries}
\label{sec:2boundary}

We now consider the problem of the joint distribution of $\ell_1,\ell_2$, of
the total extension $\ell=\ell_1+\ell_2$ 
and of the shape conditioned to $\ell$ and to the aspect ratio, see Fig. \ref{fig:extensions}. We study
the massless case. \\

\subsection{PDF and density of $\ell_1,\ell_2$ and of $\ell,p$} 
\label{subsec:PDF12} 

Consider the driving by a kick of amplitude $f(x)$ with a bounded support and 
we ask about the probability that the avalanche has remained in the
interval $[r_1,r_2]$ with $r_1<r_2$ (which includes the support of $f(x)$). It is equal to
\bea
P(S(r_1)=0, S(r_2)=0) = e^{ \int dx f(x) \tilde{u}_{r_1,r_2}(x) }
\eea
where $\tilde{u}_{r_1,r_2}(x)$ is  the solution of the massless instanton equation 
\begin{equation}
\tilde{u}''(x) + \tilde{u}(x)^2 = 0 \quad , \quad \tilde{u}(x=r_1) = \tilde{u}(x=r_2) = - \infty 
\label{Instanton2s}
\end{equation}
on the interval $x \in [r_1,r_2]$.
The solution of this problem, as discussed in \cite{Delorme}, is an elliptic Weirstrass function
\bea
&& \tilde{u}_{r_1,r_2}(x) = - \frac{6}{(r_2-r_1)^2}  {\cal P}_0(z) \quad , \quad z= \frac{x-r_1}{r_2-r_1} \\
&& {\cal P}_0(z) := {\cal P}(z; g_2=0, g_3=g_3^0:= \frac{\Gamma(1/3)^{18}}{(2 \pi)^6} ) 
\eea 
Here the value of $g_3$ is such that
the real period is unity, i.e. $2 \Omega(g_2=0,g_3^0)=1$, see \eqref{halfper}. Two important properties
of ${\cal P}_0$ are
\bea
{\cal P}_0(1-z)={\cal P}_0(z)  \quad , \quad {\cal P}_0(z) \simeq_{z \to 0} \frac{1}{z^2} 
+ \frac{g_3^0 z^4}{28} + O(z^{10})
\eea 
Because of these properties, one immediately sees that near the edges of the interval $[r_1,r_2]$ 
one recovers to leading order the single edge solution
\bea
\tilde{u}_{r_1,r_2}(x) \simeq - \frac{6}{(r_2-x)^2} -  \frac{3 g_3^0}{14} \frac{(r_2-x)^4}{(r_2-r_1)^6} + O((r_2-x)^{10})
\eea 
plus subleading corrections.

If we consider now a kick of amplitude $f$ at $x=x_s=0$ (we define the origin as the position of the seed),
we have the joint CDF of $\ell_1,\ell_2$ (as defined in Fig. \ref{fig:extensions})
\bea
P(\ell_1<l_1, \ell_2<l_2) = e^{ f \tilde{u}_{r_1,r_2}(0) } = e^{ - 6 f \frac{1}{(l_1+l_2)^2}  {\cal P}_0(\frac{l_2}{l_2+l_1} ) }  \quad , \quad l_1=-r_1 \quad l_2=r_2
\eea 
and of course $r_1<0$ and $r_2>0$ since the interval must contain the seed. The joint PDF is
\bea
P(\ell_1,\ell_2) = \partial_{\ell_1} \partial_{\ell_2} e^{ - 6 f \frac{1}{\ell^2}  {\cal P}_0(p) } \quad , \quad \ell=\ell_1+\ell_2 \quad , \quad p = \frac{\ell_2}{\ell_1+\ell_2} 
\eea 
where $p \in [0,1]$ is the aspect ratio. This expression is complicated but its small $f$ behavior, i.e.
the density, is a bit simpler. The density is
\bea
\rho(\ell_1,\ell_2) = \partial_f|_{f=0^+} P(\ell_1,\ell_2) = \partial_{\ell_1} \partial_{\ell_2} 
\tilde{u}_{r_1=-\ell_1,r_2=\ell_2}(0) = - 6 \partial_{\ell_1} \partial_{\ell_2} [ \frac{1}{\ell^2}  {\cal P}_0(p) ]
\eea 
Clearly it is better to trade the variables $(\ell_1,\ell_2)$ for
the pair $(\ell,p)$. The reverse transformation is $\ell_1= (1-p) \ell$ and $\ell_2=p \ell$. The jacobian, and the partial derivative operators are 
\bea \label{der1} 
&& d\ell_1 d\ell_2 = \ell d\ell dp \quad , \quad \partial_{\ell_1}=\partial_\ell - \frac{p}{\ell} \partial_p \quad , \quad 
\partial_{\ell_2}=\partial_\ell + \frac{1-p}{\ell} \partial_p \\
&&  \partial_{\ell_1} \partial_{\ell_2} =  \frac{1}{\ell^2} [ (2p-1) \partial_p + p(p-1) \partial^2_p ]
+ \frac{1- 2 p}{\ell} \partial_p \partial_\ell + \partial_\ell^2 \\
&&~~~~~~~~ = \frac{1}{\ell} \partial_p [ p(p-1) \frac{1}{\ell} \partial_p + (1- 2 p)  \partial_\ell ] 
+ \frac{2}{\ell} \partial_\ell+  \partial_\ell^2 \label{der1last} 
\eea 
Since $\rho(\ell_1,\ell_2) d\ell_1 d\ell_2 = \rho(\ell,p) d\ell dp$, we have 
$\rho(\ell,p) = \ell \rho(\ell_1,\ell_2)$ (with some abuse of notation, since
we denote by $\rho$ both functions, while they are different functions of their arguments). 
Hence
\bea
 \rho(\ell,p) =  - 6 \ell \partial_{\ell_1} \partial_{\ell_2} [ \frac{1}{\ell^2}  {\cal P}_0(p) ] 
&=& - 6 \left( \partial_p [ p(p-1) \frac{1}{\ell} \partial_p + (1- 2 p)  \partial_\ell ] 
+ 2 \partial_\ell + \ell \partial_\ell^2 \right) [ \frac{1}{\ell^2}  {\cal P}_0(p) ] \\
& = & - \frac{6}{\ell^3} \left( \partial_p [  ( p(p-1) {\cal P}'_0(p) + 2(2 p-1)  {\cal P}_0(p) ) ]
+ 2 {\cal P}_0(p) \right) \label{middle} \\
& = & -\frac{6}{\ell^3}  \left((p-1) p {\cal P}_0''(p)+(6 p-3) {\cal P}_0'(p)+6 {\cal P}_0(p) \right) \label{end} 
\eea
One can check that the differential operator in $p$ in \eqref{end} vanishes when applied to
$1/p^2$ and $1/(1-p)^2$. The reason for these "zero modes" is that 
near any of the two edges, as mentionned above, one recovers the one boundary problem, where
nothing depends on $\ell_1$ or $\ell_2$ (repectively). Hence it is better to replace
${\cal P}_0 \to \tilde {\cal P}_0={\cal P}_0(p)-1/p^2 - 1/(1-p)^2$ which is now a 
regular function (vanishing quartically at each boundary $p=0,1$). This allows e.g. to
integrate by part in the line \eqref{middle} and obtain the density of total extension alone
\bea
\rho(\ell) = \int_0^1 dp \rho(\ell,p) = - \frac{6}{\ell^3} (-4+2 \int_0^1 dp \tilde P_0(p) ) = \frac{8 \sqrt{3} \pi}{\ell^3} 
\eea 
using that $[( p(p-1) \tilde {\cal P}'_0(p) + 2(2 p-1)  \tilde {\cal P}_0(p) )]_0^1=-4$.\\

In summary one can write the joint density of total extension $\ell$ and of aspect ratio
as
\bea \label{resrholp} 
\rho(\ell,p) = \frac{8 \pi \sqrt{3}}{\ell^3} \tilde P(p) \quad , \quad 
\tilde P(p) = - \frac{\sqrt{3}}{4 \pi} \left( \partial_p [  ( p(p-1) {\cal P}'_0(p) + 2(2 p-1)  {\cal P}_0(p) ) ]
+ 2 {\cal P}_0(p) \right)
\eea 
where $\tilde P(p)$ is the PDF of the aspect ratio, i.e. it is normalized to unity $\int_0^1 dp \tilde P(p)=1$.

\subsection{Mean shape conditionned to total extension and aspect ratio} 
\label{subsec:aspect} 

In this section we obtain the mean shape conditioned to the total extension and aspect ratio.
As we show, it amounts to solve a Lam\'e equation.\\

We again restrict to a local kick of amplitude $f$ at $x_s=0$ (taking the seed position
to be the origin). We want to study the joint PDF $P(S(r_1) =0,S(r_2)=0, S(x_0))
=P(\ell_1<-r_1,\ell_2<r_2,S(x_0))$ whose Laplace transform is given by
\be
\int_0^{+\infty} dS(x_0) e^{\lambda S(x_0)} P(\ell_1<-r_1,\ell_2<r_2,S(x_0))
= e^{f \tilde{u}^{\lambda, x_0}_{r_1,r_2}(0)}
\label{2point}
\ee
where we denote $\tilde{u}^{\lambda, x_0}_{r_1,r_2}(x)$ the solution of the massless instanton equation
with a source localized in $x_0$
\begin{equation}
\tilde{u}''(x) + \tilde{u}(x)^2 =  - \lambda \delta(x-x_0) \quad , \quad \tilde{u}(x=r_1) = \tilde{u}(x=r_2) = - \infty 
\label{Instanton2s}
\end{equation}
on the interval $x \in [r_1,r_2]$. It is easy to see that 
\bea \label{instsolu11} 
\tilde{u}^{\lambda, x_0}_{r_1,r_2}(x) &=& \frac{-6}{(r_2-r_1)^2} {\cal P}(\frac{x-r_1}{r_2-r_1}; g_2=0, g_3^-) \quad , \quad x < x_0 \\
&=& \frac{-6}{(r_2-r_1)^2} {\cal P}(\frac{r_2-x}{r_2-r_1}; g_2=0, g_3^+) \quad , \quad x > x_0 \nn
\eea 
The parameters $g_3^{\pm}$ are determined by the conditions of continuity
and of proper derivative jump at $x=x_0$. They read 
explicitly
\bea \label{selfc1} 
&&  {\cal P}(z_0, g_2=0, g_3^-)  =  {\cal P}(1-z_0, g_2=0, g_3^+) \\
&& {\cal P}'(z_0, g_2=0, g_3^-) + {\cal P}'(1-z_0, g_2=0, g_3^+) = - \hat \lambda \nn \\
&& z_0 = \frac{x_0-r_1}{r_2-r_1} \quad , \quad 1-z_0 = \frac{r_2-x_0}{r_2-r_1}
\quad  , \quad  \hat \lambda = \frac{\lambda}{6} (r_2-r_1)^3 \nn
\eea 
We note that for $\lambda=0$ one has
\bea
\lambda=0 \Rightarrow  g_3^-=g_3^+ = g_3^0  = \frac{\Gamma(1/3)^{18}}{(2 \pi)^6}  \quad , \quad
\tilde{u}^{0, x_0}_{r_1,r_2}(x) = \frac{-6}{(r_1-r_2)^2} {\cal P}_0(\frac{r_2-x}{r_2-r_1})
= \frac{-6}{(r_1-r_2)^2} {\cal P}_0(\frac{x-r_1}{r_2-r_1})
\eea
since ${\cal P}_0'(\frac{1}{2}) ={\cal P}'(\frac{1}{2}, g_2=0, g_3^0) =0$. \\

Brute force expanding in powers of $\lambda$ allows to get the mean shape as well as
its higher moments. This is detailed in the Appendix. For the mean shape it is equivalent
but easier to write, as we did in previous sections, a linear "Schrodinger" equation. Let us
define the expansion in $\lambda$ of the solution of the instanton equation
\bea \label{exp22} 
 \tilde{u}^{\lambda, x_0}_{r_1,r_2}(0) = \tilde{u}_{r_1,r_2}(x) + \lambda \tilde u^{(1)}_{r_1,r_2}(x) + O(\lambda^2) \quad ,\quad \tilde{u}_{r_1,r_2}(x) = - \frac{6}{(r_2-r_1)^2}  {\cal P}_0(z) \quad , \quad z= \frac{x-r_1}{r_2-r_1} 
\eea 
Inserting into \eqref{Instanton2s} we see that the $O(\lambda)$ term, $\tilde u^{(1)}_{r_1,r_2}(x)$ is the solution of the linear equation
\bea
\tilde u^{(1) \prime \prime}(x) - \frac{12}{(r_2-r_1)^2}  {\cal P}_0( \frac{x-r_1}{r_2-r_1}) \tilde u^{(1)}(x) = - \delta(x-x_0)
\eea 
which is again the zero energy Green function $\tilde u^{(1)}_{r_1,r_2}(x)=\langle x| \frac{1}{H} | y \rangle$
of the "Schrodinger" equation, associated to the Hamiltonian $H=-\partial_x^2 + V(x)$ with the potential 
$V(x)=\frac{12}{(r_2-r_1)^2}  {\cal P}_0( \frac{x-r_1}{r_2-r_1})$.\\

This problem is known as the Lame problem. Let us use the reduced coordinates 
$z=(x-r_1)/(r_2-r_1)$ and $z_0=(x_0-r_1)/(r_2-r_1)$ and write  $\tilde u^{(1)}(x) = \frac{1}{(r_2-r_1)^2} \psi(z)$,
then $\psi(z)$ satisfies 
\bea \label{Lame}
- \psi''(z) + 12 {\cal P}_0(z) \psi(z) = (r_2-r_1)^3 \delta(z-z_0) 
\eea 
on the interval $z \in [0,1]$.
Two linearly independent solutions of the homogeneous equation are, using extensively the symmetry $ {\cal P}_0(1-z)= {\cal P}_0(z)$ and the properties of the Weirstrass function (see Appendix \ref{app:W}) 
\bea \label{defpsi12} 
\psi_1(z) = 2 {\cal P}_0(z) + z {\cal P}'_0(z) \quad , \quad \psi_2(z)=\psi_1(1-z) = 2 {\cal P}_0(z) + (z-1) {\cal P}'_0(z)
\eea 
Note that near the edge at $z=0$ (that is $x=r_1$)
\bea \label{asymptpsi}
&& \psi_1(z) \simeq \frac{3 g_3^0 z^4}{14}+\frac{3 {g_3^0}^2 z^{10}}{2548}+O\left(z^{16}\right) \\
&& \psi_2(z) \simeq  \frac{2}{z^3}-\frac{g_3^0 z^3}{7}+O\left(z^4\right)
\eea 
and thus each function neatly corresponds, near the edge, to one of the two solutions with power laws exponents $a=-3,4$ found
in the single boundary problem, in \eqref{proper}. Note that they exchange behaviors at the other edge $z=1$ (that is $x=r_2$)
since $\psi_2(z)=\psi_1(1-z)$. Their Wronskien is
\bea
\psi'_1(z) \psi_2(z) - \psi'_2(z) \psi_1(z) = 3 g_3^0
\eea 
Thus the solution of \eqref{Lame} (continuous and with the proper derivative discontinuity at $z=z_0$) is
{ \bea
\psi(z) = \frac{(r_2-r_1)^3}{3 g_3^0} \left( \psi_1(z_0) \psi_2(z) \theta(z-z_0) +  \psi_2(z_0) \psi_1(z) \theta(z_0-z) \right)
\eea
}
So finally one has
{ \bea
\tilde u_{r_1,r_2}^{(1)}(x) = \frac{r_2-r_1}{3 g_3^0} \left( \psi_1(z_0) \psi_2(z) \theta(z-z_0) +  \psi_2(z_0) \psi_1(z) \theta(z_0-z) \right) \quad , \quad z = \frac{x-r_1}{r_2-r_1} \quad , \quad z_0 = \frac{x_0-r_1}{r_2-r_1}
\eea \\
}
%

Let us start with the cumulative conditional mean shape, which is easier to calculate, using (see Appendix \ref{sec:rel})
\be \label{ddd} 
 \langle S(x_0) \rangle_{\ell_1<- r_1, \ell_2<r_2} = f \tilde u_{r_1,r_2}^{(1)}(0) 
\ee
Since $x=0$ (position of the seed) corresponds to $z=1-p$ we obtain for $x_0<0$ (i.e. $z_0<1-p$)
\bea
\langle S(x_0<0) \rangle_{\ell_1<- r_1, \ell_2<r_2} 
&=& \frac{f}{3 g_3^0} (r_2-r_1) \psi_1(z_0) \psi_1(p) \quad , \quad p= \frac{r_2}{r_2-r_1} \quad , \quad 
z_0= \frac{x_0-r_1}{r_2-r_1} \label{mainresult210} 
 \\
& = & \frac{f}{3 g_3^0} (r_2-r_1)
(2   {\cal P}_0(z_0)+ z_0 {\cal P}_0'(z_0) )
(2 {\cal P}_0(p)   + p {\cal P}_0'(p))  \label{mainresult21} 
\eea
and for $x_0>0$ (i.e. $z_0>1-p$)
\bea
\langle S(x_0>0) \rangle_{\ell_1<- r_1, \ell_2<r_2} 
&=& \frac{f}{3 g_3^0} (r_2-r_1) \psi_2(z_0) \psi_2(p) \quad , \quad p= \frac{r_2}{r_2-r_1} \quad , \quad 
z_0= \frac{x_0-r_1}{r_2-r_1} \label{mainresult220} 
 \\
& = & \frac{f}{3 g_3^0} (r_2-r_1)
(2   {\cal P}_0(z_0)+ (z_0-1) {\cal P}_0'(z_0) )
(2 {\cal P}_0(p)   + (p-1) {\cal P}_0'(p))  \label{mainresult22} 
\eea

In the limit $r_1 \to - \infty$, using the asymptotic behaviors \eqref{asymptpsi}
we recover the result for the single boundary at $x=r_2$, i.e. \eqref{condmeanint}.
Let us sketch it for $x_0<0$, in \eqref{mainresult21} one has $p \to 0$ and $1-z_0=(r_2-x_0)/(r_2-r_1) \to 0$,
thus using \eqref{asymptpsi}, the first line in \eqref{mainresult21} becomes
$\frac{f}{3 g_3^0} (r_2-r_1) \times \frac{2}{(1-z_0)^3} \times \frac{3 g_3^0 p^4}{14} = \frac{1}{7} f \frac{r_2^4}{(r_2-x_0)^3}$ as all factors $r_2-r_1$ cancel. It works similarly for $x_0>0$. 
Note that the results \eqref{mainresult210}, \eqref{mainresult220} can also be 
obtained by a direct expansion of \eqref{instsolu11} and of the self-consistent equation \eqref{selfc1} 
in powers of $\lambda$,
see Appendix \ref{app:2bshape}.

We can now study the behavior near the edges of the avalanche. Let us look at the
edge at $x=r_2$, hence at $x_0>0$ near $r_2$ and at \eqref{mainresult22}. 
Since $z_0=1 - \frac{r_2-x_0}{r_2-r_1}$ we can rewrite
\bea
\langle S(x_0>0) \rangle_{\ell_1<- r_1, \ell_2<r_2} 
&=& \frac{f}{3 g_3^0} (r_2-r_1) \psi_1(\frac{r_2-x_0}{r_2-r_1}) \psi_2(p)  \simeq \frac{f}{14} \frac{(r_2-x_0)^4}{(r_2-r_1)^3}  \psi_2(p)
\eea
To compare with the single boundary result \eqref{condmeanint}, we rewrite it in the form
\bea
&& \langle S(x_0) \rangle_{\ell_1<- r_1, \ell_2<r_2}  = \frac{f}{7} \frac{(r_2-x_0)^4}{r_2^3} 
\phi(\frac{r_2}{r_2-r_1}) \\
&& \phi(p) = \frac{1}{2} p^3 \psi_2(p) = \frac{1}{2} p^3  (2 {\cal P}_0(p)  + (p-1)  {\cal P}_0'(z)) 
\label{phiphi} 
\eea 
where $\phi(p)$ is a smoothly decaying function from $\phi(0)=1$ to $\phi(1)=0$
\bea
&& \phi(p) = 1-\frac{p^6 \Gamma \left(\frac{1}{3}\right)^{18}}{896 \pi ^6}+\frac{3 p^7 \Gamma
   \left(\frac{1}{3}\right)^{18}}{1792 \pi ^6}-\frac{5 p^{12} \Gamma
   \left(\frac{1}{3}\right)^{36}}{41746432 \pi ^{12}}+O\left(p^{13}\right) \\
   && = \frac{3 (1-p)^4 p^3 \Gamma \left(\frac{1}{3}\right)^{18}}{1792 \pi ^6} + O((1-p)^5)
\eea 
For $r_1 \to -\infty$, the aspect ratio $p \to 0$, and using $\phi(0)=1$
one recovers \eqref{condmeanint}. We see that for all values of $p$ the
mean shape vanishes at the edge with the same power $4$. However the 
coefficient depends on the aspect ratio $p$ as described by $\phi(p)$.
Hence conditioning on $\ell_1 < - r_1$ 
changes the coefficient of the
power $(r_2-x_0)^4$ of the mean (cumulative) conditioned shape 
near the edge, with respect to no conditioning to the other edge $r_1=-\infty$. \\

Let us now calculate the mean shape. Let us start again from
\eqref{2point}. Taking $\partial_{r_1} \partial_{r_2}$ on both sides we obtain the same equation
for the joint PDF $P(\ell_1,\ell_2,S(x_0))$, and keeping only the term linear in $f$ in the 
limit $f=0^+$ we obtain the following equation for the joint density $\rho(\ell_1,\ell_2,S(x_0))
= \partial_f|_{f=0} P(\ell_1,\ell_2,S(x_0))$
\bea
\int dS(x_0) e^{\lambda S(x_0)} \rho(\ell_1,\ell_2,S(x_0)) =   \partial_{\ell_1} \partial_{\ell_2}  
\tilde u^{\lambda, x_0}_{-\ell_1,\ell_2}(0)
\eea 
For more details on these and the manipulation below see Appendix \ref{sec:rel}.
It is more convenient to use the variables $\ell,p$, and using $\rho(\ell,p) = \ell \rho(\ell_1,\ell_2)$
(see remark above Eq. \eqref{end}),
rewrite the above for the joint density $\rho(\ell,p,S(x_0))$ as
\bea \label{LTpl} 
\int dS(x_0) e^{\lambda S(x_0)} \rho(\ell,p,S(x_0)) =  \ell \partial_{\ell_1} \partial_{\ell_2}  
\tilde u^{\lambda, x_0}_{-\ell_1,\ell_2}(0)
\eea 
Expanding to first order in $\lambda$ and using \eqref{exp22} we find that 
the mean shape conditioned to the couple $\ell,p$ is given by, in the limit $f=0^+$ (we recall
that the seed is in position $x_s=0$)
\bea \label{res10}
\langle S(x_0) \rangle_{\ell,p} = \frac{1}{\rho(\ell,p)} 
\ell \partial_{\ell_1} \partial_{\ell_2}  \tilde u^{(1)}_{-\ell_1,\ell_2}(0)
\eea 
Similarly to obtain the mean shape conditioned only on $\ell$, we can integrate over $p$ both sides of \eqref{LTpl} 
and then divide by $\rho(\ell)$. This leads to
\bea \label{res2}
\langle S(x_0) \rangle_{\ell} = \frac{1}{\rho(\ell)} \int_0^1 dp 
\, \ell \partial_{\ell_1} \partial_{\ell_2}  \tilde u^{(1)}_{-\ell_1,\ell_2}(0)
\eea 

To continue the calculation from \eqref{res10}, \eqref{res2} we need to calculate the second derivative of the solution to the instanton equation. Let us start with $x_0<0$, i.e. for $z_0 <  1-p$, using \eqref{ddd} and \eqref{mainresult210}. We obtain
\bea \label{w1}
&& \ell \partial_{\ell_1} \partial_{\ell_2} u^1_{-\ell_1,\ell_2}(0) = \frac{1}{3 g_3^0} G(p,z_0) \quad , \quad x_0<0 \quad , \quad z_0<1-p \\
&& G(p,z_0) = 
 (p-1) p \psi_1\left(z_0\right) \psi_1''(p)+\left(z_0-1\right) z_0 \psi_1(p) \psi_1''\left(z_0\right)+\left(p \left(2
   z_0-1\right)-z_0+1\right) \psi_1'(p) \psi_1'\left(z_0\right) \nn
\eea
Note that the derivative is at fixed $x_0$. 
Note that $G(p,z_0)=G(z_0,p)$ is a symmetric function of its arguments. 
The calculation for $x_0>0$, i.e. for $z_0 >  1-p$ using \eqref{mainresult220} gives exactly the same
result replacing $\psi_1$ by $\psi_2$ hence using that $\psi_2(z)=\psi_1(1-z)$ it gives
\bea \label{w2}
&& \ell \partial_{\ell_1} \partial_{\ell_2} u^1_{-\ell_1,\ell_2}(0) = \frac{1}{3 g_3^0}  G(1-p,1-z_0) \quad , \quad x_0>0 \quad , \quad z_0>1-p 
\eea
Hence we obtain from \eqref{res10} the mean shape conditioned to both the total extension
and the aspect ratio
\bea \label{shapeboth} 
&& \langle S(x_0) \rangle_{\ell,p} = \frac{1}{\rho(\ell,p)} \hat G(p,z_0) 
\quad , \quad p= \frac{\ell_2}{\ell_2+\ell_1} \quad , \quad 
z_0= \frac{x_0+\ell_1}{\ell_2+\ell_1} \\
&& \hat G(p,z_0) =  \frac{1}{3 g_3^0 } ( G(p,z_0) \theta(z_0<1-p) + 
G(1-p,1-z_0) \theta(z_0>1-p) )  \label{shapeboth2} 
\eea 
where we recall that $G(p,z_0)$ is defined in \eqref{w1}, $\psi_{1,2}$ are defined in \eqref{defpsi12} 
and $\rho(\ell,p)$ was obtained in \eqref{resrholp}.

To obtain $\langle S(x_0) \rangle_{\ell}$ from \eqref{res2} one needs to integrate over $p$ in 
\eqref{w1} and \eqref{w2}, more precisely one has
\bea \label{mean3} 
 \langle S(x_0) \rangle_{\ell} = \frac{f(z_0)}{\rho(\ell)} \quad , \quad f(z_0)= \int_0^1 dp \, \hat G(p,z_0) \quad , \quad 
z_0= \frac{x_0+\ell_1}{\ell} \
\eea 
It turns out that this integral can be performed explicitly. Since it is not trivial however, we will
use a trick. We note that, using the symmetry between $p$ and $z_0$ in the
definitions of $G$ and $\hat G$ we have that the following marginals define
the same function $f$
\bea
\int_0^1 dp \, \hat G(p,z_0) = f(z_0) \Leftrightarrow  \int_0^1 dz_0 \, \hat G(p,z_0) = f(p)
\eea 
Hence the function $f(p)$ can be equivalently determined from the knowledge of
the following conditional mean total size
\bea
\langle S \rangle_{\ell,p} = \int dx_0 \langle S(x_0) \rangle_{\ell,p} 
= \frac{\ell}{\rho(\ell,p)} \int_0^1 dz_0 \hat G(p,z_0) = \frac{\ell}{\rho(\ell,p)} f(p) \label{Srel} 
\eea 
We now turn to the direct calculation of $\langle S \rangle_{\ell,p}$ from which
we will obtain the function $f(p)$, hence the desired result for the mean
shape at fixed extension from \eqref{mean3}.\\

It is useful for the following to introduce the primitive of $\psi_1(z)=2 {\cal P}_0(z) + z {\cal P}'_0(z)$ 
which has an explicit form
\bea
&& \Psi_1(z) = \int_0^z  \psi_1(z') dz' = z {\cal P}_0(z) - \zeta_0(z) 
\eea
where $\zeta_0(z)=\zeta(z;0,g_3^0)$ is the zeta-Weierstrass function 
which satisfies
\bea \label{zeta0prop} 
&& \zeta_0'(z)=- {\cal P}_0(z) \quad , \quad \zeta_0(z) + \zeta_0(1-z) = \frac{2 \pi}{\sqrt{3}} 
\eea 
The fact that $ \zeta_0(z) + \zeta_0(1-z)$ is constant is a consequence of 
${\cal P}_0(z) = {\cal P}_0(1-z)$ (by differenciation w.r.t. $z$). The constant
is known since $\zeta_0(\frac{1}{2}) =  \frac{\pi}{\sqrt{3}}$.

\subsection{Joint density of size and extension}
\label{subsec:jointsizeext} 

Let us now determine directly the joint density of size and extension
which allows to obtain $\langle S \rangle_{\ell,p}$ and the function $f(z_0)$ in \eqref{mean3}.
The Laplace transform of the conditional distribution of the total avalanche size $S$ (with a seed at $x=0$) can be obtained as
\bea \label{LTS} 
\int dS e^{\mu S} P(\ell_1<- r_1, \ell_2 < r_2,S) = e^{f \tilde u_{r_1,r_2}^\mu(0)} 
\eea 
where $\tilde u_{r_1,r_2}^\mu(x)$ is the solution of 
\bea
\tilde u''(x)  + \tilde u(x)^2 = - \mu  \quad , \quad \tilde u(r_1)=-\infty  \quad , \quad \tilde u(r_2)=-\infty
\eea
on the interval $x \in [r_1,r_2]$. 
The solution is
\bea \label{solumu} 
\tilde u_{r_1,r_2}(x) = - \frac{6}{(r_2-r_1)^2}  {\cal P}(\frac{x-r_1}{r_2-r_1}; g_2=- \frac{1}{3} \mu (r_2-r_1)^4,g_3)
\eea 
where $g_3=g_3(\mu)$ is determined by the condition that the period is unity, i.e. 
$2 \Omega(g_2,g_3)=1$ equivalently
\bea
{\cal P}'(\frac{1}{2};g_2=- \frac{1}{3}  \mu (r_2-r_1)^4,g_3) =0
\eea 
Expanding this equation in powers of $g_2$ and $g_3$ near $g_2=0$ and
$g_3=g_3^0$ and matching the powers one obtains the following expansion
of $g_3(\mu)$ in powers of $g_2$ 
\bea \label{g3exp} 
&& g_3 = g_3(\mu) = g^0_3   -2 X g_2 - \frac{X^2}{g_3^0} g_2^2
+ (\frac{7}{432 g_3^0} - \frac{56}{27} \frac{X^3}{(g_3^0)^2}) g_2^3 + O(g_2^4) \quad , \quad 
X := \zeta \left(\frac{1}{2};0, g^0_3 \right) =  \frac{\pi}{\sqrt{3}} 
\eea
where we have used that ${\cal P}'_0(\frac{1}{2})=0$ and
${\cal P}_0(\frac{1}{2})=(g_3^0/4)^{1/3}$. Here $\zeta$ is the $\zeta$ Weierstrass function
mentioned in the previous subsection and $\zeta_0(z)=\zeta(z;0,g_3^0)$ 
satisfies the properties \eqref{zeta0prop} with $\zeta_0(\frac{1}{2}) =  \frac{\pi}{\sqrt{3}}$
as mentioned there.

From \eqref{LTS}, the formula for the conditional average of $e^{\mu S}$ is
\bea
\langle e^{\mu S}  \rangle_{\ell_1<- r_1, \ell_2 < r_2} = 
e^{f (\tilde u_{r_1,r_2}^\mu(0) - \tilde u_{r_1,r_2}^{\mu=0}(0))}  
\eea 
Expanding first to linear order in $\mu$, and then using the solution \eqref{solumu}
together with the first term $O(g_2)$ in the expansion \eqref{g3exp},
we obtain the cumulative conditional mean total size as 
\bea \label{cumcondtot}
&& \langle S  \rangle_{\ell_1<- r_1, \ell_2 < r_2} = f \partial_\mu|_{\mu=0} \tilde u_{r_1,r_2}^\mu(0) = 2 f (r_2-r_1)^2  g(p) \\
&& g(p)=g(1-p) = \frac{1}{6 g_3^0} 
( 2 {\cal P}_0(p)^2 -\frac{4 \pi  {\cal P}_0(p) }{\sqrt{3}}+ {\cal P}'_0(p)
 (\zeta_0(p) -\frac{2 \pi  p}{\sqrt{3}}) ) 
 \quad , \quad {p=\frac{r_2}{r_2-r_1} }
 \\
 && = \frac{p^2}{20}-\frac{\pi  p^4}{14 \sqrt{3}}+\frac{3 p^8 \Gamma \left(\frac{1}{3}\right)^{18}}{394240
   \pi ^6}+O\left(p^9\right)
\eea 
where $p$ is the aspect ratio of the avalanche.
In the limit $r_1 \to -\infty$, i.e. $p \to 0$, and using the above asymptotics, we recover
nicely the result \eqref{momS1} for a single boundary. The higher moments 
can be similarly calculated and the second moment is displayed in the Appendix \ref{app:higher}.

Let us compare with the result  \eqref{mainresult21}, \eqref{mainresult22},
which can be written as (with $z_0=\frac{x_0-r_1}{r_2-r_1}$)
\bea
\langle S(x_0) \rangle_{\ell_1<- r_1, \ell_2<r_2}
= \frac{f}{3 g_3^0} (r_2-r_1) [ \psi_1(z_0) \psi_1(p)  \theta(z_0<1-p) + 
\psi_1(1-z_0) \psi_1(1-p)  \theta(z_0>1-p) ]
\eea 
By definition of the total size of the avalanche one must have 
\bea
\langle S \rangle_{\ell_1<- r_1, \ell_2<r_2} = (r_2-r_1) \int_0^1 dz_0 \langle S(x_0) \rangle_{\ell_1<- r_1, \ell_2<r_2}
\eea 
The two results are found to be consistent, since indeed we have checked that 
\bea
g(p) = \frac{1}{6 g_3^0} ( \psi_1(p) \int_0^{1-p} dz_0 \psi_1(z_0) 
+  \psi_1(1-p) \int^1_{1-p} dz_0 \psi_1(1-z_0) )
\eea 
\\

Let us now calculate the conditional mean total size, $\langle S \rangle_{\ell,p}$,
conditioned on both $\ell$ and $p$. Through the same manipulations as in the previous
subsection, one has, in the limit $f=0^+$
\bea
&& \langle e^{\mu S} \rangle_{\ell_1,\ell_2} = \frac{1}{\rho(\ell_1,\ell_2)} 
\partial_{\ell_1} \partial_{\ell_2}  \tilde u_{-\ell_1,\ell_2}^\mu(0) \quad , \quad 
\langle e^{\mu S} \rangle_{\ell,p} = \frac{1}{\rho(\ell,p)} 
\ell \partial_{\ell_1} \partial_{\ell_2}  \tilde u_{-\ell_1,\ell_2}^\mu(0)
\eea 
Expanding to first order in $\mu$, and using \eqref{cumcondtot}, we obtain the mean total
size conditioned to both the total extension $\ell$ and the aspect ratio $p$ as
\bea \label{Slp} 
&& \langle S \rangle_{\ell,p} = \frac{1}{\rho(\ell,p)} 
\ell \partial_{\ell_1} \partial_{\ell_2}  \partial_\mu|_{\mu=0} 
\tilde u_{-\ell_1,\ell_2}^\mu(0) =  2 \frac{1}{\rho(\ell,p)} 
\ell \partial_{\ell_1} \partial_{\ell_2}  [\ell^2 g(p)]  \\
&& = \frac{2 \ell}{\rho(\ell,p)} ( 6 g(p) + \partial_p[ p(p-1) g'(p) + 2 (1-2 p) g(p) ] ) \nn \\
&& = \frac{2 \ell}{\rho(\ell,p)} (2 g(p) + (1-2 p) g'(p) + p (p-1) g''(p) ) \nn 
\eea 
where we have used \eqref{der1last}, and we recall that $\rho(\ell,p)$ is given by \eqref{resrholp}.

We now obtain the mean total size conditioned to the total extension $\ell$ as follows
\bea
&& \langle S \rangle_\ell = \frac{2}{\rho(\ell)} \int_0^1 dp \, \ell \partial_{\ell_1} \partial_{\ell_2}  [\ell^2 g(p)] 
 = \frac{\ell^3}{8 \sqrt{3} \pi} \times 12 \ell  \int_0^1 dp g(p) = \frac{\sqrt{3}}{2 \pi} \ell^4 \int_0^1 dp g(p)
\eea
where we used the second line in \eqref{Slp}, and noted that upon integration over $p$ 
the total derivative term does
not contribute since $g(p)$ vanishes quadratically at $p=0,1$. Using that
$\int_0^1 dp g(p) = 0.0026720030751$ we find
\bea \label{SS} 
\langle S \rangle_\ell = 0.00073657566 \times \ell^4
\eea 

\subsection{Mean shape conditionned to total extension} 
\label{subsec:shapext} 

We can now compare \eqref{Srel} and \eqref{Slp} to obtain the function $f(p)$ as
\bea
f(p) = 2 (2 g(p) + (1-2 p) g'(p) + p (p-1) g''(p) ) 
\eea
which using \eqref{mean3} gives the mean shape conditionned to total extension
\bea \label{mean4} 
 && \langle S(x_0) \rangle_{\ell} 
= \ell^3 {\sf s}(z_0) \quad , \quad 
z_0= \frac{x_0+\ell_1}{\ell} 
 \\
&& {\sf s}(z_0) 
 = \frac{1}{4 \pi \sqrt{3}}  (2 g(z_0) + (1-2 z_0) g'(z_0) + z_0 (z_0-1) g''(z_0) )  
 \eea 
as a function of the scaled and centered position $z_0$ normalized to be $0$ on the lower boundary and $1$ on the upper boundary. The function $g$ is defined in \eqref{cumcondtot}.
This leads to the result for the mean shape presented in \eqref{scal1}, \eqref{sexpl} and
\eqref{defg2}. Since $g(z_0)=g(1-z_0)$ we see that ${\sf s}(z_0)$ is symmetric around $z_0=1/2$.
The shape function has the following asymptotics near the edge $z_0=0$ and near the central point $z_0=1/2$
\bea \label{edge11} 
 && {\sf s}(z_0) =  \frac{z_0^3}{21}-\frac{z_0^4}{28}
 -\frac{3  \sqrt{3} \Gamma  \left(\frac{1}{3}\right)^{18} z_0^7}{98560 \pi ^7}
   +\frac{3 \sqrt{3}  \Gamma
   \left(\frac{1}{3}\right)^{18} z_0^8}{112640 \pi ^7}+O\left(z_0^9\right) \\
&& = 0.00182091 - 0.0314909 (z_0 - \frac{1}{2})^2 + .. 
\eea
We also see that 
\bea
\langle S \rangle_{\ell} =  \ell^4 \int_0^1 dz_0 \,  {\sf s}(z_0) =  \frac{\sqrt{3}}{2 \pi} \ell^4 \int_0^1 dp g(p)
\eea 
is consistent with \eqref{SS} since we have $\int_0^1 dz_0 {\sf s}(z_0)=0.00073657566$. 
Finally, using \eqref{Srel}, \eqref{mean4} and \eqref{mean3} we see that the 
mean total size at fixed extension and aspect ratio can also be written as
\bea \label{mts} 
&& \langle S  \rangle_{\ell,p} = \frac{{\sf s}(p)}{\tilde P(p)} \ell^4 
\eea 
where ${\sf s}(z)$ is the mean shape function \eqref{mean4} and $\tilde P(p)$ 
the PDF of the aspect ratio, given in \eqref{resrholp}.

\section{Avalanches upon driving by a point (''driving by the tip")}
\label{sec:tip}

\subsection{Model}

The model is a variant of the BFM in one dimension, 
where the driving is now exerted through a spring tied to a single point $x=0$, through the 
driving term $m_0^2 \delta(x) (w(t)- u(x,t))$. It is described by the equations
of motion \eqref{BFMpos} and \eqref{BFMdefTip}. As the driving function
$w(t)$ is increased, avalanches occur, with the special feature here that
the seed is always at $x=x_s=0$: the avalanches always start at $x=0$ and 
propagate from there (on both sides of $x=0$).

We will denote $\hat L_{m_0}$ the internal cutoff length associated to $m_0$, with 
$\hat L_{m_0}=1/m_0^2$, not to be confused with $L_m=1/m$ the internal cutoff length
associated to the bulk mass $m$.
We will focus mainly on zero bulk mass, $m^2=0$, In some
cases we will also consider the case $m^2>0$ (which provides a cutoff for the avalanches),
in which case we use the units $S_m,\eta_m$. For instance $\hat L_{m_0}$ is then expressed
in units $L_m=1/m$ (and the natural ratio of length is $\hat L_{m_0}/L_m=m/m_0^2$). 

We start by calculating the PDF of the local jump at the tip, of the total size,
and next we study the avalanche extension and shape.

\subsection{Joint distribution of the local jump at the tip and of the total size} 

Let us apply a local kick $\dot w(t) = w \delta(t)$ at the tip.
This leads to an avalanche of total size $S$ and with a local jump $S_0 := S(x=0)$
at the position of the tip. We first want to calculate the joint PDF, $P_w(S_0,S)$,
of $S$ and $S_0$. We start from the formula for the Laplace transform. From the 
general method described in Section \ref{sec:method}, it is given by
\bea
\overline{ e^{\mu S + \lambda_0 S_0} } = e^{ m_0^2 w \tilde u(x=0)}
\eea 
where $\tilde u(x)=\tilde u^{\lambda,\mu}(x)$ is the solution of the following
instanton equation:
\bea \label{instanton-tip} 
&& \tilde u''(x) - A \tilde u(x) - m_0^2 \delta(x) \tilde u(x) + \tilde u(x)^2 = - \mu - \lambda_0 \delta(x) 
\eea 
where we use $A=0$ for the massless case and $A=1$ for the massive case.
We can move the delta function term from the l.h.s. to the right hand side, and define a
new source parameter $\lambda$ as
\bea
&& \lambda = \lambda_0 - m_0^2 \tilde u(0)
\eea 
This allows to write the solution of (\ref{instanton-tip}) by analogy with \eqref{solu10} as
\bea  \label{solu2massless} 
&& \tilde u^{\lambda,\mu}(x)  = \frac{A- \beta^2}{2} 
+  \frac{6 \beta^2 (1-z^2) e^{- \beta |x|} }{(1 + z + (1-z) e^{- \beta |x|})^2} \quad , \quad 
\beta = (A^2 - 4 \mu)^{1/4} 
\eea 
where, from the discontinuity of the derivative at $x=0$, $z$ is the solution of
\bea \label{eqz}
&& \frac{\lambda}{\beta^3} = 3  z (1-z^2)
\eea 
which is continuously connected to $z=1$ for $\lambda=0$. 
Note that $\tilde u(x=0)=\beta^2 (1-\frac{3}{2} z^2)+ \frac{A}{2}$.\\

The joint PDF of $S$ and $S_0$ is thus given by the inverse Laplace transform
\bea \label{325} 
P_w(S,S_0) = LT^{-1}_{-\mu \to S, - \lambda_0 \to S_0} e^{ m_0^2 w \tilde u(0)}
\eea 
where $m_0^2 \tilde u(0)$ as a function of $\lambda_0,\mu$ 
is determined by eliminating $\lambda$ and $z$
in the system:
\bea
&& \lambda = \lambda_0 - m_0^2 \tilde u(0) = \lambda_0 - m_0^2 \beta^2 (1- \frac{3}{2} z^2) 
- m_0^2 \frac{A}{2}  \\
&& \frac{\lambda}{\beta^3} = 3 z (1-z^2) \quad , \quad 
\beta = (A - 4 \mu)^{1/4} 
\eea 

It is equivalent, but more convenient to state it as follows 
\bea \label{inv1} 
P_w(S,S_0) = LT^{-1}_{-\mu \to S, - \lambda_0 \to S_0} e^{ \lambda_0 w - \lambda w}
\eea 
where $\lambda$ as a function of $\lambda_0,\mu$ 
is determined by eliminating $z$
in the system:
\bea \label{syst1} 
&& \lambda = \lambda_0 - m_0^2 \beta^2 (1- \frac{3}{2} z^2) - m_0^2 \frac{A}{2}   \\
&& \lambda = 3 \beta^3 z (1-z^2) \quad , \quad 
\beta = (A- 4 \mu)^{1/4} 
\eea

\subsubsection{Moments and joint moments}
\label{sec:moments} 

This system can first be studied in presence of a bulk mass, i.e. setting $A=1$. Since in that case the
moments of $S,S_0$ do exist, one can perform an expansion 
in powers of $\lambda_0,\mu$, which is then analytic (and corresponds to an expansion
around $\beta=1$ and $z=1$). One obtains
\bea \label{moments1} 
&& \langle S_0 \rangle =  \frac{m_0^2}{m_0^2+2} w  \quad , \quad  \langle S \rangle = \frac{2 m_0^2}{m_0^2+2} w \quad , \quad \langle S_0^2 \rangle = \frac{m_0^2 w \left(3 \left(m_0^2+2\right) m_0^2 w+4\right)}{3 \left(m_0^2+2\right){}^3} \\
&& \langle S^2 \rangle = \frac{4 m_0^2 w \left(m_0^4 (3 w+1)+6 m_0^2 (w+1)+12\right)}{3 \left(m_0^2+2\right){}^3} \quad , \quad \langle S S_0 \rangle = \frac{2 m_0^2 w \left(3 \left(m_0^2+2\right) m_0^2 w+m_0^2+6\right)}{3 \left(m_0^2+2\right){}^3}
\eea 
We note that as $m_0 \to +\infty$, $\langle S_0^p \rangle \to w^p$ as $S_0 \to w$ i.e.
it becomes deterministic. In the limit $w \to 0^+$ the above formulae give the moments over the density $\rho(S,S_0) = 
\partial_w|_{w=0^+} P_w(S,S_0)$
\bea \label{moments2} 
&& \langle S_0 \rangle_\rho =  \frac{m_0^2}{m_0^2+2}   \quad , \quad \langle S \rangle_\rho = \frac{2 m_0^2}{m_0^2+2}  \quad , \quad \langle S_0^2 \rangle_\rho =  \frac{4 m_0^2}{3 \left(m_0^2+2\right){}^3} \quad , \quad  \langle S^2 \rangle_\rho = \frac{4}{3} \left(1-\frac{8}{\left(m_0^2+2\right){}^3}\right) \\
&&  \langle S S_0 \rangle_\rho = \frac{2 m_0^2 \left(m_0^2+6\right)}{3 \left(m_0^2+2\right){}^3}
\eea 

\subsubsection{Distribution of total size $P_w(S)$: crossover between two limit forms}
\label{sub:totalcross} 

Let us first study $P_w(S)$, hence we set $\lambda_0=0$ in \eqref{inv1}, \eqref{syst1},
and consider for simplicity the massless case $A=0$. 
Then $z$ is the root of:
\bea
f(z) := \frac{3 z (1-z^2)}{1- \frac{3}{2} z^2} = \frac{- m_0^2}{\beta}  \quad , \quad \beta = (- 4 \mu)^{1/4}
\eea 
Since the function $f(z)$ is strictly increasing from $-\infty$ to $+\infty$ on the
interval $]\sqrt{2/3} ,+\infty[$ which contains $z=1$ (with $f(1)=0$), there
is a unique (and physical) solution on this interval $z=f^{-1}(\frac{- m_0^2}{\beta})$.
Defining the function $g(v)=3 f^{-1}(-1/v) (1- (f^{-1}(-1/v))^2)$ 
one can write
\be \label{PwSLapl}
P_w(S) = LT^{-1}_{-\mu \to S} e^{ - w (-4 \mu)^{3/4} g((- 4 \mu)^{1/4}/m_0^2) }
 \quad , \quad g(0)=\sqrt{2/3} 
\quad , \quad g(v) \simeq_{v \to +\infty} \frac{1}{2 v} 
\ee 
which leads to the two limits:

\begin{itemize}

\item $m_0^2 \to 0$, equivalently $S \ll 1/m_0^8$, then 
\bea
P_{w}(S) = LT^{-1}_{-\mu \to S} e^{- \frac{w}{2} m_0^2 (-4 \mu)^{1/2}} 
= \frac{m_0^2 w e^{-\frac{m_0^4 w^2}{4 S}}}{2 \sqrt{\pi } S^{3/2}} = \frac{1}{m_0^4 w^2} 
{\cal L}_{1/2}(S/m_0^4 w^2)
\eea 
which is the standard PDF in the case of imposed force driving, i.e. Eq. \eqref{PDFS}
with $\int dx f(x)=m_0^2 w$ in the present case, taken in the massless limit. The distribution
${\cal L}_{1/2}$ is the stable Levy law of index $1/2$, see Appendix \ref{app:Levy}.

\item $m_0^2 \to +\infty$, equivalently $S \gg 1/m_0^8$, then 
\bea
P_{w}(S) = LT^{-1}_{-\mu \to S} e^{- w \sqrt{\frac{2}{3}} (-4 \mu)^{3/4}} 
= \frac{3^{2/3} }{(4w)^{4/3}} 
{\cal L}_{3/4}(3^{2/3} S/ (4w)^{4/3})
\eea 
with ${\cal L}_{3/4}$ being the stable Levy law of index $3/4$, see Appendix \ref{app:Levy}
where its explicit expression is given in terms of hypergeometric functions. 
It is thus a non trivial (infinitely divisible) distribution, quite different from the
standard law Eq. \eqref{PDFS} for force imposed driving. The associated
density, $\rho(S)=\partial_{w}|_{w=0^+} P_w(S)$ has a simple expression
\bea \label{LT1} 
 \rho(S) = - \frac{1}{S} LT^{-1}_{-\mu \to S} \partial_\mu \sqrt{\frac{2}{3}} (-4 \mu)^{3/4} = \frac{\sqrt{3}}{\Gamma(1/4) S^{7/4}} \sim S^{- \tau_{\text{loc. driv.}}} \quad \text{with} \quad \tau_{\text{loc. driv.}}= \frac{7}{4}
\eea 
as obtained in \cite{Delorme}. 

\end{itemize} 

\subsubsection{Calculation of the PDF of the local jump $P_w(S_0)$}
\label{subsec:localtip} 

We now calculate the PDF, $P_w(S_0)$, of the local jump $S_0=S(x=0)$ at the position of the tip $x=0$
following a kick $w$ at the tip. We recall that the driving is "by the tip", i.e. through the 
driving term $\dot f(x,t)=m_0^2 \delta(x) (w \delta(t)- \dot u(x,t))$. It is thus different from the 
imposed force driving $\dot f(x,t) = f \delta(x-x_s) \delta(t)$ and $P_w(S_0)$ computed here should
not be confused with $P_{\{f \delta(x)\}}(S_0)$ calculated there, e.g. for $x_s=0$ in
\eqref{localforcelocal1}, \eqref{dens1}. The two PDF's coincide only in the limit
$m_0 \to 0$ with $f = w m_0^2$ fixed, as discussed below. For simplicity we treat here the massless case $m=0$, the massive case being detailed in the Appendix \ref{app:tiplocal}.

To calculate $P_w(S_0)$ we set $\mu=0$ in \eqref{inv1}, \eqref{syst1} (for $A=0$).
It is easy to see that in that limit $\beta \to 0$, and $z \sim z_0/\beta$
and we have
\bea
&& P_{w}(S_0) = LT^{-1}_{-\lambda_0 \to S_0} e^{(\lambda_0 + 3 z_0^3) w} \\
&& - 3 z_0^3 - \frac{3}{2} m_0^2 z_0^2 = \lambda_0 
\eea 
Let us write
\bea
S_0 P_w(S_0) = 
\int \frac{d \lambda_0}{2 i \pi} e^{- \lambda_0 S_0} \partial_{\lambda_0} e^{ w (\lambda_0 + 3 z_0^3)} 
\eea 
We can trade the variable $\lambda_0$ for the variable $z$ and write (we work up to a global sign)
\bea
&& S_0 P_w(S_0) = 
\int \frac{d z_0}{2 i \pi} e^{(3 z_0^3 + \frac{3}{2} m_0^2 z_0^2) S_0} 
\partial_{z_0} e^{- \frac{3}{2} m_0^2 z_0^2 w} \\
&& = 3 w m_0^2  \int \frac{d z_0}{2 i \pi} z_0 e^{3 z_0^3 S_0+ \frac{3}{2} m_0^2 z_0^2 (S_0-w)}  \\
&& = 3 w m_0^2 \partial_c \Phi(a,b,c)|_{a=9 S_0, b= \frac{3}{2} m_0^2 (S_0-w) , c= 0 } 
\eea 

This leads to the result 
\bea \label{PS00} 
P_w(S_0) &=& \frac{1}{3^{1/3} S_0^{5/3}} w m_0^2 e^{ \frac{2}{3} y^3} 
(- y {\rm Ai}(y^2) - {\rm Ai}'(y^2) ) \quad , \quad y = \frac{m_0^2 (S_0-w)}{2 \times 3^{1/3} S_0^{2/3}} 
\eea 
We can compare with \eqref{localforcelocal1}: we see that indeed for $m_0 \to 0$ at fixed
$f=m_0^2 w$ the two formula coincide. However the above result describes the full crossover
as $m_0$ is increased, which we now analyze. First one checks that $\langle S_0 \rangle=w$.
Then one can also give the density $\rho(S_0)=\partial_{w} P_w(S_0)|_{w=0^+}$ 
\bea \label{denstip0} 
\rho(S_0) = \frac{1}{3^{1/3} S_0^{5/3}} m_0^2 e^{ \frac{2}{3} y^3}  
(- y {\rm Ai}(y^2) - {\rm Ai}'(y^2) ) \quad , \quad y = \frac{m_0^2 S_0^{1/3}}{2 \times 3^{1/3}} 
\eea 
which exhibits the single crossover scale $S_0 \sim m_0^{-6}$. This density crosses 
over from
\be
\rho(S_0) \simeq \frac{m_0^2}{3^{2/3} \Gamma(\frac{1}{3}) S_0^{5/3}}  
\quad , \quad S_0 \ll m_0^{-6} 
\ee
to 
\bea
\rho(S_0)  \simeq \sqrt{\frac{3}{2 \pi }} \frac{1}{m_0^3 S_0^{5/2}}
\quad , \quad S_0 \gg m_0^{-6} \label{dens02n} 
\eea 

One can be a bit more precise: since $P_w(S_0) \to \delta(S_0-w)$ in the
limit $m_0 \to \infty$, there are really two regions in the PDF. Using that 
\bea \label{Airyas2} 
&& - y {\rm Ai}(y^2) - {\rm Ai}'(y^2)  \simeq_{y \to +\infty}  \frac{1}{8 \sqrt{\pi} y^{5/2}} e^{- \frac{2}{3} y^3} \\
&& - y {\rm Ai}(y^2) - {\rm Ai}'(y^2)  \simeq_{y \to -\infty}  \frac{\sqrt{-y}}{\sqrt{\pi}} e^{\frac{2}{3} y^3}
\eea 
we see that whenever $m_0^2 (S_0-w) \gg S_0^{2/3}$ one has
\bea \label{tail1} 
P_w(S_0)  \simeq \sqrt{\frac{3}{2 \pi }} \frac{w}{m_0^3 (S_0-w)^{5/2}}
\eea 
while whenever $m_0^2 (w-S_0) \gg S_0^{2/3}$ one has 
\bea \label{tail2} 
P_w(S_0)  \simeq \frac{1}{\sqrt{6 \pi }} \frac{m_0^3 w \sqrt{w-S_0}}{S_0^2} e^{- \frac{m_0^6 (w-S_0)^3}{18 S_0^2}} 
\eea 
While the first regime contains the tail leading to the density \eqref{dens02n},
the second regime describes the cutoff function at small $S_0$. In the limit of
large, but not infinite $m_0^2$, the fluctuations scale as $S_0-w \sim m_0^{-2}$
and more precisely if one defines 
\be
S_0-w = 2 \times 3^{1/3} w^{2/3} \times  \frac{s}{m_0^2} 
\ee
then one sees from \eqref{PS00} that the random variable $s=O(1)$ has a limit probability density $p(s)$
as $m_0 w^{2/3} \to + \infty$
\bea
p(s) =   2 e^{ \frac{2}{3} s^3} 
(- s {\rm Ai}(s^2) - {\rm Ai}'(s^2) ) 
\eea 
and one can check that $\int_{-\infty}^{+\infty} ds p(s)=1$. This distribution
describes the convergence to $P_w(S_0) \to \delta(S_0-w)$, with the
right and left tail regimes \eqref{tail1} and \eqref{tail2}, respectively.

\subsubsection{Calculation of the joint PDF of the total and local size $P_w(S,S_0)$}
\label{subsec:crossPP} 

It turns out that one can calculate {\it explicitly} the joint PDF $P_w(S,S_0)$, to which we
now turn. We consider simultaneously the massive, $A=1$, and massless, $A=0$, cases.
From \eqref{inv1}, \eqref{syst1} we can write
\bea
S_0 P_w(S,S_0) = \int \frac{d \mu}{2 i \pi} e^{- \mu S} 
\int \frac{d \lambda_0}{2 i \pi} e^{- \lambda_0 S_0} \partial_{\lambda_0} e^{ w Z} 
\eea 
where $Z = m_0^2 \beta^2 (1- \frac{3}{2} z^2) + m_0^2 \frac{A}{2} $ and
$\lambda_0-Z= 3 \beta^3 z (1-z^2)$. Next we can trade the variable $\lambda_0$ 
for the variable $z$ and, 
upon the change $z \to z/\beta$ and rearranging, we obtain
\bea
&& S_0 P_w(S,S_0) 
 = e^{m_0^2 \frac{A}{2} (w-S_0)} \int \frac{d \mu}{2 i \pi} e^{- \mu S} 
\int \frac{dz}{2 i \pi} e^{- \beta^2 [ S_0 (m_0^2   + 3 z ) - w m_0^2 ] }  e^{S_0 ( m_0^2 \frac{3}{2} z^2 + 3 z^3)  } 
\partial_{z} e^{ -  \frac{3}{2} w m_0^2 z^2 } 
\eea 
We can now use that
\bea
LT^{-1}_{-\mu \to S} \, e^{- B (\sqrt{A-4 \mu}- A)} = \frac{B}{\sqrt{\pi} S^{3/2}}  
e^{- \frac{B^2}{S} + A B - A \frac{S}{4} } 
\eea 
with $\beta^2=\sqrt{A-4 \mu}$. Hence, performing the integration over $\mu$, and the derivation over $z$, we obtain
\bea
&& S_0 P_w(S,S_0) = \frac{3 w m_0^2 }{\sqrt{\pi} S^{3/2}} e^{A ( \frac{m_0^2}{2} (w-S_0) -  \frac{S}{4})
- \frac{(S_0-w)^2 m_0^4}{S}} \\
&& 
~~~~~~~~~~~~ \times \int \frac{dz}{2 i \pi} 
[ 3 z^2 S_0 + (S_0 -w) m_0^2 z ] e^{- \frac{6 S_0 (S_0-w) m_0^2}{S} z + (-\frac{9 S_0^2}{S} + \frac{3}{2} m_0^2 (S_0-w) ) z^2 + 3 S_0 z^3 } 
\eea 
To evaluate this integral we rewrite it using the function $\Phi(a,b,c)$ defined
in \eqref{Phi} (also in Appendix A of \cite{Delorme}) as
%
\bea
&& S_0 P_w(S,S_0) =
\frac{3 w m_0^2 }{\sqrt{\pi} S^{3/2}} e^{A ( \frac{m_0^2}{2} (w-S_0) -  \frac{S}{4})}
e^{- \frac{(S_0-w)^2 m_0^4}{S}} \\
&& ~~~~~~~~~~~~  \times 
[ 3 S_0 \partial_b + (S_0 -w) m_0^2 \partial_c ] \Phi(a,b,c)|_{a=9 S_0, b=-\frac{9 S_0^2}{S} + \frac{3}{2} m_0^2 (S_0-w) ,
c=- \frac{6 S_0 (S_0-w) m_0^2}{S}} \nonumber 
\eea
and using again \eqref{Phi} 
we finally obtain the joint PDF
\bea \label{jointPw}
&& P_w(S,S_0) = B \left(y \text{Ai}\left(y^2\right) - \text{Ai}'\left(y^2\right) \right) \quad , \quad y = \frac{m_0^2 S \left(S_0-w\right)+6 S_0^2}{2 \sqrt[3]{3} S S_0^{2/3}} \\
&& B = \frac{2\ 3^{2/3} m_0^2 S_0^{1/3} w }{\sqrt{\pi } S^{5/2}}
\exp \left(-\frac{3 m_0^2 S_0^2
   \left(S_0-w\right)}{S^2}+\frac{m_0^6 \left(S_0-w\right){}^3}{36 S_0^2}-\frac{m_0^4
   \left(S_0-w\right){}^2}{2 S}-\frac{6 S_0^4}{S^3}
+ \frac{A}{4} (2 m_0^2(w-S_0)-S) \right) \nn
\eea 
We have checked numerically that the moments \eqref{moments1} are recovered in the
massive case. We have also checked numerically that integrating \eqref{jointPw}
over $S$ one recovers
$P_w(S_0)$ in \eqref{PS00} in the massless case.

Note that it takes the scaling form
\bea
P_w(S,S_0;A) = m_0^{14} \tilde P_{w m_0^6}(m_0^8 S, m_0^6 S_0; A/m_0^8) 
\eea
where $\tilde P$ is obtained from $P$ setting $m_0=1$ in formula \eqref{jointPw}.
Hence the characteristic crossover scales are 
$S \sim m_0^{-8}$, $S_0 \sim w \sim m_0^{-6}$ 
and one can define the reduced scales $\tilde S = m_0^8 S$ and $\tilde S_0 = m_0^6 S$.

We can now study the joint sizes density $\rho(S,S_0) = \partial_w|_{w=0} P_w(S,S_0)$
which is given by
\bea \label{jointrhow}
&& \rho(S,S_0) = 
B \left(y \text{Ai}\left(y^2\right) - \text{Ai}'\left(y^2\right) \right) \quad , \quad  y = S_0^{1/3}  \frac{m_0^2 S +6 S_0}{2 \sqrt[3]{3} S} \\
&& B = \frac{2\ 3^{2/3} m_0^2 S_0^{1/3}  \exp \left(-\frac{3 m_0^2 S_0^3}{S^2}
+\frac{m_0^6 S_0}{36}-\frac{m_0^4 S_0^2}{2 S}
-\frac{6 S_0^4}{S^3} - A ( S_0 \frac{m_0^2}{2} +  \frac{S}{4})
\right)}{\sqrt{\pi } S^{5/2}} \nn
\eea 
and obeys the scaling form $\rho(S,S_0;A) =  m_0^{20} \tilde \rho(m_0^8 S, m_0^6 S_0; A/m_0^8)$,
where $\tilde \rho$ is obtained from $\rho$ setting $m_0=1$ in formula \eqref{jointrhow}.

{\bf Remark}. In the limit $m_0 \to +\infty$, in the massless case ($A=0$) the equations
\eqref{inv1}, \eqref{syst1} can be solved in an expansion in $1/m_0^2$ as
\bea
&& z = \sqrt{\frac{2}{3}} + \frac{1}{m_0^2} ( \frac{\beta}{3} - \frac{\lambda_0}{\sqrt{6} \beta^2}) + .. \\
&& \lambda_0 - \lambda = \lambda_0 - \sqrt{\frac{2}{3}} \beta^3 
+ \frac{1}{m_0^2} (\beta^4 - \sqrt{\frac{3}{2}} \beta \lambda_0) + \dots
\eea
Hence the value at zero of the instanton solution \eqref{solu2massless}, i.e.
$\tilde u(x=0)=\beta^2 (1-\frac{3}{2} z^2) \simeq (\lambda_0 - \sqrt{\frac{2}{3}} \beta^3)/m_0^2$ vanishes for $m_0^2 \to +\infty$.
So in this limit the instanton equation becomes effectively equivalent to
solving $\tilde u''(x) + \tilde u(x)^2 = - \mu$ with $\tilde u(x=0)=0$, where
the derivative discontinuity is free to adjust. This is true
because $S_0=w$ in that limit, hence $\lambda_0$ is not needed. Thus it amounts to solve the 
instanton equation on a half-axis with Dirichlet boundary conditions
and one can check that the r.h.s. in $\exp( m_0^2 w \tilde u(x=0))$ in \eqref{325} 
remains finite when $m_0^2 \to +\infty$. It can be obtained from the discontinuity of the solution on the half-axis, i.e. one has $\lim_{m_0 \to +\infty} m_0^2 w \tilde u(x=0) = \tilde u'(x=0)$.

\subsection{Distribution of the extension for avalanches driven by the tip: 
crossover as $m_0$ is varied} 
\label{subsec:tipext} 

We now ask what is the probability that the avalanche reaches the distance
$\ell_2$ from the tip, i.e. from the seed at $x=0$, i.e the driving point (on one side, say $x>0$). 
Here we restrict on the case where the bulk mass is zero $m=0$. 
The associated CDF is
\bea
P(\ell_2 < r) = e^{ m_0^2 w \tilde u_r(x=0)}  \label{f1} 
\eea 
where $\tilde u_r(x)$ is the solution of the instanton equation
\bea \label{instanton-tip3} 
&& \tilde u''(x) - m_0^2 \delta(x) \tilde u(x) + \tilde u(x)^2 =  0 \quad , \quad \tilde u(r)=- \infty 
\eea 
with $\tilde u(- \infty)=0$. 
This is very similar to the problem studied in Section \ref{sec:1boundary}, i.e.
Eq. \eqref{instanton-tip3} (and boundary condition) is identical to \eqref{eqlambda1} with 
$\lambda=- m_0^2 \tilde u(x=0)$ and $x_0=0$. We thus know that the solution is \eqref{solu1side}, i.e.
\bea \label{solu1side2} 
\tilde u_r(x) = - \frac{6}{(\rho - x)^2} \theta(-x) - 6 {\cal P}(r-x,0,g_3) \theta(x)  \label{f2} 
\eea 
with the conditions of continuity at $x=0$ and proper derivative jump are analogous to
\eqref{cond1} and \eqref{cond2}
\bea
&& \frac{1}{\rho^2} =  {\cal P}(r,0,g_3) \label{cond1n} \quad , \quad  -  \frac{2}{\rho^3} - {\cal P}'(r,0,g_3) = \frac{m_0^2}{\rho^2} 
\label{cond2n}
\eea 
where we have replaced $\lambda/6=- m_0^2 \tilde u(0)/6=m_0^2/\rho^2$. The equations
\eqref{cond2n} determine the unknown parameters $g_3,\rho$ as a function of $r,m_0$. 
Using ${\cal P}'(z)^2 = 4 {\cal P}(z)^3 - g_3$  we obtain
\bea
g_3=- \frac{m_0^4}{\rho^4} - 4 \frac{m_0^2}{\rho^5} 
\eea 
so in fact, there is a single unknown parameter $\rho$. Since $1/m_0^2$ is an (internal) length
scale we can (for $m_0 \neq 0,+\infty$) absorb $m_0$ by defining the dimensionless
variables
\be
\tilde r = r m_0^2 \quad , \quad 
\tilde \rho = \rho m_0^2 
\ee
The relation between $\tilde \rho$ and $\tilde r$ can then be written as
\bea
&& 1= {\cal P}(\frac{\tilde r}{\tilde \rho},0, - \tilde \rho^2 - 4 \tilde \rho)   \quad , \quad \tilde r = h(\tilde \rho) := \tilde \rho  \int_1^{+\infty} \frac{dy}{
\sqrt{4 y^3 + \tilde \rho^2 + 4 \tilde \rho}} = \, _2F_1\left(\frac{1}{6},\frac{1}{2};\frac{7}{6};-\frac{\tilde \rho^2 + 4 \tilde \rho}{4}\right) \tilde \rho \label{eqf1} 
\eea 
where we have used the homogeneity property of the Weierstrass function
$\mathcal{P}(\lambda z; \lambda^{-4} g_2, \lambda^{-6} g_3) = \lambda^{-2} \mathcal{P}(z; g_2, g_3)$
as well as \eqref{defint}.
From \eqref{f1}, \eqref{f2} the CDF of $\ell_2$ reads
\bea
P(\ell_2 < r) = e^{ - 6 m_0^2 w/\rho^2 } =  e^{ - 6 m_0^6 w \phi(\tilde r) } \quad , \quad \tilde r = m_0^2 r
\eea 
The function $\phi(r)$, which is $\phi=1/\tilde \rho^2$ as a function of $\tilde r$
can be obtained from \eqref{eqf1}.
At small $\tilde r$, it is easy to invert $\tilde r=h(\tilde \rho)$
as a series at small $\tilde \rho,\tilde r$ and calculate $\phi(\tilde r)=1/\tilde \rho^2$.
At large $\tilde r$, it is more convenient to express $1/\tilde r^3$ in an expansion in small $\phi=1/\rho^2$, which starts as $1/r^3 \sim \phi$
and then invert the series to obtain $\phi(r)$. This gives, respectively, the small $r$ (divergence)
and large $r$ (vanishing) series expansions
\bea \label{detailedphi}
&& \phi(r)= \frac{1}{r^2}-\frac{1}{7 r}+\frac{43}{2548}-\frac{555 r}{338884}+\frac{75573
   r^2}{616768880}-\frac{102969 r^3}{19119835280}-\frac{474602517
   r^4}{2446306444734880}+O\left(r^5\right) \\
&& \phi(r) = \frac{2 \Gamma \left(\frac{1}{3}\right)^3 \Gamma \left(\frac{7}{6}\right)^3}{\pi ^{3/2}
   r^3}-\frac{6 \Gamma \left(\frac{1}{3}\right)^3 \Gamma
   \left(\frac{7}{6}\right)^3}{\pi ^{3/2}
   r^4}+O\left(\left(\frac{1}{r}\right)^{9/2}\right) \label{detailedphi2}
\eea 
We thus obtain the power law decay of the CDF of the upper extension $\ell_2$
at large $\tilde r$, i.e. $r \gg 1/m_0^2$, in the original variables
\bea \label{detailed1} 
P(\ell_2 <  r) = 1-\frac{12 w \Gamma \left(\frac{1}{3}\right)^3 \Gamma
   \left(\frac{7}{6}\right)^3}{\pi ^{3/2} r^3}+\frac{36 w \Gamma
   \left(\frac{1}{3}\right)^3 \Gamma \left(\frac{7}{6}\right)^3}{m_0^2 \pi ^{3/2}
   r^4}+O\left(\left(\frac{1}{r}\right)^{9/2}\right)
\eea 
At small $\tilde r=m_0^2 r^2 \ll 1$, the CDF has the usual leading order essential singularity 
$P(\ell_2 <  r) \simeq e^{ - \frac{6 m_0^2 w}{r^2}  }$, the same as for the force driving by a kick
$f = m_0^2 w$. 

From the CDF one obtains the PDF $P(\ell_2) = \partial_r P(\ell_2 <  r)|_{r=\ell_2}$.
We will give here the density 
\bea
\rho(\ell_2) := \partial_{w}|_{w=0^+} P(\ell_2) = - 6 m_0^8 \phi'( m_0^2 \ell_2)
\eea
and we find
\bea
&& \rho(\ell_2) \simeq \frac{12 m_0^2}{\ell_2^3}   \quad , \quad   \ell_2 \ll m_0^{-2} \\
&&  \rho(\ell_2) \simeq \frac{36 \Gamma \left(\frac{1}{3}\right)^3 \Gamma \left(\frac{7}{6}\right)^3}{\pi ^{3/2}
   \ell_2^4}  = \frac{99.2466..}{ \ell_2^4}  \quad , \quad   \ell_2 \gg m_0^{-2}  \label{impdisp}
\eea 
Thus the density $\rho(\ell_2)$ exhibits a crossover between 
the "imposed force" result for $\ell_2 \ll m_0^{-2}$, i.e. $\partial_f|_{f=0^+} P(\ell_2)=
12/\ell_2^3$ obtained in \eqref{PDFl2} (the $m_0^2$ factor comes from the different definition of
the density - per unit force versus per unit displacement - there)
and a new, "imposed displacement" regime $\ell_2 \gg m_0^{-2}$
with a different $\kappa_{\rm imp. displ.} = 4$ exponent. 
Clearly the limit $m_0 \to +\infty$ is well defined: in that limit only the 
large $\ell_2$ region remains and the density becomes 
a pure power law given by \eqref{impdisp}. If a small but non-zero
bulk mass is added, an upper cutoff for large $\ell_2 \sim 1/m$ will appear.

For completeness let us indicate that one can equivalently write the CDF and the PDF in the parametric forms
\bea
&& P(m_0^2 \ell_2 < \frac{\, _2F_1\left(\frac{1}{6},\frac{1}{2};\frac{7}{6};-\frac{\phi^2 + 4 \phi^{5/2}}{4 \phi^3}\right)}{\sqrt{\phi}}) 
=  e^{ - 6 m_0^6 w \phi } \\
&& 
P(\ell_2=m_0^{-2} 
\frac{\, _2F_1\left(\frac{1}{6},\frac{1}{2};\frac{7}{6};-\frac{\phi^2 + 4 \phi^{5/2}}{4 \phi^3}\right)}{\sqrt{\phi}} ) 
=  6 m_0^8 w \frac{3 \phi (1+ 4 \sqrt{\phi})}{1+ m_0^2 \ell_2 (1 + 5 \sqrt{\phi})} e^{ - 6 m_0^6 w \phi } 
\eea 
which allows easy plotting as $\phi$ is varied on $[0,+\infty[$. 
Indeed one can show directly from \eqref{invert1} that
\be
\frac{d\phi}{dr} = - \frac{3 \phi (1+ 4 \sqrt{\phi})}{1+ r (1 + 5 \sqrt{\phi})}
\ee

%
%
%
%
%
%
%

Note that here we have studied only $\ell_2$, which is how far the avalanche
reaches on the side $x>0$. One can extend the calculation, once again,
to the JPDF of $\ell_1$ and $\ell_2$. For finite $m_0$ these are correlated
and their JPDF is non-trivial. We now study the limit $m_0=+\infty$ where
simplifications occur. 

\subsection{Mean avalanche shape in the limit $m_0=+\infty$}
\label{subsec:shapetiptotal} 

It turns out that one can solve directly the case $m_0=+\infty$, where the
driving is purely by imposing the displacement of the tip. Indeed from the previous
subsection we found $\tilde u_r(0) = - 6 m_0^4 \phi(m_0^2 r)$. Since $\phi(z) \sim 1/z^3$ at large $z$,
we see that $\tilde u_r(0) \sim m_0^{-2}$ as $m_0 \to +\infty$, i.e. the instanton
solution vanishes at $x=0$, however the product $m_0^2 w \tilde u(0)$ remains finite
(hence the limit of the CDF is well defined). 

\subsubsection{Limit $m_0=+\infty$ and CDF of $\ell_2$}
\label{subsec:cdf} 

Thus to study the case $m_0=+\infty$ we can simply look for the solution of the
instanton equation with the following boundary conditions (we focus here on the case
of no bulk mass, $m=0$) 
\bea \label{instanton-tip4} 
&& \tilde u''(x) + \tilde u(x)^2 =  0   \quad , \quad \tilde u(r)=-\infty \quad , \quad \tilde u(0)=0 \label{eqm0} 
\eea 
on the interval $[0,r]$. The solution is 
\bea
\tilde u(x) = \tilde u_r(x) := - 6 {\cal P}(r-x,0,g_3)  \label{solu00} 
\eea 
and the additional condition which determines $g_3$ as a function of $r$
\bea
{\cal P}(r,0,g_3) = 0 ~~ \Leftrightarrow ~~ {\cal P}(1,0, r^6 g_3) = 0 \label{cond00}
\eea 
Let us now introduce the root ($g_3 <0$) to the equation 
\bea
{\cal P}(1,0, g_3) = 0 \quad , \quad \int_0^{+\infty} \frac{dy}{\sqrt{4 y^3 - g_3}} = 1 \label{eqroot} 
\eea 
Using the identity, valid for any $g_3$
\bea
\int_0^{+\infty} \frac{dy}{\sqrt{4 y^3 + |g_3|}} 
= \frac{2^{1/3} \Gamma \left(\frac{1}{3}\right) \Gamma \left(\frac{7}{6}\right)}{\sqrt{\pi } |g_3|^{1/6}} 
\eea 
the root of \eqref{eqroot} is found to be 
\bea \label{g3sdef} 
g_3^* = - \frac{256 \pi ^3 \Gamma \left(\frac{1}{3}\right)^6}{\Gamma \left(-\frac{1}{6}\right)^6}
=  -\frac{4 \Gamma \left(\frac{1}{3}\right)^6 \Gamma \left(\frac{7}{6}\right)^6}{\pi ^3} = - 30.40090507702060
\eea
as can also be inferred from e.g. \cite{Zeroes1}. Hence
$g_3$ determined by \eqref{cond00} is simply $g_3=g_3^*/r^6$ and 
we finally obtain the solution of \eqref{eqm0} as
\bea
\tilde u_r(x) = - 6 {\cal P}(r-x,0,g_3^*/r^6) = - \frac{6}{r^2} {\cal P}_*(1-\frac{x}{r})  \label{soluf00} 
\eea
where from now on we define
\be
{\cal P}_*(z) = {\cal P}(z,0,g_3^*) \label{defPstar} 
\ee

Which formula should now be used to obtain, e.g. the CDF of $\ell_2$. A priori 
one should use \eqref{f1} in the limit $m_0 \to +\infty$. One would like to
avoid this step to work directly at infinite $m_0$. One then notes that (see Remark
in the previous subsection)
\bea
\lim_{m_0 \to +\infty} m_0^2 \tilde u(x=0)= \tilde u_r'(x=0) 
\eea
from the instanton equation \eqref{instanton-tip3}, using that the discontinuity of
the derivative $\tilde u_r'(x=0^+) - \tilde u_r'(x=0^-) = m_0^2 \tilde u(x=0)$, and
that on the $x<0$ side $\lim_{m_0 \to +\infty} \tilde u(x)=0$ (see below for a more
detailed derivation). Hence in the limit $m_0 \to +\infty$ the CDF is given by
\bea
P(\ell_2 < r) = e^{w \tilde u_r'(0)}  \label{eqeq} 
\eea 
where here $\tilde u_r'(0)=\tilde u_r'(0^+)$. 
This derivative can be evaluated explicity from \eqref{soluf00} using that the
Weierstrass function satisfies ${\cal P}'^2=4 {\cal P}^3 - g_2 {\cal P} - g_3$,
hence ${\cal P}_*'(1)  = - (- g_3^*)^{1/2}$ and then
\bea \label{impdisp0}
\tilde u'(0) = \frac{6}{r^3} {\cal P}_*'(1) = - \frac{12}{r^3} 
\frac{\Gamma \left(\frac{1}{3}\right)^3 \Gamma \left(\frac{7}{6}\right)^3}{\pi ^{3/2}}
\eea 
Thus we find
\bea
P(\ell_2 < r) = e^{- \frac{12 w}{r^3} 
\frac{\Gamma \left(\frac{1}{3}\right)^3 \Gamma \left(\frac{7}{6}\right)^3}{\pi ^{3/2}}} 
\eea 
and
\bea \label{impdisp2}
\rho(\ell_2)= \frac{18 \sqrt{- g_3^*}}{\ell_2^4} =  \frac{36}{\ell_2^4} 
\frac{\Gamma \left(\frac{1}{3}\right)^3 \Gamma \left(\frac{7}{6}\right)^3}{\pi ^{3/2}}
\eea 
which is exactly the result \eqref{impdisp}, thus validating the present approach. 

Let us add some precisions about the $m_0=+\infty$ limit. Consider the
problem with two boundaries and denote $\tilde u^{m_0}_{r_1,r_2}(x)$ the solution
of the instanton equation \eqref{instanton-tip3} which tends to $-\infty$ for $x=r_1,r_2$
(with $r_1<0<r_2$). The solution studied in the previous subsection is thus
$\tilde u^{m_0}_{r_1=-\infty,r_2}(x)$. We surmise that in the limit $m_0=+\infty$
\be
\lim_{m_0 \to +\infty} \tilde u^{m_0}_{r_1,r_2}(x) = \tilde u^{\infty}_{r_2}(x) \theta(x) 
+ \tilde u^{\infty}_{r_1}(x) \theta(-x) 
\ee
where $\tilde u^{\infty}_{r>0}(x)$ is the solution of \eqref{instanton-tip4},
and $\tilde u^{\infty}_{r<0}(x)$ is the solution of the same equation
but defined on the negative $x<0$ half-axis. In particular $\tilde u^{\infty}_{r_1}(0^-)
= \tilde u^{\infty}_{r_2}(0^+)=0$. Now from the condition of derivative discontinuity
arising from \eqref{instanton-tip3} one has, in the limit 
\be
\lim_{m_0 \to +\infty} m_0^2 \tilde u^{m_0}_{r_1,r_2}(x) = \tilde u^{\infty \prime}_{r_2}(0^+) -  
 \tilde u^{\infty \prime}_{r_1}(0^-) \label{der33} 
\ee
Thus we have (using $x \to -x$ symmetry)
\bea
P(\ell_1 < -r_1 , \ell_2 < r_2) = e^{w \tilde u^{\infty \prime}_{r_2}(0^+)} e^{- w
 \tilde u^{\infty \prime}_{r_1}(0^-) } = e^{w \tilde u^{\infty \prime}_{r_2}(0^+)} 
 e^{w \tilde u^{\infty \prime}_{-r_1}(0^+)} 
 = P(\ell_1 < -r_1) P(\ell_2 < r_2) 
\eea 
which is perfectly consistent that in this limit the two "half-space" completely decouple. 
The total avalanche extension $\ell=\ell_1+\ell_2$ is just the sum of two independent random variables.
In the previous section, we studied $r_1=-\infty$ which leads to the
vanishing solution $\tilde u^{\infty \prime}_{r_1}(0^-)=0$. That justifies 
\eqref{eqeq}.

\subsubsection{The shape}

\label{subsec:shapetip} 

We now study the mean avalanche shape directly in the limit $m_0=+\infty$. We focus once
again on the half-space $x \geq 0$. We must now
study the following instanton equation.
\bea \label{instanton-tip5} 
&& \tilde u''(x)+ \tilde u(x)^2 =  - \lambda \delta(x-x_0)   \quad , \quad \tilde u(r)=-\infty \quad , \quad \tilde u(0)=0
\eea 
on the interval $[0,r]$. We can call $\tilde u(x)=u_r^{\lambda}(x)$ the solution. 
We can use again \eqref{der33} which extends in presence of the source.
Since the source acts only for $x>0$ the instanton solution again vanishes for $x<0$.
Using Appendix \ref{sec:rel} one has
\bea \label{aa2} 
\langle e^{\lambda S(x_0)} \rangle_{\ell_2<r} 
= e^{w (u_r^{\lambda \prime}(0)- u_r^{0 \prime}(0))}
\eea 
Hence we need to calculate the derivative $u_r^{\lambda \prime}(0)$. 
The solution of \eqref{instanton-tip5} is now
\bea
&& \tilde u_r^\lambda(x) = - \frac{6}{\rho^2} {\cal P}(1- \frac{x}{\rho},0,g_3^*)  \quad , \quad 0 < x < x_0 \\
&& \tilde u_r^\lambda(x) = - \frac{6}{r^2} {\cal P}(1- \frac{x}{r},0,g_3)  \quad , \quad x_0 < x < r
\eea 
note that it vanishes at $x=0$ and diverges at $x=r$ as required. The
matching conditions which determine the unknown parameters $\rho$ and $g_3$ as
a function of $r, x_0,\lambda$
\bea \label{consistency2} 
&& \frac{1}{\rho^2} {\cal P}(1- \frac{x_0}{\rho},0,g_3^*) = \frac{1}{r^2} {\cal P}(1- \frac{x_0}{r},0,g_3)  \\
&& \frac{6}{r^3} {\cal P}'(1- \frac{x_0}{r},0,g_3) - \frac{6}{\rho^3} {\cal P}'(1- \frac{x_0}{\rho},0,g_3^*) + \lambda=0 
\eea 
For $\lambda=0$ the solution is $\rho=r$ and $g_3=g_3^*$ and one recovers the
solution of the previous section. We now expand
\bea 
&& g_3= g_3^* + g_3^{(1)} \lambda + g_3^{(2)} \lambda^2 +O(\lambda^3)  \\
&& \rho= r + r^{(1)} \lambda + r^{(2)} \lambda^2 + O(\lambda^3)
\eea
and solve the consistency equations \eqref{consistency2} order by order in $\lambda$.
Then we insert the result in the quantity that we need to calculate
\bea \label{toexpand2} 
&& \tilde u_r^{\lambda \prime}(0) = \frac{6}{\rho^3} {\cal P}_*'(1) 
= - \frac{6}{\rho^3} (-g_3^*)^{1/2} = - \frac{A}{\rho^3} 
\quad , \quad A= 6 (-g_3^*)^{1/2} = 12
\frac{\Gamma \left(\frac{1}{3}\right)^3 \Gamma \left(\frac{7}{6}\right)^3}{\pi ^{3/2}}
\eea

The explicit calculation sketched above leads to
\bea
\partial_{\lambda}|_{\lambda=0} \tilde u^{\lambda \prime}(0) = 
\frac{A}{6 g_3^*} (2 {\cal P}_*\left(y_0\right)+ y_0 {\cal P}_*'\left(y_0 \right)) 
\quad , \quad y_0=1-z_0=1- \frac{x_0}{r}
\eea 
where we used the identity ${\cal P}_*'^2=4 {\cal P}_*^3 - g_3^*$. From this we obtain
the cumulative conditional mean shape as
\bea
&&  \langle S(x_0) \rangle_{\ell_2<r}  = 
w \partial_\lambda|_{\lambda=0} u_r^{\lambda \prime}(0) = w f(y_0=1- \frac{x_0}{r}) \\
&& f(y_0) = \frac{-1}{(-g_3^*)^{1/2}} (2 {\cal P}_*\left(y_0\right)+ y_0 {\cal P}_*'\left(y_0 \right))  
\quad , \quad g_3^*=-\frac{4 \Gamma \left(\frac{1}{3}\right)^6 \Gamma \left(\frac{7}{6}\right)^6}{\pi ^3} \\
&& ~~~~~~~= \frac{3}{14} |g_3^*|^{1/2} y_0^4 - \frac{3}{2548}  |g_3^*|^{3/2} y_0^{10} + O(y_0^{16}) 
\eea 
where in the last line we show the behavior at small $y_0$. We recall that ${\cal P}_*$
is given by \eqref{defPstar}. We note that the scaling function $f$ satisfies 
\be
\int_0^1 dy_0 f(y_0) = \frac{1}{|g_3^*|^{1/2}} \zeta(1;0,g_3^*) = 0.219308.. \label{intf} 
\ee
The mean shape at fixed extension is then, e.g. from Appendix \ref{sec:rel} see \eqref{last},
\bea
\langle S(x_0) \rangle_{\ell_2}  = w f(y_0) + \frac{1}{\ell_2 \rho(\ell_2)} (1-y_0) f'(y_0)  \quad , \quad y_0=1-\frac{x_0}{\ell_2}  \label{exact4}
\eea 
In the limit of small $w$ we can neglect the first term and using the density $\rho(\ell_2)=18 |g_3^*|^{1/2}/\ell_2^4$ from \eqref{impdisp2} we obtain the mean shape at fixed extension $\ell_2$ in the limit $w=0^+$
\bea
&& \langle S(x_0) \rangle_{\ell_2}  = \ell_2^3 \, {\sf s}(y_0)  \quad , \quad y_0=1-\frac{x_0}{\ell_2} \\
&&  {\sf s}(y_0) =  - \frac{1}{6 |g_3^*|} (1-y_0) ( 2 y_0 {\cal P}_*(y_0)^2 + {\cal P}_*'(y_0) ) \label{defs0} 
 \eea 
where $g_3^*$ is given in \eqref{g3sdef} and ${\cal P}_*$
in \eqref{defPstar}. The scaling function ${\sf s}(y_0)$ is plotted in 
Fig. \ref{fig:shapes}.
We find that the shape vanishes at $y_0=0$, i.e. $x_0=\ell_2$ and at $y_0=1$, i.e. $x_0=0$.
The latter is a consequence of the exact result that $S(x=0)=w$ 
(it can be checked on the exact formula \eqref{exact4}), 
from the imposed displacement
driving at $m_0=+\infty$. 
We find that the shape vanishes near the avalanche edge at $x=r$ as
\bea \label{se1} 
{\sf s}(y_0) = \frac{y_0^3}{21}-\frac{y_0^4}{21}-\frac{5 y_0^9 \left(\Gamma
   \left(\frac{1}{3}\right)^6 \Gamma \left(\frac{7}{6}\right)^6\right)}{1911 \pi ^3}+\frac{5 y_0^{10} \Gamma \left(\frac{1}{3}\right)^6 \Gamma \left(\frac{7}{6}\right)^6}{1911 \pi^3}+O\left(y0^{15}\right)
\eea 
We note that the leading term $\langle S(x_0) \rangle_{\ell_2}  \simeq \frac{1}{21} (\ell_2-x_0)^3$
is the same as the one in \eqref{leadingD} and in \eqref{edge11}, hinting a a universal
behavior near the edge (since the present tip driven avalanches are quite different from the
standard ones studied in previous sections). 

The behavior near the driving tip $x=0$ is linear (cusp)
\bea \label{st1} 
{\sf s}(y_0) = \frac{z_0}{6 \sqrt{|g_3^*|}}-\frac{z_0^3}{3}+\frac{2 z_0^4}{3}-\frac{1}{3} \sqrt{|g_3^*|}
   z_0^6+\frac{1}{2} \sqrt{|g_3^*|} z_0^7-\frac{5 |g_3^*| z_0^9}{28}+\frac{5 |g_3^*|
   z_0^{10}}{21}+O\left(z_0^{11}\right) \quad , \quad z_0=1-y_0=\frac{x}{\ell_2} 
\eea \\

From the mean shape at fixed extension $\ell_2$ we can extract the first conditional moment
of the total size $S$ of the avalanche.
From \eqref{exact4} and \eqref{defs0} we see that (using also \eqref{intf}) 
\bea
\int_0^1 dy_0 {\sf s}(y_0) = \frac{1}{18 |g_3^*|^{1/2}} \int_0^1 dy_0 f(y_0) = 
\frac{- \zeta(1;0,g_3^*)}{18 g_3^*} 
\eea 
which implies the first moment 
\bea \label{firstfirst}
&& \langle S \rangle_{\ell_2}  = \int_0^{\ell_2} dx_0 \langle S(x_0) \rangle_{\ell_2}
= \ell^4 \frac{- \zeta(1;0,g_3^*)}{18 g_3^*} = 0.0022097287584 \, \ell^4
\eea

\subsubsection{The second shape}

\label{subsec:shapetip2} 

We now calculate the second shape, i.e. $\langle S(x_0)^2 \rangle_{\ell_2}$. For this
purpose we need to expand to next order in $\lambda$ the equations \eqref{toexpand2}
and insert in $\tilde u_r^{\lambda \prime}(0) =  - \frac{A}{\rho^3}$ from \eqref{consistency2}.
Using \eqref{aa2} we obtain first the cumulative second shape
\be
\langle S(x_0)^2 \rangle_{\ell_2<r}  = 
w \partial^2_\lambda|_{\lambda=0} u_r^{\lambda \prime}(0) + O(w^2) 
=  w \, r^3 f_2(y_0=1- \frac{x_0}{r}) + O(w^2) 
\ee
with 
\bea
&& f_2(y_0)=
\frac{1}{9 (-g_3^*)^{3/2}} 
\bigg( g_3^* \left(y_0-1\right) y_0 \\
&& +{\cal P}_*\left(y_0\right) \left(\left(1-2 \left(y_0-1\right)
   \left(3 y_0^2 {\cal P}_*\left(y_0\right)-1\right)\right) {\cal P}_*'\left(y_0\right)+2 {\cal P}_*\left(y_0\right)
   \left(\left(5-8 y_0\right) y_0 {\cal P}_*\left(y_0\right)+4\right)\right)
\bigg)
\eea
%

From Appendix \ref{sec:rel}, see \eqref{beforelast}, we have
\bea
\langle e^{\lambda S(x_0)} \rangle_{\ell_2} 
= \frac{\partial_r u_r^{\lambda \prime}(0) }{\partial_r u_r^{0 \prime}(0)} e^{w (u_r^{\lambda \prime}(0)- u_r^{0 \prime}(0))}|_{r=\ell_2} \simeq_{w \to 0^+} \frac{1}{\rho(\ell_2)} \partial_r u_r^{\lambda \prime}(0)|_{r=\ell_2} 
\eea 
Hence we have $\langle S(x_0)^2 \rangle_{\ell_2} = \frac{1}{\rho(\ell_2)} \partial_r 
\partial^2_{\lambda}|_{\lambda=0}  u_r^{\lambda \prime}(0)|_{r=\ell_2}$. 
This leads to the second shape for $w \to 0^+$ as
\bea \label{s2s2} 
&& \langle S(x_0)^2 \rangle_{\ell_2} = \ell_2^6 \, {\sf s}_2(y_0) \quad , \quad y_0=1- \frac{x_0}{\ell_2} \\
&& {\sf s}_2(y_0)  = \frac{1}{54
   |g_3^*|^2} \bigg( -|g_3^*| \left(y_0-1\right) y_0 + {\cal P}_*\left(y_0\right) \bigg( 4 |g_3^*| \left(y_0-1\right){}^2 y_0^2
   \\
   && +\left(2 \left(y_0-1\right)
   \left(8 y_0-7\right) y_0 {\cal P}_*\left(y_0\right)-4 y_0+5\right) {\cal P}_*'\left(y_0\right)
    +2
   {\cal P}_*\left(y_0\right) \left(y_0 {\cal P}_*\left(y_0\right) \left(14 y_0 \left(y_0-1\right){}^2
   {\cal P}_*\left(y_0\right)-6 y_0+3\right)+4\right)\bigg)  \bigg) \nn
\eea 
where the scaling function is now obtained as ${\sf s}_2(y_0)=\frac{1}{\rho(\ell_2)} ( 3 f_2(y_0)+ (1-y_0)f_2'(y_0)$. It is plotted in Fig. \ref{fig:shapes}.
The behavior near the edge is
\bea
{\sf s}_2(y_0)  = 
\frac{y_0^6}{273}-\frac{38 y_0^7}{5733}+\frac{5 y_0^8}{1764}-\frac{|g_3^*| y_0^{12}}{5586}+\frac{155
   |g_3^*| y_0^{13}}{435708}-\frac{55 |g_3^*| y_0^{14}}{321048}+O\left(y_0^{15}\right)
\eea 
and near the tip 
\bea
{\sf s}_2(y_0)  = \frac{z_0^2}{18 |g_3^*|}-\frac{7 z_0^4}{36 \sqrt{|g_3^*|}}+\frac{z_0^5}{9
   \sqrt{|g_3^*|}}+\frac{z_0^6}{2}+O\left(z_0^7\right) \quad , \quad z_0=1-y_0=\frac{x}{\ell_2} 
\eea 

We note that the ratio
\bea
\frac{\langle S(x_0)^2 \rangle_{\ell_2} }{\langle S(x_0)\rangle_{\ell_2}^2} =  \frac{{\sf s}_2(y_0)}{{\sf s}_1(y_0)^2}
\eea 
plotted in Fig. \ref{fig:ratio},
varies between $\frac{21}{13}=1.61538$ for $x$ near $\ell_2$, and $2$ for $x$ near the tip $x=0$.\\

{\bf Remark}. We notice that the leading behavior for the second moment near the edge is
\be
\langle S(x_0)^2 \rangle_{\ell_2} \simeq \frac{1}{273} (\ell_2-x_0)^6
\ee 
This is exactly the same behavior as found in \eqref{momom} with the same
amplitude. This is not fortuitous. Indeed one can show that for the present tip
driven avalanches one also has $S(x_0) \simeq (\ell_2-x_0)^3 \sigma$ 
where the random variable $\sigma$ has the same PDF as in \eqref{genfunct}, which thus appears to be
universal (at least within the BBM with short-range elasticity). To see that, compare 
Eqs. \eqref{consistency2} with Eqs. \eqref{cond1}, \eqref{cond2}. It is easy to 
see that in both systems of equations, $\rho-x_0$ and 
$r-x_0$ vanish similarly, and under the ansatz \eqref{ansatz1}, 
the two systems become asymptotically equivalent. Indeed
in both cases the scaling regime is such that 
$\lambda$ hence $g_3$ are very large. On the l.h.s of the first equation
in \eqref{consistency2} $g_3^*$ is fixed so as $\rho-x_0 \to 0$ the l.h.s.
becomes identical to the l.h.s in \eqref{cond1}. The r.h.s can also
be made to coincide using the homogeneity property of the
Weierstrass function and changing $g_3 \to g_3/r^6$. Similarly the 
second equation in \eqref{consistency2} becomes identical to the
one in \eqref{cond2} by the same manipulation. Since $g_3$ is eliminated
between the two equations its rescaling is immaterial. The two systems are
then equivalent. One has
\bea
&& \langle e^{\lambda S(x_0) } \rangle_{\ell_2=r} = \frac{\partial_r u^\lambda_r(0)}{\partial_r u^0_r(0)}
=  \frac{\partial_r \rho^{-2}}{\partial_r r^{-2}} \simeq \partial_r \rho \\
&& \langle e^{\lambda S(x_0) } \rangle_{\ell_2} = \frac{\partial_r u^{\lambda,\prime}_r(0)}{\partial_r u^{0,\prime}_r(0)}
=  \frac{\partial_r \rho^{-3}}{\partial_r r^{-3}} \simeq \partial_r \rho 
\eea 
since $\rho \simeq r$ when $x_0 \to r$. Hence the generating functions are identical
in this regime, and the PDF's of the random variable $\sigma$ are identical.

\subsubsection{Direct calculation of the conditional moments of the total avalanche size} 
\label{subsec:conddirect} 

We can also calculate the conditional moments of the total avalanche size $S$.
We need to solve now 
\bea \label{instanton-tip5} 
&& \tilde u''(x)+ \tilde u(x)^2 =  - \mu \theta(x)   \quad , \quad \tilde u(r)=-\infty \quad , \quad \tilde u(0)=0
\eea 
on the interval $[0,r]$. We can call $\tilde u(x)=u_r^{\mu}(x)$ the solution. The solution
reads 
\bea
\tilde u^\mu_r(x) =  - \frac{6}{r^2} {\cal P}(\frac{r-x}{r}; g_2=- \frac{1}{3} \mu r^4 , g_3(\mu))
\eea 
where $g_3(\mu)$ is such that, where in the second equation we use \eqref{defint}
\bea
{\cal P}(1; g_2=- \frac{1}{3} \mu r^4 , g_3(\mu)) = 0 ~~ \Leftrightarrow ~~ \int_0^{+\infty} \frac{dy}{\sqrt{4 y^3 + \frac{1}{3} \mu r^4 y - g_3(\mu)}} = 1
\eea 
with $g_3(0)=g_3^*$. Using mathematica and ${\cal P}_*(1)=0$, ${\cal P}'(1)=-(-g_3^*)^{1/2}$ one easily finds
\be
g_3(\mu)= -|g_3^*|+\frac{1}{3} \mu  r^4 X 
-\frac{\mu ^2 r^8 \left(\sqrt{|g_3^*|}-2 X^2\right)}{72
   |g_3^*|} 
+ \frac{7 \mu ^3 r^{12} \left(-9 \sqrt{|g_3^*|} X+|g_3^*|+16 X^3\right)}{11664
   |g_3^*|^2} + O(\mu^4)  , \quad X= \zeta(1;0,g_3^*)
\ee
We aim to calculate
\bea
&& \tilde u^{\mu \prime}_r(0) =  \frac{6}{r^3} {\cal P}'(1, g_2=- \frac{1}{3} \mu r^4 , g_3(\mu)) \\
&& = -\frac{6 \sqrt{|g_3^*|}}{r^3}+\frac{\mu  r X}{\sqrt{|g_3^*|}}-\frac{\mu ^2
   \left(r^5 \left(\sqrt{|g_3^*|}-4 X^2\right)\right)}{24
   |g_3^*|^{3/2}}+\frac{\mu ^3 r^9 \left(-90 \sqrt{|g_3^*|} X+7
   |g_3^*|+220 X^3\right)}{3888 |g_3^*|^{5/2}}+O\left(\mu ^4\right)
\eea 
from which we obtain the conditional generating functions, see Appendix \ref{sec:rel} 
\be
\langle e^{\mu S} \rangle_{\ell_2<r}  =  e^{w (u_r^{\mu \prime}(0)- u_r^{0 \prime}(0))}
= 1+ w (u_r^{\mu \prime}(0)- u_r^{0 \prime}(0)) + O(w^2) 
\quad , \quad \langle e^{\mu S} \rangle_{\ell_2}|_{w=0^+} = \frac{1}{\rho(\ell_2)} 
\partial_r u_r^{\mu \prime}(0)|_{r=\ell_2}
\ee
where $\rho(\ell_2)$ is given in \eqref{impdisp2}.
This leads to the lowest moments 
\bea
\left\{\ \langle S \rangle_{\ell_2<r} ,\langle S^2 \rangle_{\ell_2<r} , \langle S^3 \rangle_{\ell_2<r}  \right\}
= w \left\{\frac{r X}{\sqrt{|g_3^*|}},-\frac{r^5 \left(\sqrt{|g_3^*|}-4
   X^2\right)}{12 |g_3^*|^{3/2}},\frac{r^9 \left(-90 \sqrt{|g_3^*|} X+7
   |g_3^*|+220 X^3\right)}{648 |g_3^*|^{5/2}}\right\} \\
= w \{0.2193081289 r, 0.0001665224164 r^5 , 5.244855047*10^{-7} r^9\} + O(w^2)
\eea 
as well as 
\bea \label{cc} 
 \left\{\ \langle S \rangle_{\ell_2} ,\langle S^2 \rangle_{\ell_2} , \langle S^3 \rangle_{\ell_2}  \right\}|_{w=0^+}
&=&
\left\{\frac{\ell_2^4 X}{18 |g_3^*|},-\frac{5 \ell_2^8 \left(\sqrt{|g_3^*|}-4
   X^2\right)}{216 |g_3^*|^2},\frac{\ell_2^{12} \left(-90 \sqrt{|g_3^*|} X+7
   |g_3^*|+220 X^3\right)}{1296 |g_3^*|^3}\right\} \nn \\
   &=&  \{0.002209728758 \ell_2^4, 8.389323605 \times 10^{-6} \ell_2^8, 4.756201414 \times 10^{-8} \ell_2^{12}\}
\eea 
We see that the result for the first conditional moment $\langle S \rangle_{\ell_2}$ is in agreement
with the formula \eqref{firstfirst} obtained there by integrating the mean shape over $x_0$.

\section{Avalanches in the BFM in higher dimension $d>1$}
\label{sec:higherd}

We now study the BFM in space dimension $d>1$, as defined by the equations of motion \eqref{BFMpos1} and
\eqref{BFMdef1}. 

\subsection{Probability that an avalanche does not contain the neighborhood of a given point}
\label{sub:point} 

One can ask about the PDF of the local size $S_0=S(x=0)$ in dimension $d$, and in particular
about the probability that an avalanche does not reach the point $x=0$, i.e. that $S_0=0$. 
We focus on the massless case. To this aim one must
solve the instanton equation with the source $\lambda(x)= \lambda \delta^d(x)$ 
\bea
\nabla^2_x \tilde u + \tilde u^2 = - \lambda \delta^d(x)  \label{instd} 
\eea 
where $\nabla^2_x$ is the $d$-dimensional Laplacian.
By contrast with $d=1$, there is no explicit solution for arbitrary $\lambda$. 
Furthermore, although \eqref{instd} remains well defined in perturbation theory in $\lambda$,
and can be solved iteratively as such (leading to finite moments of $S_0$ upon adding a mass)  
in $d \geq 2$ there are some UV problems with the delta source, especially
if one wants to study the limit $\lambda \to -\infty$. It is safer 
to consider a smoothed out source, e.g. $\lambda(x)= \lambda a^{-d} \theta(a-|x|)$.
One can then take $\lambda \to - \infty$
and obtain the probability that the avalanche does not enter a ball of radius $a$
(a problem studied below). On the sphere $|x|=a$ one has $\tilde u(|x|=a)=-\infty$. Taking
then $a \to 0^+$ the problem converges to the instanton equation
\bea
\nabla^2_x \tilde u + \tilde u^2 = 0  \quad , \quad \tilde u(0)=-\infty  \label{instd2}
\eea 
which thus describes the probability that the avalanche does not
reach an infinitesimal ball $a=0^+$ around $x=0$.

It turns out that there is a very simple solution to \eqref{instd2}. Indeed,
looking for a rotationally invariant solution, we rewrite it in radial coordinates $r=|x|$
\bea
\frac{d^2}{d r^2} \tilde u + \frac{d-1}{r} \frac{d}{dr} \tilde u + \tilde u^2 = 0  
\eea 
and a solution to this equation, valid in any $d<4$, is
\bea
\tilde u(x) = - \frac{2 (4-d)}{x^2}  \label{simple1} 
\eea 
Using this solution we can thus obtain the probability that an avalanche following a kick at $x_s$,
i.e. with driving $\dot f(x,t)=f \delta(t)\delta^d(x-x_s)$,
does not contain the point $x_0$ (more precisely, a small ball around it, see discussion
above) 
\bea \label{s00} 
{\rm Prob}(S(x_0) =0) = {\rm Prob}(x_0 \notin \Omega) = e^{- f \frac{2 (4-d)}{(x_0-x_s)^2} }
\eea 
where $\Omega$ is the support of the avalanche. 
Although this is interesting exact result in any $d$, it does not give so much information about how far the avalanche reaches, since for $d>1$ the boundary of the avalanche can be quite complicated,
fractal and a priori not simply connected (i.e. the avalanche could exhibit "holes"). However the derivative
\bea
dx_0 \cdot \nabla_{x_0} {\rm Prob}(S(x_0) =0) 
\eea 
says something about the probability that the boundary of the avalanche crosses
the interval $[x_0,x_0+dx_0]$ at least once, and possibly more (an odd number of times). Again it decays as 
$1/|x_0-x_s|^3$ as in $d=1$. 

\subsection{Probability distribution of $\Sigma = \int d^d x \frac{S(x)}{(x-x_0)^4}$ conditioned to $S(x_0)=0$}

Another exact result can be obtained for the following observable
\bea
\Sigma := \int d^d x \frac{S(x)}{(x-x_0)^4} 
\eea
conditioned to $S(x_0)=0$.
It is associated to the following instanton equation with a source
\bea
\nabla^2_x \tilde u + \tilde u^2 = - \frac{\mu}{(x-x_0)^4} \quad , \quad \tilde u(x_0)=-\infty  \label{instd22}
\eea 
which has the simple solution
\bea
\tilde u(x) = - \frac{4-d+ \sqrt{(4-d)^2 - \mu}}{(x-x_0)^2} 
\eea  
For $\mu=0$ it reduces to the previous calculation. Hence the LT of the joint
PDF that $S(x_0)=0$ and $\Sigma$ following a kick at $x_s$, reads
\bea
\int d\Sigma e^{\mu \Sigma} P(S(x_0) =0, \Sigma) = e^{- f \frac{4-d+ \sqrt{(4-d)^2 - \mu}}{(x_s-x_0)^2} } 
\eea
From which we obtain the PDF of $\Sigma$ {\it conditioned to} $S(x_0)=0$ as
\bea \label{PSigma}
&& P(\Sigma | S(x_0) =0) = LT^{-1}_{-\mu \to \Sigma} e^{f \frac{4-d - \sqrt{(4-d)^2 - \mu}}{(x_s-x_0)^2} }  = \frac{f}{2 \sqrt{\pi} \Sigma^{3/2} (x_s-x_0)^2} e^{- \frac{\left(2 (4-d) \Sigma - \frac{f}{(x_s-x_0)^2}\right)^2}{4 \Sigma}}
\eea 
Note that conditioning to $S(x_0)=0$ is necessary for $\Sigma$ to be finite (which it is
since its LT is analytic near $\mu=0$).\\

Let us analyze this result. The first moment of $\Sigma$ equals
\be \label{Sigmamean}
 \langle \Sigma \rangle_{S(x_0) =0} = \int d^d x \frac{\langle S(x) \rangle_{S(x_0)=0}}{(x-x_0)^4}
 = \frac{1}{2 (4-d)} \frac{f}{(x_s-x_0)^2} 
\ee
First note that $\Sigma$ is dimensionless (and so is the combination $\frac{f}{(x_s-x_0)^2}$). 
Second, \eqref{Sigmamean} is a non-trivial sum rule for the mean shape $\langle S(x) \rangle_{S(x_0)=0}$
around the point $x_0$ (such that $S(x_0) =0$). The analog in $d=1$ is \eqref{condmeanint},
which we reproduce here in the (different) notations of the present section (and using $r$ rather than $x_0$)
\bea
\langle S(x) \rangle_{S(r)=0} = \langle S(x) \rangle_{\ell_2<r}  =
\frac{f}{7} \left( \frac{(r-x)^4}{(r-x_s)^3}  \theta(x_s < x < r) + 
 \frac{(r-x_s)^4}{(r-x)^3}  \theta(x < x_s) \right) \label{condmeanint2} 
\eea 
and $\langle S(x) \rangle_{S(r)=0}$ vanishes for $x>r$ (a specificity of $d=1$). 
One can check from this formula that indeed 
$\int_{-\infty}^{+\infty} \langle S(x) \rangle_{S(r)=0} = \frac{1}{6} \frac{f}{(r-x_s)^2}$,
as predicted by \eqref{Sigmamean}. The mean shape $\langle S(x) \rangle_{S(x_0)=0}$
in higher dimension is studied in the next Section.

Let us analyze the main features of the PDF of $\Sigma$ in \eqref{PSigma}.
For large driving, i.e. large $\frac{f}{(x_s-x_0)^2}  \gg 1$, the random variable $\Sigma$ equals its average $\Sigma \simeq \langle \Sigma \rangle_{S(x_0) =0}$ up to $O(\langle \Sigma \rangle_{S(x_0) =0}^{1/2})$ Gaussian fluctuations. The PDF of $\Sigma$ in \eqref{PSigma} has some similarities with 
the PDF of the total size $S= \int d^d x S(x)$, given in \eqref{PDFS}. Indeed it exhibits
the same $3/2$ exponent $P(\Sigma | S(x_0) =0) \sim \Sigma^{-3/2}$. This exponent is visible at
small driving $\frac{f}{(x_s-x_0)^2}  \ll 1$ over a large range of values of $\Sigma_0 \ll \Sigma \ll 1$, with 

(i) a cutoff at small $\Sigma \ll 1$ given by $\Sigma_0 \sim \frac{f^2}{(x_s-x_0)^4} \ll 1$. Presumably this
corresponds to small avalanches localized near the seed $x_s$ with 
$\Sigma \simeq \frac{S}{(x_s-x_0)^4}$ with $S \sim S_0 \sim f^2$. 

(ii) at larger $\Sigma$ the term $\sim e^{- (4-d)^2 \Sigma}$ in \eqref{PSigma}
cuts $\Sigma$ at values of $O(1)$.

Hence we see that although there is no mass here to cut the large avalanches, the
conditioning on $S(x_0) =0$ appears to be sufficient to cut the PDF of $\Sigma$ at
large $\Sigma$ and make the moments finite, e.g. the second moment is
\bea
\langle \Sigma^2 \rangle_{S(x_0) =0} \simeq \frac{1}{4 (4-d)^3}  \frac{f}{(x_s-x_0)^2} (1 
+ \frac{(4-d) f}{(x_s-x_0)^2})
\eea
which is finite for $d<4$.

\subsection{Mean avalanche shape around a point such that $S(x_0)=0$} 
\label{shaped} 

We ask now what is the mean shape $\langle S(x) \rangle_{S(x_0)=0}$ around a
point such that $S(x_0)=0$, for an avalanche with a seed at $x_s$. 
For simplicity we set $x_0=0$. We study the massless case, the massive case
is briefly mentionned in Appendix \ref{app:msmass}. Let us recall that 
\bea
&& \langle e^{\int d^d x \lambda(x) S(x)} \rangle_{S(x_0)=0} = e^{f (u^{\lambda}(x_s)-u^{0}(x_s))}
\eea 
where $u^{\lambda}(x)$ is the solution of
\bea
\nabla^2_x \tilde u + \tilde u^2 = - \lambda(x)  \quad , \quad \tilde u(0)=-\infty  \label{instd3}
\eea 
Here we are interested only in the mean shape, i.e. averages like
\bea
&& \langle \int d^d x' \lambda(x') S(x') \rangle_{S(x_0)=0} = f \int dx' u_1(x_s,x') \lambda(x') \label{avmean}
\eea 
hence we only need to solve \eqref{instd3} to $O(\lambda)$, and we have defined
\bea
&& u^{\lambda}(x) = - \frac{2 (4-d)}{x^2} + \int dx' u_1(x,x') \lambda(x') + O(\lambda^2) 
\eea 
Replacing into \eqref{instd3} we see that $u_1(x,x')$ satisfies
\bea \label{GF2} 
\nabla^2_x \tilde u_1 -  \frac{4 (4-d)}{x^2}  \tilde u_1  = - \delta^d(x-x')
\eea 
hence it is the zero energy Green's function associated to the Schrodinger equation in the inverse square potential in any space dimension $d$. This Green function is known
(see e.g. \cite{RegularizedProp}). One uses spherical coordinates
$x = (r,{\bf \phi})$ where ${\bf \phi}$ is a $d-1$ dimensional angular vector, and the 
representation of the Laplacian 
\bea
\nabla^2_x = r^{- \frac{d-1}{2}} \partial_r^2 r^{\frac{d-1}{2}} - \frac{L^2 + \frac{(d-1)(d-3)}{4}}{r^2} 
\eea 
where $L^2$ is the square angular momentum operator, i.e. minus the Laplace-Beltrami operator (Laplacian on the sphere). It has eigenvectors $Y_{\ell,m}({\bf \phi})$, the spherical harmonics, 
and eigenvalues $\ell (\ell+d-2)$, $\ell \geq 0$, with degeneracies
$g_\ell=(2 \ell + d-2)(\ell+d-3)!/\ell!(d-2)!$. Hence one has
\bea \label{inst00} 
u_1(r,{\bf \phi},r',{\bf \phi}) = (r r')^{- \frac{d-1}{2}} \sum_{\ell=0}^\infty \sum_{m=1}^{g_\ell} 
Y_{\ell,m}({\bf \phi}) Y_{\ell,m}^*({\bf \phi}') G_{\ell + \frac{d-2}{2}}(r,r') 
\eea 
where $G_{\ell + \frac{d-1}{2}}(r,r')$ is the Green function of the equivalent radial problem
in the sector associated to angular momentum quantum number $\ell$
\bea \label{GreenRadial} 
\left(\frac{d^2}{dr^2} - \frac{4 (4-d) + ( \ell + \frac{d-2}{2})^2 - \frac{1}{4} }{r^2} \right) 
G_{\ell + \frac{d-2}{2}}(r,r') = - \delta(r-r')
\eea 
whose solutions (with the proper continuity and derivative discontinuity conditions)
are simple power laws, i.e.
\bea
G_{\ell + \frac{d-2}{2}}(r,r') = \frac{\sqrt{r r'}}{2 a_{\ell,d}} \left( (\frac{r}{r'})^{a_{\ell,d}}  \theta(r'-r)
+ (\frac{r'}{r})^{a_{\ell,d}} \theta(r-r') \right)
\eea 
leading to the final expression
\bea
u_1(r,{\bf \phi},r',{\bf \phi}) = \sum_{\ell=0}^\infty \sum_{m=1}^{g_\ell} 
Y_{\ell,m}({\bf \phi}) Y_{\ell,m}^*({\bf \phi}')  \frac{(r r')^{\frac{2-d}{2}} }{2 a_{\ell,d}}  \left( (\frac{r}{r'})^{a_{\ell,d}}  \theta(r'-r)
+ (\frac{r'}{r})^{a_{\ell,d}} \theta(r-r') \right) 
\eea
where the exponents are
\bea
&& a_{\ell,d}=\sqrt{4(4-d) + (\ell+\frac{d-2}{2})^2}
\eea 
These non-trivial exponents are associated to a given angular sector. They can be
selected by considering observables with the same symmetry. 

One easily sees that for $d=1$, only $\ell=0$ and $\ell=1$ exist, with $g_0=1$, $g_1=1$ and $g_{\ell \geq 2}=0$.
The exponent is $a_{0,1}=a_{1,1}=7/2$ and one recovers
\bea \label{avmean2}
\langle S(x=r) \rangle_{S(0)=0} =  f 
u_1(r_s,r)  = f \left(  \frac{r_s^4}{7 r^3}  \theta(r-r_s) + \frac{r^4}{7 r_s^3}  \theta(r_s-r) \right)
\eea
which is exactly our result \eqref{condmeanint} for a single boundary in $d=1$
(in different notations). Note that at variance with \eqref{condmeanint} there can be an avalanche
on the other side $x=-r$ with a similar expression (which is why both $\ell=0$ and $\ell=1$ contribute).  \\

Let us now analyze the case $d=2$. The vector $\phi$ is then simply the polar angle $\phi \in [0,2 \pi]$.
One has $g_\ell=2-\delta_{\ell,0}$, with
\bea
Y_{\ell,1}(\phi) = \frac{1}{\sqrt{2 \pi}} e^{i \ell \phi} \quad 
\quad Y_{\ell,2}(\phi) = \frac{1}{\sqrt{2 \pi}} e^{- i \ell \phi}
\eea 
and $a_{\ell,2}=\sqrt{8 + \ell^2}$. One obtains in polar coordinate (we now use $\ell$ to denote the eigenvalue of the $L_z$ operator) 
\bea \label{avmean3}
\langle S(r,\phi) \rangle_{S(0)=0} =  
f u_1(r_s,\phi_s,r,\phi)  = 
 \frac{f}{2 \pi} \sum_{\ell=-\infty}^{+\infty }  \frac{1}{2 \sqrt{8 + \ell^2}} 
 ( (\frac{r_s}{r})^{\sqrt{8 + \ell^2}}  \theta(r-r_s)
+ (\frac{r}{r_s})^{\sqrt{8 + \ell^2}} \theta(r_s-r) ) e^{i \ell (\phi_s-\phi)}
\eea
Several features can be pointed out. First, note the invariance by $r \to 1/r$, valid in $d=2$. 
Second, one can ask how the mean shape
vanishes near the point where $S(x_0=0)=0$. For $r \ll r_s$ the series is dominated
by the term $\ell=0$ hence we find
\bea \label{avmean4}
\langle S(r,\phi) \rangle_{S(0)=0} \simeq_{r \to 0}  \frac{f}{8 \pi \sqrt{2}} (\frac{r}{r_s})^{2 \sqrt{2}} 
+ O((\frac{r}{r_s})^{3} \cos(\phi_s-\phi)) \label{exponent1} 
\eea
Note that $2 \sqrt{2}=2.8284..$, thus the correction term is quite large (the difference between the exponents is small). Similarly for large $r$
\bea \label{avmean4}
\langle S(r,\phi) \rangle_{S(0)=0} \simeq_{r \to +\infty}  \frac{f}{8 \pi \sqrt{2}} (\frac{r_s}{r})^{2 \sqrt{2}} 
+ O((\frac{r_s}{r})^{3}) \label{exponent2} 
\eea
For general $r$ we do not know of a simple formula for the series (although a few integral representations
can be obtained). However for $r/r_s$ close to unity, one expects
that one can replace a sum by an integral, i.e. for $r<r_s$, we obtain, with $\ell = 2 \sqrt{2} \sinh u$
\bea
&& \langle S(r,\phi) \rangle_{S(0)=0} \simeq 
\frac{f}{4 \pi} \int_{-\infty}^{+\infty }  \frac{d\ell}{\sqrt{8 + \ell^2}} 
 e^{- \ln(\frac{r_s}{r}) \sqrt{8 + \ell^2} + i \ell (\phi_s-\phi)}
 \\
 && = \frac{f}{4 \pi} \int_{-\infty}^{+\infty } du e^{- 2 \sqrt{2} \ln(\frac{r_s}{r}) \cosh u + i 2 \sqrt{2}  (\phi_s-\phi)
 \sinh u }  \label{approx} 
\eea
For $\phi=\phi_s$ we find, using that $\int_{-\infty}^{+\infty } du \exp(- b \cosh u) = 2 K_0(b)$ (GR 3.337),
that for $r_s-r \to 0^+$
\bea
\langle S(r,\phi_s) \rangle_{S(0)=0} \simeq  \frac{f}{2 \pi}  K_0(2 \sqrt{2} \ln(\frac{r_s}{r}) )
\simeq_{r \to r_s^-} -  \frac{f}{2 \pi} \ln( \frac{r_s-r}{r_s} e^{\gamma_E} \sqrt{2} ) \label{approx2} 
\eea 
which shows that the mean shape diverges logarithmically near the seed as $x \to x_s^-$ (this comes from
small scales, see discussion in Appendix). We recalled that in $d=1$, i.e. in \eqref{avmean2},
 it has a jump discontinuity $-f$ of the first derivative at the seed. Note that the approximation \eqref{approx}
is not so bad even for small $r/r_s$ since \eqref{approx2} yields the correct exponent
\eqref{exponent1} in that limit (but an incorrect prefactor). \\

Consider now the following angular averaged mean shape 
\bea \label{ave} 
\langle \int_0^{2 \pi} d\phi \cos(\phi \ell)  S(r,\phi) \rangle_{S(0)=0} = f 
 \frac{\cos( \ell \phi_s)}{2 \sqrt{8 + \ell^2}} 
 ( (\frac{r_s}{r})^{\sqrt{8 + \ell^2}}  \theta(r-r_s)
+ (\frac{r}{r_s})^{\sqrt{8 + \ell^2}} \theta(r_s-r) ) 
\eea 
which selects a given angular momentum $\ell$ and corresponds to a pure
power law. Hence each exponent $a_{\ell,d}$ can be observed by suitably
averaging the shape around the point $x_0=0$ (conditioned to $S(x_0)=0$). 

This analysis can be extended to any dimension $d<4$. Denoting $\phi$ a $d-1$ dimensional unit vector, the shape is 
\bea \label{avmean5}
\langle S(r,\phi) \rangle_{S(0)=0} =  f 
u_1(r_s,\phi_s,r,\phi)  = f 
\sum_{\ell=0}^\infty \sum_{m=1}^{g_\ell} 
Y_{\ell,m}({\bf \phi}_s) Y_{\ell,m}^*({\bf \phi})  \frac{(r_s r)^{\frac{2-d}{2}} }{2 a_{\ell,d}}  \left( (\frac{r_s}{r})^{a_{\ell,d}}  \theta(r'-r)
+ (\frac{r}{r_s})^{a_{\ell,d}} \theta(r-r') \right) 
\eea
For $r \ll r_s$ the series is dominated
by the term $\ell=0$ hence we find
\bea \label{avmean6}
\langle S(r,\phi) \rangle_{S(0)=0} \simeq_{r \to 0} f C_d  r_s^{b_d^-} r^{b_d^+} \quad , \quad b_d^\pm = \frac{2-d}{2} \pm a_{0,d} 
=  \frac{1}{2} \left(2-d \pm \sqrt{d^2-20 d+68}\right)
\label{exponent2} 
\eea
and similarly
\bea \label{avmean6}
\langle S(r,\phi) \rangle_{S(0)=0} \simeq_{r \to +\infty}  f C_d  r_s^{b_d^+} r^{b_d^-} \quad , \quad b_d^\pm = \frac{2-d}{2} \pm a_{0,d} 
=  \frac{1}{2} \left(2-d \pm \sqrt{d^2-20 d+68}\right)
\label{exponent3} 
\eea
with the non-trivial exponents
\bea
&& b^+=4 \quad , \quad b^-=-3 \quad \text{for} \quad d=1 \\
&& b^+=2 \sqrt{2}  \quad , \quad b^-=- 2 \sqrt{2} \quad \text{for} \quad d=2 \\
&& b^+= \frac{1}{2} (\sqrt{17}-1) \quad , \quad b^-= - \frac{1}{2} (1+\sqrt{17}) \quad \text{for} \quad d=2 \\
&& b^+ \to 0 \quad , \quad b^- \to -2 \quad \text{for} \quad d \to 4
\eea 
Again, one can select a given angular momentum $\ell$ and a pure power law with
exponent 
\bea
b_{\ell,d}^\pm = \frac{2-d}{2} \pm a_{\ell,d} 
=  \frac{2-d}{2} \pm \sqrt{4(4-d) + (\ell+\frac{d-2}{2})^2} 
\eea
by suitably
averaging the shape as in \label{ave} around the point $x_0=0$ (conditioned to $S(x_0)=0$). 

\subsection{Probability that an avalanche does not reach a distance $R$ from the seed}
\label{subsec:reach} 

Consider a simply connected domain ${\cal D}$ containing the point $x=x_s$.
To obtain the probability that an avalanche started at $x=x_s$ remains inside
${\cal D}$, one must solve the instanton equation with the boundary 
condition $\tilde u(x)= - \infty$ for all points $x$ in the complementary of the
domain ${\cal D}^c=\mathbb{R}^d-{\cal D}$. However, heuristically, this appears to be equivalent
to solving the same equation with the boundary 
condition $\tilde u(x)= - \infty$ on the boundary of the domain, $x \in \partial {\cal D}$.
This is a property of the local elastic kernel that we consider here, which we
already used in $d=1$. We can safely assume it to remain true in $d>1$ (under
some conditions of regularity of the domain and its frontier). \\

Let us now calculate the probability that
an avalanche starting at $x=x_s$ does not reach the sphere $r=R$ centered at point $x=0$ (in spherical coordinates around $x=0$). Here the seed 
can be either inside the sphere $|x|=r <R$ (and then it gives the probability that the avalanche
remains inside the sphere), or
outside $|x|=r > R$ (and then it gives the probability that the avalanche remains
outside the sphere). This requires to find the solution, for $r \in [0,R[ \, \cup \, ]R,+\infty[$, denoted $\tilde u_R(r)$, of 
\bea
\frac{d^2}{d r^2} \tilde u + \frac{d-1}{r} \frac{d}{dr} \tilde u + \tilde u^2 = 0 \quad , \quad \tilde u(R)=-\infty 
\quad , \quad \tilde u(+\infty)=0
\eea 
which is smooth and bounded everywhere, except on the sphere where it diverges. 
It is easy to see that for $d<4$
\bea \label{expected1} 
\tilde u_R(r) \simeq_{r \to +\infty} - \frac{2(4-d)}{r^2} \quad , \quad \tilde u_R(r) \simeq_{r \to R^\pm} - \frac{6}{(r-R)^2} 
\eea 
i.e. the behavior far from the sphere is similar to setting $R=0$, i.e. \eqref{simple1}, while the
behavior near the sphere is similar to the single boundary $d=1$ problem (since from close distance, the sphere looks like a hyperplane). The solution $\tilde u_R(r)$ interpolates between these limit behaviors.
Clearly it takes the scaling form
\bea
\tilde u_R(r) = \frac{1}{R^2} \tilde u(r/R) \label{scaling1} 
\eea
where we simply denote $\tilde u_{R=1}(r)=\tilde u(r)$ the solution with $R=1$. In $d=1$ we know its explicit form
\bea \label{d1} 
\tilde u(r)  = - \frac{3}{2} {\cal P}_0(\frac{1-r}{2}) \, \theta(1-r) - \frac{6}{(r-1)^2} \theta(r-1) 
\eea 
where ${\cal P}_0(z)$ is defined in \eqref{defP0} and satisfies the properties \eqref{propP0},
however we do not know the explicit solution for $d>1$.
Interestingly, the scaling form \eqref{scaling1} already leads to some information.
We will analyze two observables.\\


{\bf (i) The probability that an avalanche is confined within a ball of radius $R$
centered on its seed}, is given by choosing $x_s=0$. Calling $r_{\max}=\max_{x \in \Omega} |x-x_s|$ the 
maximal distance of any point of the avalanche to the seed we have
\bea
{\rm Prob}(r_{\max} < R) = e^{f \tilde u_R(x_s=0) } = e^{\frac{f}{R^2} \tilde u(0) }
\eea 
In $d=1$, and in the notations of the previous sections, 
it is $r_{\max}=\max(\ell_1,\ell_2)$. The PDF of $r_{\max}$, $P(r_{\max})$ for
a kick of strength $f$, and the associated density $\rho(r_{\max})$ are 
thus given by (we recall that $\tilde u(0)<0$)
\bea
&& P(r_{\max}) = f \frac{- 2 \tilde u(0)}{r_{\max}^3} \, e^{\frac{f}{r_{\max}^2} \tilde u(0) } \quad , \quad  \rho(r_{\max}) = \frac{-2 \tilde u(0)}{r_{\max}^3} 
\eea
To obtain the constant $\tilde u(0)$, we expand around $r=0$. One can write
\bea
\tilde u(r) = \tilde u(0) f_d(- r^2 \tilde u(0)) 
\eea 
with $f(0)=1$ and $f(z)$ satisfies 
\bea
- 4 z f''(z) -2 d f'(z) + f(z)^2 = 0
\eea 
The Taylor expansion $f(z)=\sum_{n \geq 0} f_n z^n$ obeys the recursion 
$f_{n+1} = \frac{1}{4(n+1)(n+ \frac{d}{2})} \sum_{p=0}^n f_p f_{n-p}$ with 
$f_0=1$. The lowest coefficients are $f^{d=1}_n=\left\{1,\frac{1}{2},\frac{1}{12},\frac{1}{72},\cdots \right\}$,
$f^{d=2}_n=\left\{1,\frac{1}{4},\frac{1}{32},\frac{1}{288},\cdots\right\}$,
$f_n^{d=3}= \left\{\frac{1}{6},\frac{1}{60},\frac{11}{7560},\cdots\right\}$.
One finds that the coefficients grow as $f_n \simeq 24 (n+1) a^{-(n+1)}$, leading to $f(z) \simeq 24 \frac{a}{(z-a)^2}$ near $z=a$. This matches the expected behavior \eqref{expected1},
$\tilde u(r) =  - 6 \frac{1}{(r-1)^2}$, near $r=1$, upon the choice $a=- \tilde u(0)$.
Obtaining the $f_n$ recursively we can thus determine $a$ and we obtain 
\bea
&& \tilde u(0)|_{d=1} \approx -8.84752 \quad , \quad \tilde u(0)|_{d=2}  \approx - 12.5634
\quad , \quad \tilde u(0)|_{d=3}  \approx - 15.7179
\eea
A check is obtained in $d=1$, where the exact value, from \eqref{d1}, is 
$\tilde u(0) = - \frac{3}{2} {\cal P}_0(\frac{1}{2},\Gamma(1/3)^{18}/(64 \pi^6))
=-8.8475159..$. \\

{\bf (ii) The probability that an avalanche with a seed at $r=r_s$ avoids the ball $B(0,R)$}.
To obtain this probability, we can first study the case of large $r_s$, 
and look for an expansion of $\tilde u(r)$ at large $r$, around the
leading term, from \eqref{expected1}, 
$\tilde u(r) \simeq_{r \to +\infty} - \frac{2(4-d)}{r^2}$. The equation for
the first subleading term is similar to the one studied in \eqref{GF2}
hence one finds anomalous exponents, as encountered above in Section \ref{shaped}.\\

It is useful instead to write $\tilde u(r)$, for $r>1$, in the form
\bea \label{form1} 
&& \tilde u(r) =-  \frac{2(4-d)}{r^2} (1 + v(A r^{2+b})) 
\eea
where $A>0$ is for now arbitrary. It leads to
\bea \label{eqdiffv} 
(b+2) z \left((b+d-4) v'(z)+(b+2) z v''(z)\right)+2 (d-4) v(z)^2+2 (d-4) v(z) = 0
\eea 
and we can look for a solution such that $v(z) \simeq z$ at small $z$. This implies
$4 (-4 + d) + b (-2 + b + d)=0$ which leads to $b=b_d^{\pm}$ 
as given by \eqref{exponent3}. Since we want $2+b<0$
the solution is
\bea \label{bvalue2}
b = b^-_d = \frac{1}{2} \left(2-d-\sqrt{(d-20) d+68}\right)
\eea 
which leads to $b=-3$ in $d=1$, $b= - 2 \sqrt{2}$ in $d=2$ and 
$b= -\frac{1}{2}(1+ \sqrt{17})$ in $d=3$. For $d \to 4^{-}$ we find $b \to -2^{-}$. 

It is useful to trade $d$ for $d= -b-\frac{8}{b+4}+6$, and call $v(z)=v_b(z)$ the 
solution. Then $v_b(z)$ satisfies
\bea \label{eqdiffv2} 
(b+2) (b+4) z^2 v_b''(z)-2 b \left(-z v_b'(z)+v_b(z)^2+v_b(z)\right) = 0  \quad , \quad v(0)=0 \quad v'(0)=1
\eea
where we can fix $v'(0)=1$ since $A$ is as yet undetermined.
It is then easy to generate the Taylor series $v(z)=z + \sum_{n\geq 2} v_n z^n$.
One finds e.g. $v_2=\frac{b}{b^2+7 b+8}$ and 
\bea \label{seriesv}
&& v(z)|_{d=2} = z + \frac{1}{17} \left(7+4 \sqrt{2}\right) z^2+\frac{1}{238} \left(59+41 \sqrt{2}\right)
   z^3+O\left(z^4\right) \\
&& v(z)|_{d=3}  = z+  \frac{1}{12} \left(5+\sqrt{17}\right) z^2+\frac{1}{177} \left(46+11 \sqrt{17}\right)
   z^3+O\left(z^4\right)
\eea 
The undetermined constant $A$ in \eqref{form1} will be such that 
$v(z)$ diverges at $z=A$, so that $\tilde u(r) \simeq  -6/(r-1)^2$
for $ r \to 1^+$. Extrapolation of the radius of convergence of the
series gives $A_{d=1} \approx 1.99$, $A_{d=2} \approx 1.97$, $A_{d=3} \approx 1.9$.
In $d=1$ one knows that $\tilde u(r)= \frac{-6}{(r-1)^2}$ exactly.
This is consistent with the above with $A=2$ and $v(z)=\frac{z(4-z)}{(2-z)^2}= z+\frac{3 z^2}{4}+\frac{z^3}{2}+\frac{5 z^4}{16}+O\left(z^5\right)$. \\

Hence, using the scaling form \eqref{scaling1},
we can summarize our results as follows. The probability that the avalanche (of support denoted $\Omega$)
with a seed at a distance $r=r_s>R$ from the point $0$, avoids the ball $B(0,R)$ is
\bea
 {\rm Prob}( B(0,R) 
 \cap \Omega = \emptyset) &=& e^{f \tilde u_R(r_s)} =
 e^{\frac{f}{R^2} \tilde u(r_s/R)} =   \exp\left( -  \frac{2 (4-d) f}{r_s^2} \left[1 + v\left(A (\frac{r_s}{R})^{2+b_d^-}
\right)  \right] \right) \quad , \quad r_s > R \\
 &\simeq & e^{-  \frac{6 f}{(r_s-R)^2} } \quad , \quad 0< r_s-R \ll R 
\eea 
where the exponent $b_d^-$, the function $v(z)$ and the constant $A$ depend 
only on the dimension $d$ and are given by \eqref{bvalue2}
and \eqref{eqdiffv2},\eqref{seriesv} (with $2+b_d^-<0$ and $A \approx 2$ was estimated above). 
Note that if we consider the limit of a small ball, $R \to 0$, we recover
the result of Section \ref{sub:point} which is thus validated.\\

Let us denote $r_{\min}=\min_{x \in \Omega} |x|$, $0 < r_{\min}<r_s$, the minimal distance 
between the origin $x=0$ and any point of the
avalanche (started at $r_s$) and denote its PDF as $P_{r_s}(r_{\min})$.
Then one has 
\bea
 {\rm Prob}( B(0,R)  \cap \Omega = \emptyset )  = {\rm Prob}(r_{\min} > R)
\eea
and taking a derivative w.r.t. $R$ gives the PDF, for any $R>0$
\bea \label{pdfrmin} 
P_{r_s}(r_{\min}=R) &=& - \frac{d}{dR} {\rm Prob}(r_{\min} > R) = f \rho_{r_s}(R) {\rm Prob}(r_{\min} > R) 
\eea
where $\rho_{r_s}(r_{\min}=R)= \partial_f P_{r_s}(r_{\min}=R)|_{f=0^+}$ is the density
\bea \label{pdfrmin2} 
 \rho_{r_s}(r_{\min}=R) &=& \frac{2 A (2+b_d^-)(d-4)}{R^{3+b_d^-} r_s^{-b_d^-}} \, v'\left(A (\frac{r_s}{R})^{2+b_d^-} \right) \quad , \quad 0< R<r_s \\
& \simeq & \frac{2A (2+b_d^-)(d-4)}{R^{3+b_d^-} r_s^{-b_d^-}}  \quad , \quad 0 < R \ll r_s \\
& \simeq & \frac{12}{(r_s-R)^3} \quad , \quad 0 < r_s-R \ll r_s
\eea 
where in $d=1$ the last expression is exact, and agrees with \eqref{PDFl2} 
since there $r_{\min}=r_s - \ell_2$. Note that for $1 \leq d <4$ the exponent $b_d^-<-2$ (it is given in \eqref{bvalue2}),
but that $3 + b_d^- \geq 0$ and
increases from $0$ to $1$ as $d$ increases from $d=1$ to $d=4$. The density of $r_{\min}$ thus
diverges both for $r_{\min}=R \to 0$ (weakly and for $d>1$) and for $r_{\min} \to r_s$. The second
divergence corresponds to small avalanches. The first divergence of the density of $r_{\min}$ with exponent $3+ b_d^-$ is presumably related to the probability that
a large avalanche in $d>1$ contains a "hole" around $x=0$ of size $r_{\min}$. 
Finally note that the probability of the event $r_{\min}>0$ is not unity but equals to
\be
\int_{0^+}^{+\infty} dr_{\min} P_{r_s}(r_{\min}) = e^{-  \frac{2 (4-d) f}{r_s^2}}= {\rm Prob}(S(0)=0) 
\ee 
as it should, i.e. the probability 
$P_{r_s}(r_{\min})$ has a delta function piece at $r_{\min}=0$ of amplitude
$1-e^{-  \frac{2 (4-d) f}{r_s^2}}$. 

\subsection{Span and shape in $d>1$}
\label{subsec:spanhigherd} 

Let us now study the shape in $d>1$ dimension conditionned to the
span, see Fig. \ref{fig:2dspan}. Let us decompose the coordinate as $x=(x_1,x_\perp)$ 
where $x_1 \in \mathbb{R}$ and $x_\perp \in \mathbb{R}^{d-1}$. 
We want to impose that the avalanche starting at the seed $x_s=(x^s_{1},x^s_\perp)$ does not reach the
level $x_1=r$ (with $x^s_1<r$). Let us use the (somewhat abusive) shorthand notation $S(r)$ 
\bea
S(r) \equiv S_{\rm tot}(r) = \int d^{d-1} x_\perp S(r,x_\perp)
\eea 
for the total size along the transverse space (in Fig. \ref{fig:2dspan} one has $d=2$ and $x_\perp=x_2$).
Consider a point $y=(y_1,y_\perp)$ below the level $r$, i.e. $y_1<r$,
and the local size of the avalanche $S(y)=S(y_1,y_\perp)$. 
We are interested in the PDF of $S(y)$, denoted $P(S(y) , S(r)=0)$, joint with the 
event $S(r)=0$, to ensure that
the level $r$ has not been reached. 
Its Laplace transform is 
\bea \label{507} 
\int dS(y) e^{\lambda S(y)} P(S(y) , S(r)=0) = e^{f \tilde u^\lambda(x_s)}
\eea 
where $\tilde u^\lambda(x)$ is the solution of 
\bea
\nabla_x^2 \tilde u(x) + \tilde u(x)^2 = - \lambda \delta(x_1-y_1) \delta^{d-1}(x_\perp-y_\perp) 
\quad , \quad \tilde u(r,x_\perp)=-\infty \quad , \quad \tilde u(-\infty,x_\perp) = 0
\eea 
on the strip $x_1 \in ]-\infty, r[$. We denote $\tilde u_r(x_1,x_\perp;y_1,y_\perp)$ the solution of this
problem. We will calculate only the mean shape, hence we look for a solution
perturbative in $\lambda$ as
\bea
\tilde u_r(x_1,x_\perp;y_1,y_\perp)=  - \frac{6}{(r-x_1)^2} + \lambda \tilde u_{1r}(x_1,x_\perp;y_1,y_\perp)
 + O(\lambda^2) 
\eea
Hence $\tilde u_1(x)= \tilde u_{1r}(x_1,x_\perp;y_1,y_\perp)$ is the solution of
\bea
\nabla_x^2 \tilde u_1(x) - \frac{12}{(r-x_1)^2} \tilde u_1(x) = -  \delta(x_1-y_1) \delta^{d-1}(x_\perp-y_\perp) 
\eea 
Let us Fourier transform w.r.t. the transverse space and define
\bea
\tilde u_{1r}(x_1,x_\perp;y_1,y_\perp) = \int \frac{d^{d-1}q}{(2 \pi)^{d-1}} 
e^{i q (x_\perp-y_\perp)}  \tilde u_{1r}(x_1,y_1;q)
\eea 
where $\tilde u_{1r}(x_1,y_1;q)$ is now solution of 
\bea
(\partial_{x_1}^2 - q^2) \tilde u_{1r}(x_1,y_1;q) - \frac{12}{(r-x_1)^2} \tilde u_{1r}(x_1,y_1;q) = 
-  \delta(x_1-y_1) 
\eea 
The independent solutions of the homogeneous equation depend only on $|q|$ and they are
\bea
&&  \sqrt{r-x_1} K_{7/2}(|q| (r-x_1)) \sim_{q \to 0} \frac{1}{|q|^{7/2} (r-x_1)^3} \\
&&  \sqrt{r-x_1} I_{7/2}(|q| (r-x_1)) \sim_{q \to 0} |q|^{7/2} (r-x_1)^4
\eea 
Hence we obtain the solution
\bea \label{resu1} 
&& \tilde u_{1r}(x_1,y_1;q) =  \tilde F_{|q|}(r-y_1,r-x_1) \theta(x_1<y_1<r) + \tilde F_{|q|}(r-x_1,r-y_1) \theta(y_1<x_1<r)  \\
&& \tilde F_q(a,b) = \sqrt{a b} I_{7/2}(q a) K_{7/2}(q b) \quad , \quad 0<a<b
\eea
which is a symmetric function in $(x_1,y_1)$, 
and one can check that it has the correct derivative 
discontinuity $[\tilde u_1]_{x_1=y_1^-}^{x_1=y_1^+}=-1$. 
The function $\tilde F_q(a,b)$ has the following asymptotics for small and large $q$
\bea
&& \tilde F_q(a,b) = \frac{1}{7} \frac{a^4}{b^3} \bigg( 
1+\frac{1}{90} q^2 \left(5 a^2-9 b^2\right)+\frac{q^4 \left(33 b^4-22 b^2 a^2+5
   a^4\right)}{3960}+O\left(q^{6}\right) \bigg) \\
&& = e^{-b q} \left( \frac{\cosh (a q)}{q} + \frac{6}{q^2}  ( \frac{\cosh (a q)}{b} -\frac{\sinh (a
   q)}{a} ) + O(\frac{1}{q^3}) \right)
\eea \\

Denoting $\ell_2= \max_{x \in \Omega} x_1$ the ``upper span'' (or one sided span),
see Fig. \ref{fig:2dspan}, we thus obtain by expanding \eqref{507} to $O(\lambda)$ 
%
\bea \label{eq1} 
&& \int dS(y) S(y) P( S(y) , \ell_2 < r-x_1^s ) 
= f e^{ - \frac{6 f}{(r-x_1^s)^2}} \int \frac{d^{d-1}q}{(2 \pi)^{d-1}}  e^{i q(x_\perp^s - y_\perp)} 
\tilde u_{1r}(x_1^s,y_1;q)  \nn
\eea 
where ${\rm Prob}(\ell_2 < r-x_1^s ) = e^{ - \frac{6 f}{(r-x_1^s)^2}}$, and 
$\rho(\ell_2)=\frac{12}{\ell_2^3}$, independently of the space dimension $d$,
a property of the BFM previously discussed. Thus we obtain the
shape conditioned to the span $\ell_2 < r-x_1^s$ as
\bea
\langle S(y_1,y_\perp) \rangle_{\ell_2<r-x_1^s} = f 
\int \frac{d^{d-1}q}{(2 \pi)^{d-1}}  e^{i q(x_\perp^s - y_\perp)} 
\tilde u_{1r}(x_1^s,y_1;q)
\eea 
where $\tilde u_{1r}(x_1,y_1;q)$ is given in \eqref{resu1}.
This gives the result \eqref{resSy1}, where 
here $x_s=(x^s_1,x_\perp^s)$ denote the position of the seed.
One can write also the result for an arbitrary kick $f(x,t)=f(x_1,x_\perp) \delta(t)$ 
with support in the strip $x_1<r$ as 
\bea
\langle S(y_1,y_\perp) \rangle_{\ell_2<r-x_1^s} = 
 \int dx_1 \frac{d^{d-1}q}{(2 \pi)^{d-1}} f(x_1,q) \tilde u_{1r}(x_1,y_1;q) e^{- i q y_\perp} 
\eea 
which allows to select some specific Fourier components $q$ 
by tuning the shape of the kick.\\

In two dimension, $d=2$, one can obtain a more explicit formula. 
Indeed the inverse Fourier transform of $\tilde F_{|q|}(a,b)$ reads
\bea
\int_{-\infty}^{+\infty} \frac{dq}{2 \pi} e^{i q c} \tilde F_{|q|}(a,b) = \frac{1}{2 \pi} \Phi(\frac{(a-b)^2 + c^2}{2 a b})
\quad , \quad 0<a<b
\eea 
with, for $z>0$, 
\be \label{defPhi}  \Phi(z)  = \frac{1}{6} \left(4-15 (z+1)^2\right)+\frac{1}{4} (z+1) (5 z (z+2)+2) \log
   (1+ \frac{2}{z}) 
\ee
This holds because of the following identity (Gradstein ET I 49 (47) pp 1113)
\bea
\int_0^{+\infty} dq K_\nu(a q) I_\nu(b q) \cos(c q) = \frac{1}{2 \sqrt{ab}} 
Q_{\nu-\frac{1}{2}}(\frac{a^2 + b^2 + c^2}{2 a b}) 
\eea 
where $Q_\nu$ is the associated Legendre function. Here we 
have $\Phi(z) \equiv Q_3(1+z)$ (with a proper continuation, however).
The small $z$ and large $z$ asymptotics of $\Phi(z)$ are
\bea
 \Phi(z)    &=&  \frac{1}{6} (-3 \log z -11+3 \log 2)+O(z) \\
   &=&  \frac{2}{35 z^4}-\frac{8}{35 z^5}+\frac{40}{63 z^6}-\frac{32}{21 z^7}+\frac{112}{33
   z^8}+O(\frac{1}{z^9}) 
\eea 

Hence we find in $d=2$ the explicit formula
\bea
\langle S(y_1,y_\perp) \rangle_{\ell_2<r-x_1^s} = \frac{f}{2 \pi} \Phi \left(\frac{(y_1-x_1^s)^2  + (y_\perp-x_\perp^s)^2}{2 (r-y_1)(r-x_1^s)} \right)
\eea 
a symmetric function of $(x_1^s,y_1)$, both smaller than $r$, and 
where $\Phi(z)$ is given in \eqref{defPhi}. It exhibits a logarithmic singularity at the
position of the seed (as was also found above in Section \ref{shaped}). One can
check (e.g. numerically) that upon integration over $y_\perp$, one recovers for the total
size along the transverse space, denoted somewhat abusively 
$S(y_1) = \int d^{d-1} x_\perp S(y_1,x_\perp)$, 
\bea
\langle S(y_1) \rangle_{\ell_2<r-x_1^s} = \frac{f}{7} \frac{(r-y_1)^4}{(r-x_1^s)^3} \theta(x_1^s<y_1<r) 
+ \frac{f}{7} \frac{(r-x_1^s)^4}{(r-y_1)^3} \theta(y_1<x_1^s<r)
\eea
i.e. as for the BFM in $d=1$, as expected.\\

Let us now apply the general procedure given in 
 \eqref{last} to obtain the mean shape at fixed ``upper span'' $\ell_2$.
 We obtain, focusing on the limit $f \to 0^+$
\bea \label{gengen} 
\langle S(y_1,y_\perp) \rangle_{\ell_2=r-x_1^s} 
= \frac{\ell_2^3}{12} \int \frac{d^{d-1}q}{(2 \pi)^{d-1}}  e^{i q(x_\perp^s - y_\perp)} 
\partial_r u_{1r}(x_1^s,y_1;q)|_{\ell_2=r-x_1^s} 
\eea 
which can be evaluated in any dimension using  \eqref{resu1}.\\

In $d=2$ one obtains an explicit expression. Denoting the distance of the measurement point $y$
to the seed as $R$ (remembering that we must set $x_1^s + \ell_2=r$ in \eqref{gengen})
\be
R^2=(y-x_s)^2= (y_1-x_1^s)^2  + (y_\perp-x_\perp^s)^2= (r-y_1-\ell_2)^2 + (y_\perp-x_\perp^s)^2
\ee
we have
\bea \label{shape2d} 
\langle S(y_1,y_\perp) \rangle_{\ell_2=r-x_1^s} 
= - \frac{\ell_2 (\ell_2+ r -y_1)}{24 (2 \pi) (r-y_1)^2} R^2  \Phi'(z)|_{z=\frac{R^2}{2 \ell_2 (r-y_1)}}
\eea 
with the following expression and small and large $z$ asymptotics
\bea \label{serphiprime} 
 - \Phi'(z) &=& 
\frac{(z+1) (15 z (z+2)+2)}{2 z (z+2)}
-\frac{3}{4} (5 z (z+2)+4) \log (1+ \frac{2}{z}) \\
& =& \frac{1}{2 z}+ 3 \log z+\frac{31}{4}-3 \log 2 +O(z \log z) \\
& =& \frac{8}{35 z^5}-\frac{8}{7 z^6}+\frac{80}{21 z^7}-\frac{32}{3 z^8}+O(\frac{1}{z^9})
\eea 

The mean shape conditioned to the ``upper span'' $\ell_2$, $\langle S(y_1,y_\perp) \rangle_{\ell_2}$, must vanish when $y_1$ reaches $y_1-x_1^s=\ell_2$ since there can be no motion for $y_1-x_1^s>\ell_2$.
We now study how it vanishes. Expanding \eqref{shape2d} 
for $y_1=x_1^s + \ell_2- D$, where $D$ is the distance to the edge we find
\bea
\langle S(y_1,y_\perp) \rangle_{\ell_2}\big|_{y_1=x_1^s + \ell_2- D} =
\frac{16  \ell_2^7}{105 \pi  \left(\ell_2^2+(y_\perp-x_\perp^s)^2\right)^4} \, D^3 -\frac{16 \ell_2^6
   \left(\ell_2^2-(y_\perp-x_\perp^s)^2\right)}{105 \pi 
   \left(\ell_2^2+(y_\perp-x_\perp^s)^2\right)^5} \, D^4 +O(D^5)
   \eea
and one can check that integrating over $y_\perp$ produces exactly the
result \eqref{res1b} for the $d=1$ BFM, i.e. $\frac{D^3}{21} - \frac{D^4}{28 \ell_2}  +O(D^5)$,
as it should. Hence the mean shape still vanishes along any point of the ``boundary''
$y_1=x_1^s + \ell_2$ with the cubic power, however the amplitude
depends on the ratio of the transverse to longitudinal distance to the seed.
Note that of course this is a mean effect, for each avalanche the point
where the maximum span is realized is somewhere on the boundary $y_1=x_1^s + \ell_2$
(finding the distribution of the position of this point is a more difficult problem). \\

Finally, note also the very simple result for the mean shape at the position of the seed,
setting $R \to 0$ and $r-y_1 \to \ell_2$ one finds from \eqref{shape2d} and \eqref{serphiprime}  
\bea
\langle S(x_1^s,x^s_\perp) \rangle_{\ell_2=r-x_1^s} 
= \frac{\ell_2^2}{24 \pi} 
\eea


\subsection{Probability that an avalanche does not hit a cone in two dimension} 
\label{subsec:cone} 

One can generalize the geometries of two of the previous sections, 
and ask what is the probability that an avalanche remains confined
inside a conical domain of apex $x=0$. Here we restrict to $d=2$, and use polar coordinates
$(r, \phi)$. Consider the conical domain, or wedge,
${\cal D} = (r,\phi) \in ]0,+\infty[ \times ]0,\phi_0[$
with an opening angle $0< \phi_0 \leq 2 \pi$.
The probability that an avalanche $\Omega$, with a seed inside the wedge
at $x=(r_s,\phi_s)$, 
remains strictly confined inside the wedge (i.e. is not allowed to touch the
two radial lines at $\phi=0,\phi_0$) is
\be
{\rm Prob}(\Omega \cap {\cal D}^c = \emptyset) = {\rm Prob}(S(r,0)=S(r,\phi_0)=0, r \in [0,+\infty[) 
= e^{f \tilde u(r_s,\phi_s) }
\ee
where $\tilde u(r,\phi)$ is the solution of the instanton equation
\bea \label{instantoncone}
(\partial_r^2 + \frac{1}{r} \partial_r + \frac{1}{r^2} \partial_\phi^2 ) \tilde u + \tilde u^2 =0 \quad , \quad 
\tilde u(r,0) = \tilde u(r,\phi_0) = - \infty \quad , \quad r \geq 0 
\eea 
which diverges on each boundary. The half plane case, $\phi_0= \pi$, corresponds
to the problem of the one-sided span considered in the previous section. 
Note that the opening angle of the wedge
can be either smaller than for the half plane problem, $\phi_0 \leq \pi$, or
larger $\pi < \phi_0 \leq 2 \pi$.  
The case $\phi_0=2 \pi$ is very interesting, as it requires the avalanche 
not to touch the positive axis, i.e. $x=(r \geq 0, \phi=0)$. 
In all cases one requires that the point $x=0$ is not visited by the avalanche, as
in Section \ref{sub:point}, but the
restriction is always stronger as the avalanche is also required not
to cross the boundaries of the wedge. Hence one has the bound 
\bea \label{boundwedge} 
{\rm Prob}(S(0)=0) > {\rm Prob}(S(x)=0 , x \in {\cal D}^c)
\eea 
which will be useful below.
\\

Given the symmetry of the problem we will search for a solution
of the form (the numerical factors have been introduced for convenience)
\bea
\tilde u(r,\phi) = - \frac{6}{r^2} (\frac{1}{3} + h(\phi)) 
\eea 
The condition to be a solution of \eqref{instantoncone} is 
\bea \label{wedgecond} 
h''(\phi) = 6 h(\phi)^2 - \frac{2}{3} 
\eea 
whose general real solution is expressed as a Weierstrass elliptic function
\bea
h(\phi) = {\cal P}(\phi; g_2=\frac{4}{3} , g_3)
\eea 
where $g_3$ is for now a free real parameter. 
There is also an arbitrariness $\phi \to \phi +c$, which we have fixed
since we must require from \eqref{instantoncone} that $h(0)=h(\phi_0)=+\infty$. \\

Let us recall (see Appendix) that the Weirstrass function is periodic
on the real axis with half-period denoted $\Omega(g_2, g_3)$.
It is smooth and bounded on its fundamental domain (one period interval) and
diverges positively on both ends, i.e. at $\phi=0^+$ as $\simeq 1/\phi^2$ 
and similarly at $\phi = 2 \Omega(g_2, g_3)$. We thus require
that 
\be \label{condOmega} 
\phi_0 = 2 \Omega(g_2=\frac{4}{3} , g_3) \Leftrightarrow g_3=g_3(\phi_0) 
\ee
which allows to determine the parameter $g_3=g_3(\phi_0)$ as a function of the wedge opening
angle $\phi_0$. We denote 
\be
h_{\phi_0}(\phi) = {\cal P}(\phi; g_2=\frac{4}{3} , g_3(\phi_0))
\ee 
the corresponding solution, which satisfies the required boundary conditions.
Note the symmetry $h_{\phi_0}(\phi_0-\phi)=h_{\phi_0}(\phi)$, with $h_{\phi_0}'(\phi_0/2)=0$.
The function $h_{\phi_0}(\phi)$ attains its minimum value for $\phi=\phi_0/2$. An equivalent 
condition to determine $g_3(\phi_0)$ is thus ${\cal P}'(\phi_0/2; g_2=\frac{4}{3} , g_3(\phi_0)) =0$.
It is useful to recall that
\be
\Omega(g_2, g_3) = \int_{e_1}^{+\infty} \frac{dt}{\sqrt{4 t^3 - g_2 t-g_3}} 
\ee
where $t=e_1$ is the largest positive root of $4 t^3 - g_2 t-g_3=0$. 
\\

We already know the exact solution in one case, $\phi_0=\pi$. Indeed
then $\tilde u(r,\phi)=- 6/y^2$. This corresponds to
\be
h_\pi(\phi) = - \frac{1}{3}  + \frac{1}{\sin^2 \phi}
\ee
which, as we can check, indeed satisfies the equation \eqref{wedgecond}. 
We can use the other property of the Weierstrass function,
$g_3=4 {\cal P}(z)^3-{\cal P}'(z)^2 - g_2 {\cal P}(z)= 4 h(\phi)^3 - h'(\phi)^2 - \frac{4}{3} h(\phi)$, 
to find the value of $g_3$ corresponding to that case, and 
one finds $g_3(\pi)=\frac{8}{27}$. The roots are $e_1=\frac{2}{3}$ and double roots $e_{2,3}= - \frac{1}{3}$
and in that case the period is $\phi_0=2 \Omega(g_2=\frac{4}{3},\frac{8}{27})=\pi$. \\

Let us reexamine the bound \eqref{boundwedge}. From \eqref{s00} in $d=2$ it implies that for all $\phi \in ]0,\phi_0[$
one must have 
\bea
h_{\phi_0}(\phi) > \frac{1}{3} 
\eea 
A necessary condition is thus
\bea
{\cal P}(\Omega;\frac{2}{3},g_3) = e_1 > \frac{1}{3} 
\eea 
which is equivalent to $g_3 > - \frac{8}{27}$. One can check that 
the point $g_3=- \frac{8}{27}$ corresponds to the point where the period diverges
and the solution there is $h_{-8/27}(\phi)= \frac{1}{3} + \frac{1}{\sinh^2(\phi)}$.
Hence the interval of variation of $g_3$ is included in $]- \frac{8}{27},+\infty[$. In
fact, as we see below $g_3$ varies in $]g_3^c,+\infty[$, where $g_3^c$ is
slightly larger than $- \frac{8}{27}$. \\

Let us start with the limit of a small opening angle $\phi_0 \ll 1$. 
One easily sees that it corresponds to large positive $g_3$. In the 
large $g_3$ limit one can neglect $g_2$ (i.e. set it to zero in a
first approximation) and one is back to our well studied 
function $h_{\phi_0}(\phi) \simeq {\cal P}_0(\phi)$. Hence we
obtain 
\bea
\phi_0 \simeq (\frac{g_3^0}{g_3})^{1/6}  \quad , \quad g_3 \simeq g_3^0/\phi_0^6 
\eea 
where we recall $g_3^0=\frac{\Gamma(1/3)^{18}}{(2 \pi)^6}$. 
In this regime the cone acts as two almost parallel boundaries, 
and the results are very similar to the $d=1$ problem studied above.\\

As the angle $\phi_0$ is increased from this limit the value of $g_3$
decreases. For some standard cases, we have solved numerically
the condition \eqref{condOmega} and we find
\bea
\phi_0 &=& \frac{\pi}{2} \quad , \quad \Omega= \frac{\pi}{4} \quad , \quad  g_3 \approx 52.665 
\quad , \quad h_{\phi_0}(\frac{\phi_0}{2}) = 2.40836 \\
\phi_0 &=& \frac{2 \pi}{3} \quad , \quad \Omega= \frac{\pi}{3} \quad , \quad  g_3 \approx 
8.59156
\quad , \quad h_{\phi_0}(\frac{\phi_0}{2}) = 1.37624 \\
\phi_0 &=& \frac{3 \pi}{2} \quad , \quad \Omega= \frac{3 \pi}{4} \quad , \quad  g_3 \approx - 0.2751955 
\quad , \quad h_{\phi_0}(\frac{\phi_0}{2}) = 0.403541 
\eea
We also find that the case $g_3=0$ corresponds exactly to 
\bea
\frac{\phi_0}{2} = \Omega(\frac{4}{3},0)=\sqrt[4]{3} K(-1) 
= \frac{\sqrt[4]{3} \Gamma \left(\frac{1}{4}\right)^2}{4 \sqrt{2 \pi }}
\quad , \quad h_{\phi_0}(\frac{\phi_0}{2}) =\frac{1}{\sqrt{3}} 
\eea 
that is $\frac{\phi_0}{\pi}=1.098430697$. Note that there may be spurious roots for $g_3$
which are not obtained continuously from this solution, hence can be eliminated.\\

%
%
%
%

Finally, the interesting case $\phi_0=2 \pi$ corresponds to the value $g_3=g_3^c$
which we find numerically as $g_3^c \approx - 0.295402 = - \frac{8}{27} + .0008941$.
It corresponds to the constraint that the avalanche does not intersect 
the positive axis. The solutions for $\frac{-8}{27} < g_3 < g_3^c$ do not seem
to correspond to physical cases, since $\phi_0$ cannot be larger than $2 \pi$.
The extreme case $g_3=\frac{-8}{27}$ in a loose sense corresponds 
to the weaker constraint of not visiting $x=0$ studied previously. 
Indeed, it has infinite period and upon some rescaling by this infinite period, one can eliminate
the term $1/\sinh^2(\phi)$, recovering $h(\phi) \simeq \frac{2}{3}$ and 
thus the solution $\tilde u(x)=- \frac{4}{r^2}$ valid in this case (for $d=2$). \\

Once the solution for the wedge $\phi_0$ is obtained one can easily obtain the solution 
for the "angular extensions", $\varphi_1,\varphi_2$ which are the angular analogs of $\ell_1,\ell_2$. 
We can define, for a seed at $x_s= (r,\phi_s=0)$ 
\be
\varphi_2= \max_{x=(r,\phi) \in \Omega} \phi \quad , \quad \varphi_1= - \min_{x=(r,\phi) \in \Omega} \phi 
\ee 
Then the CDF of $(\varphi_1,\varphi_2)$ is given by
\be \label{jcdf} 
{\rm Prob}( \varphi_1 <  \phi_1 , \varphi_2 < \phi_2) =
\exp\left( - \frac{6}{r^2} (\frac{1}{3} + h_{\phi_1+\phi_2}(\phi_1) \right)
\ee 
Taking derivatives one obtains the PDF, the density, and the PDF of the total angular
extension $\varphi =\varphi_1+\varphi_2$, analogous to $\ell$.

\acknowledgements

I thank C. Le Priol, A. Rosso, T. Thiery for collaborations on related topics. 
This work was supported initially by PSL grant ANR-10-IDEX-0001-02-PSL,
then by ANR grant ANR-17-CE30-0027-01 RaMaTraF.

\appendix

\section{Weierstrass and Elliptic functions}
\label{app:W} 

Here we recall  some properties of   Weierstrass's elliptic function $\mathcal{P}$
(source is Ref. \cite{AbramowitzStegun} chapter 18, and Wolfram Mathworld).
It appears in complex analysis as the only doubly periodic function on the complex plane with a double 
pole $1/z^2$ at zero\footnote{It also appears as the second derivative of the Green function of the free field on a torus.}.
Denoting $\omega_1,\omega_2$ the two (a priori complex) primitive half-periods, every point of the lattice $\Lambda= \{ 2 m \omega_1 + 2 n \omega_2 | (n,m) \in \mathbb{Z}^2 \}$ is a pole of order $2$ for $\mathcal{P}$. It 
can be constructed for $z \in \mathbb{C}-\Lambda$ as
\be
{\cal P}(z|\omega_1,\omega_2):= \frac{1}{z^2} +
\sum_{m,n \neq (0,0)} \frac{1}{(z- 2 m \omega_1 - 2 n \omega_2)^2} - \frac{1}{(2 m \omega_1+ 2 n \omega_2)^2}
\ee
It is an even function of the complex variable $z$, with ${\cal P}(z) = {\cal P}(-z)$. Note that the choice of primitive vectors $(2 \omega_1,2 \omega_2)$ is not unique, since one can alternatively choose any linear combination. 
The conventional choice of roots $g_2$ and $g_3$ is defined from its expansion around $z=0$,
\be  \label{g2g3}
{\cal P}(z|\omega_1,\omega_2)   = \frac{1}{z^2} + \frac{g_2}{20} z^2 + \frac{g_3}{28} z^4 + \mathcal{O}(z^6)\ .
\ee  
The function $\mathcal{P}$ is alternatively denoted ${\cal P}(z|\omega_1,\omega_2)  = {\cal P}(z;g_2,g_3)$,
the notation used here, and defined in Mathematica as $\text{WeierstrassP}[z,\{g_2,g_3\}]$.
The parameters $g_2,g_3$ are expressed from the half-periods 
\be
g_2= 60 \sum_{m,n \neq (0,0)}  \frac{1}{(2 m \omega_1 + 2 n \omega_2)^4} 
\quad , \quad g_3 =  140 \sum_{m,n \neq (0,0)}  \frac{1}{(2 m \omega_1 + 2 n \omega_2)^6}\ .
\ee 
The Weierstrass elliptic function verifies the homogeneity property,
\be \label{homo} 
\mathcal{P}(\lambda z; \lambda^{-4} g_2, \lambda^{-6} g_3) = \lambda^{-2} \mathcal{P}(z; g_2, g_3)\ ,
\ee
and the non-linear differential equation
\be \label{eqdiff1} 
{\cal P}'(z)^2 = 4 {\cal P}(z)^3 - g_2  {\cal P}(z) - g_3\ .
\ee
It is thus linked to elliptic integrals. Restricting now to $g_2,g_3 \in \mathbb{R}$ 
and focusing on $z \in \mathbb{R}$ one can   choose one half-period to be real, which we
denote   $\Omega$
\footnote{The conventions are such that if $\Delta<0$, $\Omega=\omega_1$ is real and 
$\omega_2$ imaginary (for $g_3>0$ and the 
reverse for $g_3<0$), and if $\Delta<0$, $\Omega= \omega_1 \pm \omega_2$. }.
The function ${\cal P}(z)$ is then periodic in $\mathbb{R}$ of period $2 \Omega$
and diverges at all points $2 m \Omega$, $m \in \mathbb{Z}$. It is  
defined in the fundamental interval $]0,2 \Omega[$,  repeated by 
periodicity. In this interval it satisfies the symmetry $\mathcal{P}(2 \Omega-z;g_2,g_3)=\mathcal{P}(z;g_2,g_3)$.
Its values in the first half-interval, i.e.\ for $z \in [0,\Omega]$ are such that (with $y \in [e_1,\infty]$)
\be 
z=\int_y^{\infty} \frac{dt}{\sqrt{4t^3 - g_2 t - g_3}}   \Leftrightarrow y = \mathcal{P}(z;g_2,g_3) \label{defint}
\ee
where $e_1$ is the largest real root of the polynomial in $t$
\be
4t^3 - g_2 t - g_3 = 4 (t-e_1) (t-e_2) (t-e_3)\ .
\ee
The roots $e_i$ are all real if $\Delta=g_2^3- 27 g_3^2 >0$ and 
only one, namely $e_1$, is real if $\Delta<0$. Hence the period is given by
\be \label{int}
\Omega = \int_{e_1}^{\infty}\!\!\!\frac{dt}{\sqrt{4 t^3 - g_2 t - g_3}}\,, \; {\cal P}(\Omega)=e_1\,, \; {\cal P}'(\Omega)=0\ .
\ee
It is always finite, except when $e_1$ is a double root, in which case $\Delta=0$
and the period is infinite $\Omega=\infty$. 

For $g_2=0$ the integral (\ref{int}) can be  calculated explicitly using
\be
\begin{split}
\int_1^{\infty} \frac{du}{\sqrt{u^3-1}} &= \frac{\Gamma(1/3)^3}{4^{\frac{2}{3}} \pi}  = 
 \frac{- \sqrt{\pi}\, \Gamma(1/6)}{\Gamma(- 1/3)}\ ,\\
\int_{-1}^{\infty} \frac{du}{\sqrt{u^3+1}} &= \sqrt{\pi}\frac{\Gamma(1/3)}{\Gamma(5/6)}\ . 
\end{split}
\ee
The half-periods are
\be \label{halfper} 
\Omega=\left\{
\begin{array}{ll}
\dfrac{1}{4 \pi} \Gamma(1/3)^3 g_3^{-1/6} &\;\text{ when }\;g_3 >0\\
\sqrt{\pi} \dfrac{\Gamma(1/3)}{4^{\frac{1}{3}} \Gamma(5/6)}  |g_3|^{-1/6}  &\;\text{ when }\;g_3 <0
\end{array}\right.\ ,
\ee 
and the other period can be chosen as $\frac{1}{2} \Omega (1+ i \sqrt{3})$. Finally, taking another derivative of (\ref{eqdiff1}) we see that the Weierstrass function also satisfies
\be  \label{second} 
{\cal P}'' (z)= 6 {\cal P}(z)^2 - \frac{g_2}{2} \ ,
\ee  
and $\mathcal{P}(z;g_2,g_3)$ is the only solution of this differential equation which diverges at $z=0$ as in (\ref{g2g3}). The Weierstrass function allows to find solutions to the 1D instanton equation in a region with no source
\be
\tilde u''(x) - A \tilde u(x) + \tilde u(x)^2 = 0\ ,
\ee
where $A=1$ is the massive case and $A=0$ the massless case. Comparing with Eq.~(\ref{second}) we see that a family of 
solutions are
\be  \label{soluP} 
\tilde u(x) = \frac{A}{2}-6 b^2 {\cal P} \left(c+b x;\frac{A^2}{12 b^4},g_3\right)\ .
\ee  
Because of the homogeneity relation (\ref{homo}), this is a two-parameter family.
These solutions are periodic.
In the massless case $A=0$, the period of (\ref{soluP}) is $\Omega/b$ where
$\Omega$ is given by (\ref{halfper}). 

{\bf Derivatives w.r.t. $g_2,g_3$}. There are formula for derivatives w.r.t. $g_2$ and $g_3$
(Mathematica). 
They simplify for $g_2=0$ and become
\be
 \partial_{g_3} {\cal P} = \frac{2   {\cal P} +z  {\cal P}'}{6 g_3} \quad , \quad  \partial_{g_3} {\cal P}' 
=  \frac{   {\cal P}' + 2 z  {\cal P}^2}{2 g_3} 
\ee
Note that the first equation, $2   {\cal P} +z  {\cal P}' - 6 g_3 \partial_{g_3} {\cal P} = 0$, simply expresses the scale invariance. 

{\bf Expansion in series of $g_3$}. Let us again specify to $g_2=0$. Using the definition \eqref{defint} and 
expanding the integrand in series of $g_3$ and integrating over $t$, one recognizes an hypergeometric
series and one can rewrite
\bea
z = \frac{\, _2F_1\left(\frac{1}{6},\frac{1}{2};\frac{7}{6};\frac{g_3}{4
   y^3}\right)}{\sqrt{y}}
\eea 
Inverting this series we obtain the series expansion of ${\cal P}$ in powers of $g_3$ as
\bea
{\cal P}(z,{g_2=0,g_3})= \frac{1}{z^2}+\frac{g_3 z^4}{28}+\frac{g_3^2 z^{10}}{10192}+\frac{g_3^3
   z^{16}}{5422144}+\frac{3 g_3^4 z^{22}}{9868302080}+O\left(z^{26}\right) \label{Pexp1} 
\eea
Note that it is the same as the expansion in powers of $z$. More generally it can be written as
\bea
{\cal P}(z,{g_2=0,g_3})= \sum_{p=0}^{\infty} c_p g_3^p z^{6 p - 2}  \label{Pexp2} 
\eea 
where $c_p$ satisfies the recursion
\bea
c_p = \frac{1}{(p-1)(6 p +1)} \sum_{k=1}^{p-1} c_k c_{p-k}  \label{Pexp3} 
\eea 
with intial condition $c_0=1$ and $c_1=1/28$.

\section{Levy stable laws and some inverse Laplace transforms}
\label{app:Levy} 

One sided Levy stable function ${\cal L}_\alpha(x)$, defined for $0< \alpha <1$ and $x>0$, are such that
\bea
{\rm LT}_{x \to p} \, {\cal L}_\alpha(x) = e^{- p^\alpha} ~~ 
\Leftrightarrow ~~ {\cal L}_\alpha(x) = {\rm LT}^{-1}_{p \to x}  \, e^{- p^\alpha}
\eea 
They satisfy the property that they are stable under convolution, up to
a rescaling
\bea
[ {\cal L}_\alpha \star {\cal L}_\alpha ](x) = 2^{-1/\alpha} {\cal L}_\alpha(2^{-1/\alpha} x) 
\eea 
which means that the (scaled) sum of two random variables $2^{-1/\alpha} (X+X')$ 
distributed with ${\cal L}_\alpha$ is also distributed with ${\cal L}_\alpha$, i.e. it is
infinitely divisible and attractor of all i.i.d. sums of positive random variables
which distribution decays as $1/X^{1+\alpha}$ at large $X$. 

Let us recall that a PDF is infinitely divisible if it can be expressed as the 
PDF of a sum of an arbitrary number of i.i.d random variables. 
Every infinitely divisible probability distribution corresponds in a natural way to a Levy process
(i.e. a process with stationary independent increments), and their characteristic
function are parameterized by the Levy-Kintchine formula. Stable laws are particular
cases. 

Their large $x$ asymptotics can be written as a series representation
\cite{Humbert}
\bea
{\cal L}_\alpha(x) = \frac{1}{\pi} \sum_{j=1}^\infty \frac{(-1)^{j+1}}{j! x^{1+\alpha j}}
\Gamma(1+ \alpha j) \sin(\pi \alpha j) 
\eea 
and the small $x$ asymptotics is obtained from the saddle-point method as
\bea
{\cal L}_\alpha(x) \sim_{x \to 0^+}  x^\frac{-2 + \alpha}{2(1-\alpha)} \exp \left( - 
(1-\alpha) (\frac{\alpha}{x})^{\frac{\alpha}{1-\alpha}} \right)
\eea 
Its Fourier transform has the form
\bea
\int dz e^{i k x} {\cal L}_\alpha(x) = \exp(- |k|^\alpha e^{- i \frac{k}{|k|} \frac{\pi \alpha}{2}} )
\eea 

There is a general expression of $g_\alpha(x)$ for generic 
rational values $\alpha=l/k$, $\ell < k \in \mathbb{N}$,
in terms of either (i) a Meijer $G$ function (ii) a sum of $k-1$ 
generalized hypergeometric functions, 
see formula (2) and (3-4) in \cite{Penson1}.

Let us recall some simplest and useful ones
\bea
{\cal L}_{1/2}(x) = \frac{1}{2 \sqrt{\pi} x^{3/2}} e^{-1/(4 x)} 
\eea 
known as the Levy-Smirnov function, then (see e.g. \cite{Barkai})
\bea
{\cal L}_{1/3}(x) = \frac{1}{3 \pi x^{3/2}} K_{1/3}(\frac{2}{\sqrt{27 x}}) 
= \frac{\text{Ai}\left(\frac{1}{(3 x)^{1/3}}\right)}{3^{1/3} x^{4/3}}
\eea 
in terms of the modified Bessel function of the second kind,
then
\bea
{\cal L}_{2/3}(x) = \frac{\sqrt{3}}{x \sqrt{\pi} } e^{-\frac{2}{27 x^2}} 
W_{1/2,1/6}(\frac{4}{27 x^2}) 
\eea 
in terms of the Whittaker $W$ function,
then
\bea
{\cal L}_{1/4}(x) = 
\frac{\Gamma \left(\frac{1}{4}\right) \,
   _0F_2\left(;\frac{1}{2},\frac{3}{4};-\frac{1}{256 x}\right)}{4 \sqrt{2} \pi 
   x^{5/4}}+\frac{\Gamma \left(-\frac{1}{4}\right) \Gamma \left(\frac{1}{4}\right) \,
   _0F_2\left(;\frac{3}{4},\frac{5}{4};-\frac{1}{256 x}\right)}{16 \sqrt{2} \pi
   ^{3/2} x^{3/2}}-\frac{\Gamma \left(-\frac{1}{4}\right) \,
   _0F_2\left(;\frac{5}{4},\frac{3}{2};-\frac{1}{256 x}\right)}{32 \sqrt{2} \pi 
   x^{7/4}}
\eea
and finally
\bea
{\cal L}_{3/4}(x) =
\frac{3 \, _2F_2\left(\frac{5}{6},\frac{7}{6};\frac{3}{4},\frac{5}{4};-\frac{27}{256
   x^3}\right)}{8 \sqrt{\pi } x^{5/2}}+\frac{\Gamma \left(\frac{7}{4}\right) \,
   _2F_2\left(\frac{7}{12},\frac{11}{12};\frac{1}{2},\frac{3}{4};-\frac{27}{256
   x^3}\right)}{\sqrt{2} \pi  x^{7/4}}+\frac{\Gamma \left(\frac{13}{4}\right) \,
   _2F_2\left(\frac{13}{12},\frac{17}{12};\frac{5}{4},\frac{3}{2};-\frac{27}{256
   x^3}\right)}{6 \sqrt{2} \pi  x^{13/4}}
\eea
which we have rechecked carefully. 
Note that the formula given in \cite{Montroll} for $\alpha=3/4$ contains
misprints. 

Note that the paper Ref. \cite{Penson2} contains some further useful ILT formula.
For instance it shows that if one knows $f(x) = LT^{-1}_{p \to x} F(p)$
then one can write
\bea
LT^{-1}_{p \to x} \, F(p^\alpha) = \int dt f(t) \frac{1}{t^{1/\alpha}} g_\alpha(x/t^{1/\alpha})
\quad , \quad  
LT^{-1}_{p \to x} \, p^{\alpha-1} F(p^\alpha) 
= \frac{x}{\alpha} \int dt f(t) \frac{1}{t^{1+ 1/\alpha}} g_\alpha(x/t^{1/\alpha})
\eea 
and a few more identities from
which one can deduce some ILT formula (for instance the one
of $p^{-1/4} e^{-p^{1/4}}$ and so on).

%

\section{Density of local sizes, details for the massive case}
\label{app:massivePS0} 

We give here some details of derivation for the massive case 
results of Section \ref{subsec:massivePS0}.

For a general driving kick $w(x)$ the general (formal) formula for the PDF of local size at $x=0$
is 
\bea
&& P_{\{w(x) \}}(S_0) = LT^{-1}_{-\lambda=-3 z(1-z^2) \to S_0} \exp( \int dx w(x) \frac{6 (1-z^2) e^{- |x|} }{(1 + z + (1-z) e^{-|x|})^2} )
\eea
Let us sketch first the two (known) easier cases. \\

{\it Uniform kick.} We have $\int dx \tilde u^\lambda(x) = 6(1-z)$. Hence
\bea
&& S_0 P_{w}(S_0) = \int_C \frac{d \lambda}{2 i \pi} e^{- \lambda S_0} \partial_\lambda e^{6 w  (1-z_\lambda)}  = \int_C \frac{d z}{2 i \pi} e^{- 3 z(1-z^2) S_0} \partial_z e^{6 w  (1-z)} 
\\ 
&& = - 6 w \int_C \frac{d z}{2 i \pi} e^{- 3 z(1-z^2) S_0 + 6 w  (1-z)} 
 = - 6 w \Phi(a,b,c)|_{a=9 S_0,b=0,c=-3 S_0 - 6 w} 
\eea 
leading to formula \eqref{mass1} in the text.

\medskip

{\it Local kick at $x=0$.} We have $\tilde u^\lambda(0) = \frac{3}{2}(1-z^2)$. Hence
\bea
&& S_0 P_{w}(S_0) = \int_C \frac{d \lambda}{2 i \pi} e^{- \lambda S_0} \partial_\lambda e^{w \frac{3}{2}(1-z^2)}  = \int_C \frac{d z}{2 i \pi} e^{- 3 z(1-z^2) S_0} \partial_z e^{w \frac{3}{2}(1-z^2)} \\
&&  = 3 w \int_C \frac{d z}{2 i \pi} e^{- 3 z(1-z^2) S_0} z e^{w \frac{3}{2}(1-z^2)} 
 = 3 w e^{\frac{3}{2} w} \partial_c \Phi(a,b,c)|_{a=9 S_0,b=- \frac{3}{2} w,c=-3 S_0} 
\eea
leading to formula \eqref{mass2} in the text. \\

We now use the formula for the avalanche density associated to a local kick
at $x_s$. It satisfies:
\bea
\int_0^\infty dS_0 (e^{\lambda S_0}-1) \rho_{x_s}(S_0) = u^\lambda(x_s) 
\eea 
We thus search for the inverse Laplace transform of $S_0 \rho_x(S_0)$ as
the contour integral:
\bea
&& S_0 \rho_x(S_0) = \int \frac{d\lambda}{2 i \pi} \partial_\lambda u^\lambda(x) e^{- \lambda S_0} = -  \int \frac{d\lambda}{2 i \pi} u^\lambda(x) \partial_\lambda e^{- \lambda S_0} 
 = \int \frac{dz}{2 i \pi} \frac{6 (1-z^2) e^{- |x|} }{(1 + z + (1-z) e^{-|x|})^2}  \partial_z e^{- 3 z(1-z^2) S_0} \nn \\
&& = 3 S_0 \int \frac{dz}{2 i \pi} \frac{6 (1-z^2) e^{- |x|} }{(1 + z + (1-z) e^{-|x|})^2}  (3 z^2-1) e^{- 3 z(1-z^2) S_0} \nn
\eea\\
We now exponentiate the denominator:
\bea
&&  \rho_x(S_0) = \int_0^\infty dt  f(t) \quad , \quad 
f(t) = - 18 t e^{- |x|} \int \frac{dz}{2 i \pi}  (1-z^2) (1- 3 z^2)  
e^{- t (1 + z + (1-z) e^{-|x|}) - 3 z(1-z^2) S_0} \nn
\eea 
The contour integral is an Airy type integral. It can be written as:
\bea
&& f(t) = - 18 t e^{- |x| - t (1 + e^{-|x|})} (1-\partial_b)(1- 3 \partial_b) \Phi(a,b,c)|_{a=9S_0, b=0,c=-3 S_0 - t(1-e^{-|x|}) } \nn
\eea 
we obtain
after a rescaling of $t$, $t \to t 3^{2/3} s_0^{1/3}/(1-e^{-|x|})$ 
\be \label{res1} 
\rho_x(S_0) = - 12 \frac{e^{-|x|}}{(1-e^{-|x|})^2} \frac{1}{S_0} 
 \int_0^\infty dt t e^{-t 3^{2/3} S_0^{1/3} \frac{1+e^{-|x|}}{1-e^{-|x|}}} \left[ 
 \frac{1}{2} t  (t - 2 \times 3^{1/3}  S_0^{2/3} ) 
Ai\left(t + 3^{1/3} S_0^{2/3} \right) + 
Ai'\left( t + 3^{1/3} S_0^{2/3} \right) \right]
\ee
which takes the scaling form \eqref{scalingS0} with
\bea
&& \phi(s,s_0) = - 12 s^{2/3} \int_0^{+\infty} dt t e^{  - 2 \times 3^{2/3} t s^{1/3}} 
 (\frac{t (t-2 s_0)}{2} {\rm Ai}(t+s_0) + {\rm Ai}'(t+s_0) ) \\
&& = - 12 s^{2/3} \int_0^{+\infty} dt e^{  - 2 \times 3^{2/3} t s^{1/3}} 
[ \frac{d}{dt}  (\frac{t^2}{2} {\rm Ai}'(t+s_0) ) - \frac{3}{2} s_0 t^2 {\rm Ai}(t+s_0)] 
\eea
transformed into \eqref{phiss0h} after integration by parts. 

To verify \eqref{sumrule1} we first integrate \eqref{res1} 
over $x$ using that $\frac{d}{dx} \frac{1+e^{-x}}{2(1-e^{-x})} = 
- \frac{e^{-x}}{(1-e^{-x})^2}$ and then we check the identity
\bea
  &&   - \frac{1}{S_0^{4/3}} 4 \sqrt[3]{3} \int_0^{+\infty} dt e^{-3^{2/3} \sqrt[3]{S_0} t} \left(\frac{1}{2} t \left(t-2
   \sqrt[3]{3} S_0^{2/3}\right) \text{Ai}\left(t+\sqrt[3]{3}
   S_0^{2/3}\right)+\text{Ai}'\left(t+\sqrt[3]{3} S_0^{2/3}\right)\right) \\
   && =  \frac{2 \times 3^{1/3}}{S_0^{4/3}} {\rm Ai}(3^{1/3} S_0^{2/3}) 
\eea

%
%


\section{Expansion in $\lambda$, more details} 
\label{app:lambda} 

We give some more details for the expansion in $\lambda$ of Section \ref{subsec:highermoments}.
We start from \eqref{great}, with $z=r-x_0$,
and substitute \eqref{Pexp3t} to obtain $\lambda$ as a series in $g_3$ (at fixed $z$).
Then we invert the series and we find $g_3$ as a series in $\lambda$ as
\bea
&& g_3=-\frac{2 \lambda }{3 z^3}-\frac{\lambda ^2}{252}-\frac{\lambda ^3 z^3}{22932}-\frac{5
   \lambda ^4 z^6}{7842744}-\frac{4379 \lambda ^5 z^9}{399666234240}-\frac{199 \lambda ^6
   z^{12}}{953050250880}-\frac{256124087 \lambda ^7
   z^{15}}{59920808579914044672} \\
   && -\frac{3501234283 \lambda ^8
   z^{18}}{37970870279061320939520}-\frac{10873895503273 \lambda ^9
   z^{21}}{5252130776999761912354406400}+O\left(\lambda ^{10}\right) \nn
\eea
We then use that $\rho=x_0 + 1/\sqrt{{\cal P}(z,g_3)}$ where we can insert the above series and
obtain 
\bea
&& \rho = r +\frac{\lambda  z^4}{84}+\frac{\lambda ^2 z^7}{3822}+\frac{53 \lambda ^3
   z^{10}}{8133216}+\frac{1223 \lambda ^4 z^{13}}{7136897040}+\frac{3223267 \lambda ^5
   z^{16}}{693820582640640}+\frac{407163391 \lambda ^6
   z^{19}}{3170413152376404480}\\
   && +\frac{346172136673 \lambda ^7
   z^{22}}{96192871373622013046784} 
    +\frac{857301071467 \lambda ^8
   z^{25}}{8416876245191926141593600}+\frac{1352811921954613 \lambda ^9
   z^{28}}{465688928893978889562090700800}+O\left(\lambda ^{10}\right) \nn
\eea 
From this we obtain both branches of the instanton solution to higher orders than in \eqref{bothbranches}, for $x_0>0$
\bea
&& \tilde u^\lambda_r(0) = - \frac{6}{\rho^2} = -\frac{6}{r^2}+\frac{\lambda  z^4}{7 r^3}+\frac{\lambda ^2 \left(16 r z^7-13
   z^8\right)}{5096 r^4}+\frac{\lambda ^3 \left(477 r^2 z^{10}-684 r z^{11}+247
   z^{12}\right)}{6099912 r^5} \\
   && +\frac{\lambda ^4 \left(273952 r^3 z^{13}-536220 r^2
   z^{14}+355680 r z^{15}-80275 z^{16}\right)}{133222078080 r^6}+O\left(\lambda ^5\right)
\eea 
and for $x_0<0$ 
\bea
&& \tilde u^\lambda_r(0)  = - 6 {\cal P}(r,0,g_3)=-\frac{6}{r^2}+\frac{\lambda  r^4}{7 z^3}+\lambda ^2
   \left(\frac{r^4}{1176}-\frac{r^{10}}{3822 z^6}\right)+\frac{\lambda ^3 \left(2
   r^{16}-19 r^{10} z^6+57 r^4 z^{12}\right)}{6099912 z^9} \\
   && +\frac{\lambda ^4 \left(-48 r^{22}+780 r^{16} z^6-5795 r^{10} z^{12}+18200 r^4 z^{18}\right)}{133222078080 z^{12}}+O\left(\lambda ^5\right)
\eea

%
%
%
%
%
%

\section{Expectation values conditioned to avalanche extensions} 

\label{sec:rel} 

Let us give some relations used in the text to calculate from the solution of the instanton equation some expectation values conditioned 
to avalanche (one sided) extensions $\ell_1$ and $\ell_2$ 

{\bf Single boundary}. Consider an observable $S_\phi=\int dx \phi(x) S(x)$ and 
$u_r^\lambda(x)$ the solution of the instanton equation with 
$\lambda \phi(x)$ as a source and a boundary at $r$. For a 
kick at $x=0$, we start from
\bea \label{rel01} 
\int dS_\phi P(\ell_2< r , S_\phi) e^{\lambda S_\phi}= e^{f u_r^\lambda(0)}  \quad , \quad
P(\ell_2< r) = e^{f u_r^0(0)} 
\eea 
This leads to the formula for the conditional average
\bea \label{B2} 
\langle e^{\lambda S_\phi} \rangle_{\ell_2<r} = 
\frac{\int dS_\phi P(\ell_2< r , S_\phi) e^{\lambda S_\phi}}{P(\ell_2<r)}
= e^{f (u_r^\lambda(0)- u_r^0(0))}
\eea 
One can also write, taking a derivative of \eqref{rel01} w.r.t. $r$
\bea
\int dS_\phi P(\ell_2, S_\phi) e^{\lambda S_\phi}= f \partial_r u_r^\lambda(0) e^{f u_r^\lambda(0)}|_{r=\ell_2}
\quad , \quad P(\ell_2) = f \partial_r u_r^0(0) e^{f u_r^0(0)}|_{r=\ell_2}
\quad , \quad \rho(\ell_2) = \partial_r u_r^0(0)|_{r=\ell_2} \label{beforelast} 
\eea 
leading to the conditional average
\bea
\langle e^{\lambda S_\phi} \rangle_{\ell_2=r} = 
\frac{\int dS_\phi P(\ell_2 , S_\phi) e^{\lambda S_\phi}}{P(\ell_2)}|_{\ell_2=r} 
= \frac{\partial_r u_r^\lambda(0) }{\partial_r u_r^0(0)} \langle e^{\lambda S_\phi} \rangle_{\ell_2<r}
\eea 
Expanding to first order in $\lambda$, denoting $u_r^{(1)}=\partial_\lambda u_r^\lambda|_{\lambda=0}$,
one obtains
\bea
&& \langle S_\phi \rangle_{\ell_2< r} = f u_r^{(1)}(0) 
\\
&& \langle S_\phi \rangle_{\ell_2=r}
= \frac{\partial_r u_r^\lambda(0) }{\partial_r u_r^0(0)} e^{f (u_r^\lambda(0)- u_r^0(0))} 
= \frac{\partial_r u_r^{(1)}(0) }{\partial_r u_r^0(0)}
+   f u_r^{(1)}(0) = \frac{\partial_r u_r^{(1)}(0) }{\partial_r u_r^0(0)}
+  \langle S_\phi \rangle_{\ell_2< r}  \label{last}
\eea

{\bf Two boundaries}. Now we start from
\bea
\int dS_\phi P(\ell_1 < -r_1, \ell_2< r_2 , S_\phi) e^{\lambda S_\phi}= e^{f u_{r_1,r_2}^\lambda(0)}  \quad , \quad
P(\ell_1 < -r_1, \ell_2< r_2) = e^{f u_{r_1,r_2}^0(0)} 
\eea 
This leads to the formula for the conditional average
\bea
\langle e^{\lambda S_\phi} \rangle_{\ell_1 < -r_1, \ell_2< r_2} = 
\frac{\int dS_\phi P(\ell_1 < -r_1, \ell_2< r_2 , S_\phi) e^{\lambda S_\phi}}{P(\ell_1 < -r_1, \ell_2< r_2)}
= e^{f (u_{r_1,r_2}^\lambda(0)- u_{r_1,r_2}^0(0))}
\eea 

One can also write, taking two derivatives
\bea \label{complic} 
&& \int dS_\phi P(\ell_1,\ell_2, S_\phi) e^{\lambda S_\phi}= - ( f \partial_{r_1} 
\partial_{r_2} u_{r_1,r_2}^\lambda(0) + f^2  \partial_{r_1} u_{r_1,r_2}^\lambda(0) 
\partial_{r_2} u_{r_1,r_2}^\lambda(0) ) e^{f u_{r_1,r_2}^\lambda(0)}|_{r_1=-\ell_1, r_2=\ell_2} 
\eea
from which one gets $P(\ell_1,\ell_2)$ setting $\lambda=0$ and taking
a derivative w.r.t $f$ at $f=0^+$ the density
\bea
\rho(\ell_1,\ell_2) = - \partial_{r_1} \partial_{r_2} u_{r_1,r_2}^0(0)|_{r_1=-\ell_1, r_2=\ell_2}
\eea 
Since Eq. \eqref{complic} is a complicated formula we will only study its small $f$ behavior. 
In the limit $f=0^+$ one finds
\bea
\langle e^{\lambda S_\phi} \rangle_{\ell_1,\ell_2} = 
\frac{\int dS_\phi P(\ell_1,\ell_2 , S_\phi) e^{\lambda S_\phi}}{P(\ell_1,\ell_2)}
= \frac{\partial_{r_1} \partial_{r_2} u_{r_1,r_2}^\lambda(0) }{\partial_{r_1} \partial_{r_2} u_{r_1,r_2}^0(0)} |_{r_1=-\ell_1,r_2=\ell_2}
\eea

\section{expansion around $\lambda^*$}
\label{app:lambdastar} 

Here we push the expansion around $\lambda^*$ in \eqref{fls} to higher orders. 
The WeierstrassP function near a zero has a simpler expansion in $g_3$ 
\bea
&& {\cal P}(1;0,g_3)=\frac{g X}{6}-\frac{g^2 X}{72}+\frac{g^3 X}{1296}+\frac{g^4 X (12 X+65)}{31104}-\frac{g^5 (X (96 X+529))}{186624}+\frac{7 g^6 X (516
   X+2845)}{6718464}+O\left(g^7\right) \nn \\
   && g = \frac{g_3-g_3^*}{g_3^*} \quad , \quad X= {\cal P}(1;0,g_3^*)= - \sqrt{- g_3^*} 
\eea 
From this we can solve order by order in $\lambda ^*-\lambda$ the equation \eqref{cond4} (here we
denote $\tilde \lambda \to \lambda$ for notational simplicity).
We find 
\bea
A(\lambda) = \frac{(\lambda ^*-\lambda)^{1/2} }{3 \sqrt{2}}-\frac{1}{54} \left(\lambda ^*-\lambda \right)+\frac{\left(5 \lambda ^*-54\right) \left(\lambda
   ^*-\lambda \right)^{3/2}}{972 \sqrt{2} \lambda ^*}+\left(\frac{1}{54 \lambda ^*}-\frac{2}{2187}\right) \left(\lambda ^*-\lambda
   \right)^2+O\left(\left(\lambda ^*-\lambda \right)^{5/2}\right)
\eea
which gives
\bea
f(\lambda)=\frac{9 \lambda ^*}{\sqrt{2} \sqrt{\lambda ^*-\lambda }^3}+\frac{\lambda ^*-54}{12 \sqrt{2} \sqrt{\lambda ^*-\lambda }}+\left(1-\frac{2 \lambda
   ^*}{81}\right)+\frac{\left(\lambda ^* \left(35 \lambda ^*-1188\right)-1620\right) \sqrt{\lambda ^*-\lambda }}{2592 \sqrt{2} \lambda
   ^*}+O\left(\sqrt{\lambda ^*-\lambda }^2\right)
\eea 
This leads to a series expansion for $p(\sigma)$, not displayed here.

\section{Bessel functions}
\label{sec:Bessel}
We use in the text the explicit form of the half-integer modified Bessel functions
\be
I_{7/2}(x)=\frac{\left(-\frac{30}{x^3}-\frac{12}{x}\right) \sinh (x)+\left(\frac{30}{x^2}+2\right)
   \cosh (x)}{\sqrt{2 \pi } \sqrt{x}} \quad , \quad I_{5/2}(x)=
\frac{\left(\frac{6}{x^2}+2\right) \sinh (x)-\frac{6 \cosh (x)}{x}}{\sqrt{2 \pi }
   \sqrt{x}}
\ee

\section{Shape with two boundaries, expansions in powers of $\lambda$}
\label{app:2bshape} 

Here we obtain 
\eqref{mainresult210}, \eqref{mainresult220} by a direct expansion of
\eqref{instsolu11} and \eqref{selfc1} in powers of $\lambda$. We define
\bea
&& H(z) = \partial_{g_3}|_{g_3=g_3^0} {\cal P}(z; g_3) 
= \frac{2   {\cal P}_0(z)+z  {\cal P}_0'(z)}{6 g_3^0} \quad , \quad  H'(z) = \partial_{g_3}|_{g_3=g_3^0} {\cal P}'(z; g_3) =  \frac{   {\cal P}_0'(z)+ 2 z  {\cal P}_0(z)^2}{2 g_3^0} 
\eea 
Note that $H(z)$ satifies the differential equation $H''(z) = 12  {\cal P}_0(z) H(z)$.
Using the symmetries ${\cal P}_0(z_0)={\cal P}_0(1-z_0)$ and ${\cal P}_0'(z_0)= - {\cal P}_0'(1-z_0)$
one also shows the "Wronskian"-type identity
\bea
H'(1-z_0) H(z_0) + H'(z_0) H(1-z_0) = \frac{1}{12 (g_3^0)^2} (4 {\cal P}_0(z_0)^3 - {\cal P}_0'(z_0)^2 ) 
= \frac{1}{12 g_3^0} \label{identity} 
\eea 
which we use below. Let us now expand to first order in $\lambda$. We denote $\delta g_3^{\pm}=g_3^{\pm}-g_3^0$,
hence from \eqref{selfc1} we have to first order
\bea
&& H(z_0) \delta g_3^{-}  = H(1-z_0) \delta g_3^+  \quad , \quad H'(z_0) \delta g_3^-  + H'(1-z_0) \delta g_3^+  = - \hat \lambda
\eea
Solving for $\delta g_3^\pm$, we can insert the solution in the expansion of the instanton solution
\eqref{instsolu11} to first order in $\lambda$. For $x_0>0$ we find
\bea
 \tilde{u}^{\lambda, x_0}_{r_1,r_2}(0) 
&=& \frac{-6}{(r_1-r_2)^2}  {\cal P}(1-p, g_2=0, g_3^-)  \quad , \quad 1-p = - \frac{r_1}{r_2-r_1}  \\
& =& \frac{-6}{(r_1-r_2)^2} \bigg( {\cal P}_0(1-p) - \frac{\lambda}{6}
(r_2-r_1)^3
\frac{H(1-z_0) H(1-p)}{H'(1-z_0) H(z_0) + H'(z_0) H(1-z_0)} \bigg) \\
& = &\frac{-6}{(r_1-r_2)^2} \bigg( {\cal P}_0(1-p) - 2 g_3^0 \lambda 
(r_2-r_1)^3 H(1-z_0) H(1-p) \bigg)
\eea 
where we have used \eqref{identity}. Similarly for $x_0<0$ we obtain
\bea
&& \tilde{u}^{\lambda, x_0}_{r_1,r_2}(0)  = \frac{-6}{(r_1-r_2)^2} \bigg( {\cal P}_0(p) - 2 g_3^0 \lambda 
(r_2-r_1)^3 H(z_0) H(p) \bigg) \quad , \quad p =  \frac{r_2}{r_2-r_1} 
\quad , \quad x_0<0 
\eea 
Now we see, using $H(z)=\psi_1(z_0)/(6 g_3^0)$, that these equations
lead to \eqref{mainresult210}, \eqref{mainresult220}.\\

%
%
%
%
%
%
%

\section{Higher conditional moments of the total size} \label{app:higher} 

Here we give the second moment $\langle S^2  \rangle_{\ell_1<- r_1, \ell_2 < r_2}$. Performing
the expansion as explained in the text we find, in the small $f$ limit, with $p=r_2/(r_2-r_1)$ the
aspect ratio
\bea
&& \langle S^2  \rangle_{\ell_1<- r_1, \ell_2 < r_2} = f \partial^2_\mu|_{\mu=0} \tilde u_{r_1,r_2}^\mu(0) = 2 f (r_2-r_1)^6  g_2(p) \\
&& g_2(p) = g_2(1-p)= \frac{8 \pi ^6}{81 \Gamma \left(\frac{1}{3}\right)^{36}}
\times \bigg( 9 \Gamma \left(\frac{1}{3}\right)^{18} + 256 \pi ^6 \bigg(P_0(p){}^2 \left(-12 \pi ^2 p^2-9 \zeta _0(p){}^2+4 \sqrt{3} \pi  \left(3 p
   \zeta _0(p)-2\right)\right)\\
   && +P_0(p) \left(\left(6 \sqrt{3} \pi  p-9 \zeta _0(p)\right)
   P_0'(p)+28 \pi ^2\right)+4 \pi  \left(2 \pi  p-\sqrt{3} \zeta _0(p)\right) P_0'(p)-9
   P_0(p){}^3\bigg)
 \bigg) \\
&& = \frac{32 \pi ^8 p^4}{63 \Gamma \left(\frac{1}{3}\right)^{18}}-\frac{p^6}{1800}+\frac{\pi 
   p^8}{1540 \sqrt{3}}-\frac{\pi ^2 p^{10}}{22932}+O\left(p^{11}\right)
\eea 
This extends the result \eqref{cumcondtot} in the text to the second moment.
Similar manipulations as in the text allow to obtain from it 
$\langle S^2  \rangle_{\ell,p}$ and $\langle S^2  \rangle_{\ell}$.

\section{Driving at the tip, local jumps $P_w(S_0)$, massive case.} 
\label{app:tiplocal} 

Set $\mu=0$, i.e. $\beta=1$ in \eqref{inv1}, \eqref{syst1}, in the massive case $A=1$. One obtains
\bea
S_0 P_w(S_0) = \int \frac{d \lambda_0}{2 i \pi} e^{- \lambda_0 S_0} \partial_{\lambda_0} e^{ w Z} 
\eea 
where $Z =  \frac{3}{2} m_0^2 (1- z^2) $ and
$\lambda_0-Z= 3 z (1-z^2)$. Hence we can trade the variable $\lambda_0$ for the variable $z$ and write
\bea
S_0 P_w(S_0) &=& 
\int \frac{dz}{2 i \pi} e^{- S_0 [ \frac{3}{2} m_0^2  (1-  z^2)  + 3  z (1-z^2)]  } 
\partial_{z} e^{ \frac{3}{2} w m_0^2  (1-z^2) } \\
& = & 3 w m_0^2 e^{m_0^2 \frac{3}{2}  (w-S_0)} 
\int \frac{dz}{2 i \pi}  z e^{- 3 z (1-z^2) S_0 + m_0^2 (S_0-w) \frac{3}{2} z^2 }  \\
& = & 3 w m_0^2 e^{m_0^2 \frac{3}{2} (w-S_0)} 
\partial_c \Phi(a,b,c)|_{a=9 S_0, b= \frac{3}{2} m_0^2 (S_0-w) ,
c= - 3 S_0 } 
\eea 
This leads to
\bea \label{PS001} 
P_w(S_0) &=& \frac{w m_0^2}{3^{1/3} S_0^{5/3}}  e^{\frac{m_0^6 (S_0-w)^3}{36 S_0^2} + m_0^2 (w-S_0) } 
(- y {\rm Ai}(y^2+ 3^{1/3} S_0^{2/3}) - {\rm Ai}'(y^2+ 3^{1/3} S_0^{2/3}) ) \quad , \quad 
y = \frac{m_0^2 (S_0-w)}{2 \times 3^{1/3} S_0^{2/3}} \nn \\
&&
\eea 
which is the analog of \eqref{PS00} for the massive case. 
We have checked that it recovers the moments given in Section \ref{sec:moments}.
Note that in the limit
$m_0 \to +\infty$ the PDF \eqref{PS001} becomes a delta-function around $S_0=w$.

\section{Mean shape $d>1$ in the massive case and the Green function of the $1/r^2$ potential}
\label{app:msmass} 

In Section \ref{shaped} we studied the mean shape $\langle S(x) \rangle_{S(x_0)=0}$ around a
point such that $S(x_0)=0$ for the massless case. To study the massive case 
we must solve \eqref{GF2} with an additional mass term $- m^2 \tilde u_1$ on the l.h.s.
The equation \eqref{inst00} is still correct with the replacement 
$G_{\ell + \frac{d-2}{2}}(r,r') \to G_{\ell + \frac{d-2}{2}}(r,r',E)$ where, instead of \eqref{GreenRadial}, we need the Green function at finite 
energy $E$, solution of 
\bea
\left(\frac{d^2}{dr^2} + E  - \frac{\alpha+ ( \ell + \frac{d-2}{2})^2 - \frac{1}{4} }{r^2} \right) 
G_{\ell + \frac{d-2}{2}}(r,r',E) = - \delta(r-r')
\eea 
where we parametrize $E = - q^2 \leq 0$, with here $q=m$, 
and $\alpha=4(4-d)>0$. The solution is 
%
%
\bea
&& \label{massmass} 
G_{\ell + \frac{d-2}{2}}(r,r',E=-q^2) = \sqrt{r r'} ( I_{s_\ell}(q r) K_{s_\ell}(q r') \theta(r'-r)
+ K_{s_\ell}(q r) I_{s_\ell}(q r') \theta(r-r') ) \\
&& s_\ell=\sqrt{\alpha + (\ell+\frac{d-2}{2})^2} \nn
\eea 
which has the right derivative discontinuity at $r=r'$. Note that for $d=1$ one 
has $s_0=s_1=7/2$, and one finds the Bessel with index $7/2$ again.
One then obtain the mean shape in the massive case from
from $\langle S(r,\phi) \rangle_{S(0)=0} =  f 
u_1(r_s,\phi_s,r,\phi)$ and inserting \eqref{massmass} into \eqref{inst00}.

\end{widetext}

\end{document}